\documentclass[11pt]{article}
\usepackage[utf8]{inputenc}
\usepackage{geometry}
\usepackage{amsmath}
\usepackage{amssymb}
\usepackage{graphicx}
\usepackage{graphbox}
\usepackage{bm}
\usepackage{mathtools}
\usepackage{hyperref}
\usepackage{tikz}
\usepackage{setspace}
\usepackage[round]{natbib}
\usepackage{subcaption}
\usepackage{bbm}
\usepackage{mathbbol}
\usepackage{cleveref}
\usepackage{enumitem}
\usepackage{tabularx}
\usepackage{pdfpages}

\usetikzlibrary{arrows, automata}
\usepackage{fullpage}

\newcommand{\E}{\mathbbm{E}}
\newcommand{\R}{\mathbbm{R}}

\DeclareMathOperator*{\argmax}{arg\,max}
\DeclareMathOperator*{\argmin}{arg\,min}
\DeclareMathOperator{\var}{var}
\DeclareMathOperator{\cov}{cov}
\DeclareMathOperator{\cor}{cor}

\title{Comparing multilevel and fixed effect approaches in the generalized linear model setting\thanks{This research project is affiliated with the Los Angeles Education Research Institute and has benefited from its research and partnership infrastructure. He Bai was supported by the Emmy Noether Memorial Fellowship and Richter Funds, both through Reed College. Asa Ferguson was supported by 
the Paul K. Richter \& Evalyn Elizabeth Cook Richter Memorial Fund through the Reed College Science Research Fellowship award. We also thank Noah Greifer, Ian Lundberg, Nianbo Dong, Meredith Phillips, Mollie Rudnick, Emilio Chavez Herrerias, the Los Angeles Unified School District's Office of Research and Program Evaluation, and anonymous reviewers for their valuable comments and suggestions. 
} 
}
\date{\today}

\author{
	He Bai\thanks{University of Massachusetts Amherst; Email: \href{mailto:hbai@umass.edu}{hbai@umass.edu}}
		\ \ \ \ \ 
	Asa Ferguson\thanks{
		Email: \href{mailto:fergusonasaw@gmail.com}{fergusonasaw@gmail.com}
  }  
		\ \ \ \ \ 
	Leonard Wainstein\thanks{Assistant Professor, Reed College; Email: \href{mailto:lwainstein@reed.edu}{lwainstein@reed.edu}} 
		\ \ \ \ \ 
	Jonathan Wells\thanks{Assistant Professor, Grinnell College; Email: \href{mailto:wellsjon@grinnell.edu}{wellsjon@grinnell.edu}}
} 

\begin{document}

\setcounter{page}{0}
\maketitle

\thispagestyle{empty}

\vspace{-0.15in}

\begin{abstract}
We extend prior work comparing linear multilevel models (MLM) and fixed effect (FE) models to the generalized linear model (GLM) setting, where the coefficient on a treatment variable is of primary interest. This leads to three insights. (i) First, as in the linear setting, MLM can be thought of as a regularized form of FE (RegFE). This explains why group-level confounding can greatly bias MLM's treatment coefficient estimates. However, unlike the linear setting, there is not an exact equivalence between MLM, fit through (approximate) maximum likelihood estimation, and RegFE in GLMs. (ii) Second, we study a generalization of ``bias-corrected MLM" (bcMLM) to the GLM setting, and a corresponding ``bias-corrected RegFE" (bcRegFE). None of FE, bcMLM, or bcRegFE entirely solve MLM's bias problem in GLMs, but bcMLM and bcRegFE tend to show less bias than does FE. (iii) Third, as in the linear setting, MLM's default standard errors can misspecify the true intragroup dependence structure in the GLM setting, which can yield downwardly biased standard errors. A cluster bootstrap is a more agnostic alternative. We also consider a cluster-robust standard error for (bc)RegFE. Ultimately, for non-linear GLMs, we recommend bcMLM for estimating the treatment coefficient, and a cluster bootstrap for standard errors and confidence intervals. If a bootstrap is not computationally feasible, then we recommend bcRegFE with cluster-robust standard errors, or FE with cluster-robust standard errors when group sizes are larger. We apply our recommendations to estimate the effects of taking a mathematics  course in twelfth grade for students in the Los Angeles Unified School District. Our \href{https://github.com/lwainstein/mlmfe}{\texttt{mlmfe}} package easily implements our recommendations in the \texttt{R} programming language.
\end{abstract}

 \normalsize
\vspace{0.2in}

\pagebreak
\clearpage

\onehalfspacing

\newpage
\setcounter{page}{1}

\section{Introduction}\label{sec:intro}

Investigators are often confronted with data in which the observations are grouped. For example, data may describe high school students (the observations), who are clustered in schools (the groups). Or data may be collected via multilevel sampling or via panel or longitudinal data wherein observations are recorded for the same subject (e.g., a student) across multiple time periods (e.g., grades in school). This grouped data is also referred to as clustered, multilevel,  hierarchical, panel, longitudinal, or cross-sectional data. Often with this type of data, researchers are interested in estimating the effect of a ``treatment" that varies within groups. In the students-within-schools example, this treatment might be a particular class or academic program that some, but not all, students in each school are enrolled in, and investigators may be interested in estimating the effect of this treatment on student outcomes (e.g., high school graduation, GPAs, or credit accumulation). Analyzing such an effect in multilevel data poses two challenges: one of estimation and one of inference. The first, of estimation, is that it is essential to account for group-level confounding in the relationship between the treatment and the outcome of interest. For example, students in certain schools may have more access to the treatment of interest than in other schools due to school resources. Not controlling for school then risks biasing the estimated treatment effect. The second challenge, pertaining to inference, is that grouped data violates a common assumption of independence between observations -- for example, outcomes of students in the same school are likely more similar to each other than are outcomes of students from different schools. Ignoring this can lead to standard error estimates that are too small. 

Researchers often choose between two approaches to tackle these challenges: fixed effects (FE) and multilevel models (MLM). In the fixed effects approach, models may include group-level and freely varying parameters (called fixed effects) to account for group-level confounders. This approach then deals with potential dependence of observations through the choice of a variance estimator that accounts for the specific type of intragroup dependency the user believes to exist. One such variance estimator, which we give focus to here, is the ``cluster robust standard error" (\citealp{white1984asymptotic}). On the other hand, multilevel models may include the same group-level parameters as are included in a fixed effects model, but they are not freely varying. Instead, they are treated as observed values of random variables, called random effects. The distribution of these random effects provides an intragroup dependence structure that is reflected in the typical standard error estimates obtained through maximum likelihood estimation (MLE), which is also often the default method for standard error estimation for MLM.

Both of these approaches are long-standing, but \cite{hazlett2022understanding}, henceforth referred to as H\&W, showed in a review of 109 articles published from 2017 to 2019 in top education, political science, and sociology journals that there were still clear misunderstandings (as of 2019) across the applied sciences about the usage and appropriateness of MLM and FE in a given setting.\footnote{H\&W's review included the American Education Research Journal (28 articles), Educational Evaluation and Policy Analysis (8), the American Journal of Political Science (17), the American Political Science Review (13), the Journal of Politics (20), the American Journal of Sociology (13), and the American Sociological Review (10). To find the articles, they searched on
``multilevel," ``multi-level," ``hierarchical," ``random eﬀect," ``random eﬀects," ``random-eﬀect," and ``random-eﬀects." The
political science and sociology reviews covered all articles dated January 2017 through December 2018, and the
education review covered all articles dated January 2017 through April 2019. H\&W found that a large majority of the articles ignored MLM's well-studied bias concerns (see \citealp{hausman1978specification} or \citealp{clark2015should}), or used MLM's default standard error without justification of the stringent dependence structure it assumes.} H\&W clarify the specific contexts in which MLM and FE models are appropriate, providing three analytical insights in the \textit{linear} model setting. (i) First, MLMs are equivalent to FE models that are fit with a regularization method which penalizes the selection of models with large parameter values, a class of models that H\&W label ``regularized FE" (RegFE). This connection demystifies two benefits of MLM: superior predictive accuracy for the outcome in comparison to FE, and the ability to include group-level variables in the model, which FE cannot do. The connection to regularization also makes clear the well-chronicled (e.g., \citealp{hausman1978specification, clark2015should}) draw-back of MLMs: they produce biased estimates for the treatment coefficient when group-level confounding is present. (ii) Second, MLM's bias is easily corrected by what H\&W refer to as ``bias-corrected MLM" (bcMLM), which originates from a long-standing adjustment to MLMs from \cite{mundlak1978pooling}. Further, bcMLM and FE produce equivalent coefficient estimates. (iii) Third,  MLM’s default standard errors from MLE are often too small, but this can be corrected by applying cluster-robust standard errors with FE or (bc)MLM. In fact, along with coefficient estimates, the cluster robust standard error estimates from bcMLM and FE are exactly equal. 

In this paper, we extend these three analytical insights from linear models to \textit{generalized linear models} (GLMs). For analytical insight (i), we find in the GLM case that there is no longer an exact equivalence between MLM fit through (approximate) MLE, and a generalized RegFE class of models. Nevertheless, they perform similarly because they solve maximization problems associated to factors of the same objective function. 
Additionally, MLMs fit through \textit{penalized quasi-likelihood} (PQL; \citealp{green1987penalized}) are even more similar to RegFE.
Thus, MLM can still be thought of as a form of regularization,  and group-level confounding can still greatly bias MLM's coefficient estimates in the GLM case. For analytical insight (ii), generalized forms of bcMLM and FE are not necessarily equivalent in the GLM setting. Further, bcMLM may be preferable to FE, because FE has non-negligible finite-sample bias in its coefficient estimates. The bias-correction step for bcMLM can also be applied to RegFE, which we call ``bias-corrected RegFE" (bcRegFE) and may also be preferable to FE. Finally, for analytical insight (iii), MLM still makes strict assumptions on the intragroup dependence structure, leaving the default standard errors obtained by MLE vulnerable to misspecification. Further, at the time of writing, we are unaware of a comprehensive extension of cluster robust standard errors to MLE-fit (bc)MLM in the GLM setting. However, empirical results show that a cluster bootstrap performs well, providing close to nominal coverage rates for confidence intervals, particularly in settings with a large number of groups. Additionally, we consider a cluster-robust standard error for bcRegFE that also performs well with many groups. Ultimately, in a \textit{non-linear} GLM, we recommend applying bcMLM for estimation of the coefficient on the treatment variable, and a cluster bootstrap for variance estimation and inference. If a cluster bootstrap is too computationally intensive for a given dataset, we instead recommend bcRegFE with cluster robust standard errors, or FE with cluster robust standard errors when group sizes are large. Note that this differs from H\&W's recommendation in the \textit{linear} setting to use either FE \textit{or} bcMLM for estimation (given that they are exactly equal) and apply cluster robust standard errors for variance estimation and inference.

Although the literature on MLMs in the GLM framework is less extensive than that on linear MLMs, many of our findings and recommendations are not new. \cite{schunck2017within} also note that the equivalence between FE and bcMLM estimates breaks down in the GLM case, but that the estimates remain similar. \cite{brumback2010adjusting}, \cite{brumback2013adjusting},  and \cite{goetgeluk2008conditional} have investigated settings in which bcMLM is biased, and many have recommended it over FE and uncorrected MLM (e.g., \citealp{raudenbush2009adaptive}; \citealp{bell2019fixed}; \citealp{schunck2017within}). \cite{cameron2015practitioner} have also noted that a cluster bootstrap is an option for variance estimation with MLMs. However, given the widespread misunderstanding of MLM and FE that H\&W identified relatively recently, it is likely that many disciplines have not yet fully internalized these lessons for non-linear GLMs.\footnote{For example, of the 24 articles across the two education journals (the American Education Research Journal and Educational Evaluation and Policy Analysis) that H\&W reviewed and for which bias could be an issue, only one of the articles addressed or accounted for MLM's bias problem, and none of them applied bcMLM. Then, of the 24 articles in education that employed the simplest MLM with group-varying intercepts, all of the articles used the default MLM standard error without discussion, justification, or robustness checks.} Further, we are unaware of work that compares and connects MLM to regularization in the GLM framework as explicitly and rigorously as we do here.

To preview, Section~\ref{sec:background} introduces our notation, the GLM framework, and the MLM and FE models. This section also discusses parameter estimation with MLM and FE, and frames these models in a causal inference setting. Section~\ref{sec:insights} then extends H\&W's analytical insights to the GLM setting. In Section~\ref{sec:application}, we demonstrate our recommendations in a real-data setting: estimating the effects of taking a mathematics course in twelfth grade for students in the Los Angeles Unified School District. Section~\ref{sec:conclusion} concludes and discusses our recommendations in more detail. We make these recommendations easily implementable through our \href{https://github.com/lwainstein/mlmfe}{\texttt{mlmfe}} package for the \texttt{R} programming language.

\section{Background}\label{sec:background}

\subsection{Notation}\label{subsec:notation}

We largely follow the notation and terminology used in H\&W. To help the reader, Appendix~\ref{app:tables}  lists (i) the abbreviations we use pertaining to models (Table~\ref{tab.abbreviations}) and (ii) the symbols we use (Table~\ref{tab.symbols}). Additionally, we often use the example of students within schools to motivate our discussions of grouped data.

Let $g=1, \dots, G$ index the group (e.g., the school). Vectors belonging to group $g$ will be subscripted with $g$ and the $i^{\textrm{th}}$ unit (e.g., student) in group $g$ with $g[i]$. For example, $Y_g$ denotes the outcome vector of all observations in group $g$ and $Y_{g[i]}$ denotes the outcome of observation $i$ in group $g$. This notation emphasizes that group $g$ contains observation $i$. Group $g$ has size $n_g$ and $N = \sum_{g=1}^{_G} n_g$ is the total number of observations. 

Let $X_{g[i]}$ be a $p$-dimensional vector of covariates, including an intercept term. One element of $X_{g[i]}$ will be referred to as a ``treatment". In the students-within-school example, this treatment may be a particular class or academic program. The remainder of $X_{g[i]}$ then includes other potential unit-level characteristics (e.g., student demographics). Let $\beta$ denote the coefficient vector associated with $X_{g[i]}$. We define $X_g$ as the matrix of covariate vectors $X_{g[i]}$ for group $g$, and $X$ as the matrix of $X_{g[i]}$ for the entire sample.
    \begin{align*}
    	X_{g[i]} &= \begin{bmatrix} 1 \\ X_{g[i]}^{(1)} \\ \vdots \\ X_{g[i]}^{(p-1)} \end{bmatrix} \in \mathbbm{R}^{p} \ , \ X_g = \begin{bmatrix} X_{g[1]}^{\top} \\ \vdots \\ X_{g[n_g]}^{\top} \end{bmatrix}  \in \mathbbm{R}^{n_g \times p} \ , \  X = \begin{bmatrix} X_{1} \\ \vdots \\ X_{_G}  \end{bmatrix} \in \mathbbm{R}^{N \times p} \ , \ \beta = \begin{bmatrix} \beta_0 \\ \beta_1 \\ \vdots \\ \beta_{p-1} \end{bmatrix} \in \mathbbm{R}^{p}
    \end{align*}
    \normalsize
\noindent Next,  let $Z_{g[i]}$ be a $d$-dimensional vector of covariates, which will often contain a subset of the covariates in $X_{g[i]}$, along with an intercept term which functions as an indicator of membership to group $g$. The $Z_{g[i]}$ then have an associated coefficient vector $\gamma_g$ for each group $g$. Also, let $Z_g$ be the matrix of $Z_{g[i]}$ for group $g$, let $Z$ be a block diagonal matrix of the $Z_g$, and let $\gamma$ stack the $\gamma_g$ into a matrix. 
    \begin{align*}
	    Z_{g[i]} &= \begin{bmatrix} Z_{g[i]}^{(0)} \\ \vdots \\ Z_{g[i]}^{(d-1)} \end{bmatrix} \in \mathbbm{R}^{d}, \ Z_g = \begin{bmatrix} Z_{g[1]}^{\top} \\ \vdots \\ Z_{g[n_g]}^{\top} \end{bmatrix} \in \mathbbm{R}^{n_{g} \times d}, \ Z = \begin{bmatrix} Z_{1} & \dots & 0 \\ \vdots & \ddots & \vdots \\ 0 & \dots & Z_G  \end{bmatrix} \in \mathbbm{R}^{N \times Gd}, \ 
    \end{align*}
    
    \vspace{-.1in}
    
    \begin{align*}
    	    \gamma_g = \begin{bmatrix} \gamma_{0 g} \\ \vdots \\ \gamma_{(d-1) g} \end{bmatrix} \in \mathbbm{R}^{d}, \   \gamma = \begin{bmatrix} \gamma_1 \\ \vdots \\ \gamma_G \end{bmatrix} \in \mathbbm{R}^{Gd} 
    \end{align*}

    \normalsize
\noindent We let $Y_{g[i]}$ denote the outcome of interest (e.g., high school graduation, GPAs, or credit accumulation), let $Y_g$ denote the $n_g \times 1$ vector of outcomes for group $g$, and let $Y$ denote the $N \times 1$ vector containing the $Y_{g[i]}$ for the entire sample.
    \begin{align*}
	    & Y_{g[i]} \in \mathbbm{R} \ , \ Y_g = \begin{bmatrix} Y_{g[1]} \\ \vdots \\ Y_{g[n_g]} \end{bmatrix} \in \mathbbm{R}^{n_g} \ , \ Y = \begin{bmatrix} Y_{1} \\ \vdots \\ Y_{G} \end{bmatrix} \in \mathbbm{R}^{N} 
    \end{align*}
    \normalsize
Finally, we use $p(\cdot)$ to denote a joint probability density or mass function for a random vector.


\subsection{Fixed effect and multilevel generalized linear models}\label{subsec:mlmfe}

\subsubsection*{Generalized linear models}\label{subsec:glm}

We briefly review the generalized linear model (GLM) framework within our grouped data context before discussing fixed effect (FE) and multilevel models (MLMs) in the GLM context.

GLMs are specified by two pieces of information: (i) a model for the relationship between the conditional expectation of $Y_{g[i]}$ (given $X$, $Z$, and the model parameters) and  $X_{g[i]}$ and $Z_{g[i]}$, and (ii) a model for the probability distribution of $Y$ (given $X$, $Z$, and the model parameters). Let $\mu_{g[i]} = \E( Y_{g[i]} \ | \ X, Z, \beta, \gamma)$. Specification (i) in a GLM requires that 
\begin{align}\label{eq:link}
     \mu_{g[i]}  = h^{-1} (  X^{\top}_{g[i]} \beta + Z_{g[i]}^{\top} \gamma_g )
\end{align}
where $h(\cdot)$, called the link function, is an invertible function which relates $\mu_{g[i]}$ to the linear component, $X^{\top}_{g[i]} \beta + Z_{g[i]}^{\top} \gamma_g$. For example, the classical linear regression setting arises from the choice of the identity link function, $h(t) = t$, in which case $\mu_{g[i]}=X_{g[i]}^{\top}\beta+Z_{g[i]}^{\top} \gamma_g$. When $Y_{g[i]}$ is binary (e.g., high school graduation), then $\mu_{g[i]} = p(Y_{g[i]} = 1 \ | \ X, Z, \beta, \gamma)$, in which case it is prudent to choose a link function whose inverse only takes values from 0 to 1. The logit function, $h(t) = \mathrm{log} (\frac{t}{1-t})$, used for logistic regression, accomplishes this task, allowing:
    \begin{align}\label{eq:glm_lr}
     	\mu_{g[i]} &= \frac{ \mathrm{exp} (X^{\top}_{g[i]} \beta + Z_{g[i]}^{\top} \gamma_{g})}{1 + \mathrm{exp} (X^{\top}_{g[i]} \beta + Z_{g[i]}^{\top}  \gamma_{g})}
     \end{align}
When $Y_{g[i]}$ is strictly positive (e.g., course credits accumulated), it may be preferable to choose a link function that has a strictly positive inverse. For example, the log link, $h(t) = \mathrm{log} (t)$, allows $\mu_{g[i]} = \mathrm{exp} (X_{g[i]}^{\top} \beta+Z_{g[i]}^{\top} \gamma_g)$. 

Given a particular link function $h$, Specification (ii) in a GLM is a probability model, $p_{\mathrm{GLM}}$, for the conditional distribution of $Y$ (given $X$, $Z$, and the model parameters):
\begin{align}\label{eq:glm_p}
        p(Y \ | \ X, Z, \beta, \gamma) = p_{\mathrm{GLM}} (Y \ | \ X, Z, \beta, \gamma, h, \theta)
    \end{align}
where $\theta$ is a vector of parameters associated to the conditional distribution of $Y$ that must be estimated along with $\beta$ and $\gamma$. For example, if $p_{\mathrm{GLM}}$ is a normal distribution
with $Y_{g[i]} \ | \ X, Z, \beta, \gamma, h, \theta \overset{iid}{\sim} N(\mu_{g[i]}, \sigma^2)$, then $\theta = \sigma^2$.
The model $p_{\mathrm{GLM}}$ specifies how $Y_{g[i]}$ varies about $\mu_{g[i]}$, and is used directly in the method of maximum likelihood estimation to estimate model parameters $(\beta, \gamma, \theta)$, which we review in Section~\ref{subsec:estimation}. Certain choices for $p_{\mathrm{GLM}}$ are often paired with specific link functions. For example, the normal probability model is most commonly paired with the identity link function. A Bernoulli model is required if $Y_{g[i]}$ is binary, and thus the logit link is a common choice. Finally, a Poisson model paired with the log link function is common for a $Y_{g[i]}$ that only takes positive integer values.

\subsubsection*{Varying intercepts: the group fixed effects and random intercept models}

We now introduce the FE and MLM models that we focus on, which allow a different intercept for each group in the data, but no other group-varying coefficients; in particular, we consider models where $Z_{g[i]} = [1]$. The GLM for the conditional mean of $Y_{g[i]}$ from (\ref{eq:link}) then becomes
\begin{align}\label{eq:group_glm}
     	\mu_{g[i]} &= h^{-1}(X^{\top}_{g[i]} \beta + \gamma_{g})
     \end{align}
\noindent where the $\gamma_{g}$ are group-specific deviations from the overall intercept in $\beta$, unless otherwise noted. In the students-within-schools example, this model allows different intercept terms for each school, but keeps constant the other coefficients in $\beta$ across schools. Unless otherwise noted, the results demonstrated in this manuscript hold for all GLMs (i.e., with general $h$ and $p_{\mathrm{GLM}} $). However, for illustration purposes, we will often make use of the logistic regression model, with the logit link function and a Bernoulli probability model, whose model form is
    \begin{align}\label{eq:logit_int}
        p(Y_{g[i]}=1 \ | \ X, Z, \beta, \gamma) &= \frac{ \mathrm{exp} (X^{\top}_{g[i]} \beta + \gamma_{g})}{1 + \mathrm{exp} (X^{\top}_{g[i]} \beta + \gamma_{g})}
    \end{align}

The key difference between FE and MLM concerns the distributional assumptions on the parameters they estimate. Both FE and MLM treat $\beta$ as fixed (i.e., non-random), imposing no distributional assumptions on it. However, FE and MLM differ in how they model  $\gamma_g$. FE regards $\gamma_g$ as fixed parameters, similar to $\beta$, and estimates $\gamma$ and $\beta$ simultaneously through maximum likelihood estimation. We refer to this as ``group fixed effects" (Group-FE), as do H\&W in the linear setting. For identifiability, Group-FE drops one group indicator variable if the intercept is present in $X_{g[i]}$. MLMs, however, treat $\gamma_g$ as random variables following a specified distribution (often normal). We define the ``random intercept" (RI) GLM as a MLM where only the intercept is treated as a random variable:
\begin{align}\label{eq:ri}
     	\mu_{g[i]} &= h^{-1}(X^{\top}_{g[i]} \beta + \gamma_{g}), \ \ \ \gamma_g \ | \ X, Z \overset{iid} \sim N(0, \omega^2)
     \end{align}
where all $\gamma_g$ are estimated along with an intercept term in $X_{g[i]}$. The $\gamma_g$ are often referred to as ``random effects'' and the model incorporates what has been called the ``random effects assumption'' (\citealp{bell2015explaining, kim2019causal}) that $\cor(\gamma_g, X_{g[i]}) =0$ (see Appendix~\ref{app:re_assumption} for proof). 
We explain later in Section~\ref{subsec:identification} why this assumption can yield greatly biased estimates for $\beta$. Additional specifications on $\gamma_{g}$ are also prescribed, depending on the choice of the GLM probability model $p_{\mathrm{GLM}}$.  For example, in the linear model with the identity link function $h$, the model in (\ref{eq:group_glm}) can be rewritten as 
    \begin{align}\label{eq:linear}
     	Y_{g[i]} &= X^{\top}_{g[i]} \beta + \gamma_{g} + \epsilon_{g[i]}
     \end{align}
where $\E(\epsilon_{g[i]} \ | \ X, Z, \beta, \gamma) = 0$, which is equivalent to the linear FE and MLM investigated by H\&W.\footnote{H\&W discuss this model without the ``conditional independence" assumption that $\E( \epsilon_{g[i]} | X, Z, \beta, \gamma ) = 0$, but note that it would be required for the model to recover the effect of $X_{g[i]}$ on $Y_{g[i]}$.} Using the normal probability model for $p_{\mathrm{GLM}}$ is equivalent to specifying the distribution of $\epsilon_{g[i]}$, where $\epsilon_{g[i]} \ | \ X, Z \overset{iid}{\sim} N(0, \sigma^2)$. Here, the RI model not only specifies that $\gamma_g$ are normal, but also that the $\epsilon_{g[i]}$ are independent from the random intercept of any other group: $\epsilon_{g[i]} $ is conditionally independent of $ \gamma_{g'}$ given $X$ and $Z$ for all $g$, $g'$, and $i$.

\subsubsection*{Varying slopes in fixed effect and multilevel models}

The primary focus of our analysis in this paper is on Group-FE and RI GLMs, which are special cases of a broader class of GLM MLMs. In this subsection, we briefly describe specification for this broader class of models in order to provide a unified framework for our analysis. The form for the GLM is given in (\ref{eq:link}), where the $\gamma_g$ are again group-level coefficients. But in contrast to Group-FE and RI models, we allow $Z_{g[i]}$ here to include other variables in addition to an intercept term. Again, FE estimates treat both $\beta$ and $\gamma$ as fixed parameters by fitting a GLM of $Y$ on $X$ and $Z$. During parameter estimation, FE drops covariates included in both $X$ and $Z$ from either $X_{g[i]}$, or $Z_{g[i]}$ for one group.  In contrast, MLM treats $\gamma_g$ as a random vector, 
    \begin{align}\label{eq:mlm_general}
     	\mu_{g[i]} = h^{-1}(X^{\top}_{g[i]} \beta + Z^{\top}_{g[i]} \gamma_{g}), \ \ \ \gamma_g | X, Z \overset{iid}{\sim} N(0, \Omega)
     \end{align}
where $\Omega \in \R^{d \times d}$ is a covariance matrix of parameters to be estimated. The assumed distribution on $\gamma_g$ implies the more general form of the random effects assumption: that $\cor(Z_{g[i]}^{\top} \gamma_g, X_{g[i]}) = 0$ (see Appendix~\ref{app:re_assumption} for proof). 
In other words, the whole ``random effect contribution", $Z_{g[i]}^{\top} \gamma_g$, is uncorrelated with $X_{g[i]}$.

\subsection{Parameter estimation}\label{subsec:estimation}

We now compare parameter estimation methods for MLM and FE.\footnote{We focus on frequentist estimation of all models. For a review of Bayesian estimation of MLMs, see \cite{gelman2006data}.} FE typically uses MLE to estimate its parameters, so we focus on that here. For MLMs, while there is a range of estimation approaches, we also focus on (approximate) MLE-based estimation.\footnote{We describe MLE in the context of FE and MLM here, and suggest \cite{pawitan2001all} for a more general review.} We do this because our primary goal is to show the connections between MLM and FE, which uses MLE, and the regularized form of FE we consider in Section~\ref{sec:insights}, which is best understood through a MLE-based lens. However, we acknowledge other estimation approaches where appropriate, notably penalized quasi-likelihood (PQL) in Section~\ref{subsec:insight1}. 
See \cite{jiang2021linear} for a more thorough review of other estimation approaches for MLM. 

We first compare MLE estimation of $\beta$ in FE and MLM. In general, MLE optimizes a function of the \textit{fixed} model parameters called the likelihood function, which we denote $L(\cdot)$.\footnote{Equivalently, one can optimize the (negative) natural logarithm of the likelihood, which is more common. However, throughout we largely omit this detail to reduce notation.} In both FE and MLM, the likelihood is given by $p(Y | X, Z, \Theta)$ where $\Theta$ denotes the collection of fixed parameters associated to the model. Due to their differing specifications on $\gamma$, FE and MLM differ in (i) the fixed parameters they estimate beyond $\beta$, and (ii) their ultimate expressions for the likelihood. For FE, because $\gamma$ is fixed, its fixed parameters are $\Theta = (\beta, \gamma, \theta)$. The likelihood is then 
    \begin{align}\label{eq:likelihood_fe}
        L_{\mathrm{FE}} ( \beta, \gamma, \theta) &= p (Y | X, Z, \beta, \gamma, \theta)  = p_{\mathrm{GLM}} (Y | X, Z, \beta, \gamma, h, \theta)
    \end{align}
which is fully specified by the GLM assumptions (see (\ref{eq:link}) and (\ref{eq:glm_p})). This likelihood is then maximized, often using iterative (re)weighted least squares, to arrive at the MLE estimate $(\hat{\beta}_{\mathrm{FE}}, \hat{\gamma}_{\mathrm{FE}}, \hat{\theta}_{\mathrm{FE}})$.

Because MLM treats $\gamma$ as random, its fixed parameters are $(\beta, \theta, \Omega)$. Further, to obtain an expression for the conditional probability $p (Y |X, Z, \Theta)$, FE's likelihood in (\ref{eq:likelihood_fe}) is integrated with respect to the specified distribution of $\gamma$: 
    \begin{align}\label{eq:likelihood_mlm}
        L_{\mathrm{MLM}} ( \beta, \theta, \Omega ) &= p (Y | X, Z, \beta, \theta, \Omega) = \int p (Y, \gamma | X, Z, \beta,  \theta, \Omega) d \gamma \nonumber \\
        &= \int \underbrace{ p_{\mathrm{GLM}} (Y | X, Z, \beta, \gamma, h, \theta)}_{L_{\mathrm{FE}}} p(\gamma | X, Z, \Omega) d \gamma
    \end{align}
Note that the terms inside the integral in (\ref{eq:likelihood_mlm}) are fully determined by the GLM specifications, as well as MLM's distributional specification for $\gamma$. Though while this integral has a closed-form expression in the linear setting, there is not in general a closed-form expression for other GLMs. It is mainly for this reason why many estimation approaches for MLM have been explored, such as PQL
(see \citealp{jiang2021linear}). However, we focus on implementations of MLM that use numerical integration methods, specifically Laplace Approximation or Gauss-Hermite Quadrature (\citealp{kabaila2019adaptive}), to approximate the integral in (\ref{eq:likelihood_mlm}) at each step of an iterative optimization method to maximize $L_{\mathrm{MLM}} ( \beta, \theta, \Omega )$.\footnote{This is often the default in popular statistics software (e.g., \texttt{lme4} in \texttt{R}; \texttt{melogit} and \texttt{mepoisson} in \texttt{Stata}). 
Estimation through PQL is also common (e.g., \texttt{GLIMMIX} in \texttt{SAS}).
}  Given an estimate for $(\theta, \Omega)$, plugging this estimate into $L_{\mathrm{MLM}} (\beta, \theta, \Omega)$ and then maximizing the result over $\beta$ yields an estimate for $\beta$. 
There are various approaches for estimating $(\theta, \Omega)$ that come with their own intricacies. For example, in linear models, the most common approaches are a traditional, unrestricted MLE approach and the restricted MLE approach.\footnote{Unrestricted maximum likelihood estimates of  are found by maximizing $L_{\mathrm{MLM}}$ in $(\ref{eq:likelihood_mlm})$ over $(\theta, \Omega, \beta)$:
\begin{align*}
    (\hat{\theta}_{\mathrm{UML}}, \hat{\Omega}_{\mathrm{UML}}, \hat{\beta}_{\mathrm{UML}}) = \underset{\beta, \theta, \Omega}{\argmax} \ L_{\mathrm{MLM}} (\beta, \theta, \Omega)
\end{align*}
Restricted maximum likelihood estimates are found by first integrating out $\beta$ from $L_{\mathrm{MLM}}$, and then maximizing the result over only $\theta$ and $\gamma$,
\begin{align*}
    (\hat{\theta}_{\mathrm{RML}}, \hat{\Omega}_{\mathrm{RML}}) = \underset{\theta, \Omega}{\argmax} \ \int L_{\mathrm{MLM}} (\beta, \theta, \Omega) d \beta
\end{align*}
The corresponding restricted MLE of $\beta$ then follows as:
\begin{align*}
    \hat{\beta}_{\mathrm{RML}} = \underset{\beta}{\argmax} \ L_{\mathrm{MLM}} (\beta, \hat{\theta}_{\mathrm{RML}}, \hat{\Omega}_{\mathrm{RML}})
\end{align*} 
Although these two procedures yield (potentially) different estimates of $\beta$, one can also think of them both as maximizing $L_{\mathrm{MLM}}$ over $\beta$ for different fixed choices/estimates of $\theta$ and $\Omega$:
\begin{align*}
    \hat{\beta} = \underset{\beta }{\argmax} \ L_{\mathrm{MLM}} (\beta, \hat{\theta}, \hat{\Omega})
\end{align*}
where $\hat{\beta} = \hat{\beta}_{\mathrm{UML}}$ when $(\hat{\theta}, \hat{\Omega}) = (\hat{\theta}_{\mathrm{UML}}, \hat{\Omega}_{\mathrm{UML}})$, and $\hat{\beta} = \hat{\beta}_{\mathrm{RML}}$ when $(\hat{\theta}, \hat{\Omega}) = (\hat{\theta}_{\mathrm{RML}}, \hat{\Omega}_{\mathrm{RML}})$.
} However, in what follows we only consider estimates of $\beta$ (and $\gamma$) given an arbitrary estimate of $(\theta, \Omega)$, so we simply denote MLM estimates of the parameters as $(\hat{\beta}_{\mathrm{MLM}}, \hat{\theta}_{\mathrm{MLM}}, \hat{\Omega}_{\mathrm{MLM}})$, whether they come from unrestricted or restricted MLE, or another approach.

For many models, MLE produces a biased estimator (e.g., GLMs with either non-identity link functions or non-Normal response distributions). Fortunately, under mild regularity conditions, MLE produces parameter estimates that are consistent and asymptotically efficient.\footnote{These regularity conditions relate to the smoothness of the likelihood function as well as to the shape of the parameter space. See \cite{lehmanncasella1998} for more details.} Thus, MLE often yields excellent estimators when sample sizes are large.\footnote{Often, MLE's bias is of order at most $n^{-1}$, where $n$ is the \textit{effective} number of \textit{independent} observations in the sample, which may be strictly less than the sample size if response values in the sample are correlated (e.g., serial correlation or clustering). \cite{mccullaghnelder1989} provide explicit calculation of the order $n^{-1}$ bias in the case of GLMs with natural parameters and canonical link functions. 
} There are, however, cases where the bias may be considerable: when the sample size is small, or when the number of parameters is large relative to the effective number of independent observations. Because both FE and MLM apply to data that is correlated within groups, their estimates of $\beta$ may exhibit nontrivial bias. This bias is often more pronounced for FE than for MLM, since FE typically requires the estimation of a far greater number of fixed parameters (the dimension of $\gamma$ is often larger than that of $\Omega$, and grows with $G$). This has been called the ``incidental parameters problem" (\citealp{neyman1948consistent, lancaster2000incidental}), which also has implications for consistency. For example, while $\hat{\beta}_{\mathrm{FE}}$ is consistent for $\beta$ as $G$ grows and $n_g$ stays fixed in linear and Poisson models, it is not in logistic regression --- consistency in logistic regression requires $n_g$ to grow.

Next, we consider how FE and MLM estimate $\gamma$.\footnote{Because $\gamma$ is assumed random in MLM, estimates of $\gamma$ in MLM are often instead called ``predictions" (e.g., \citealp{jiang2021linear}). However, we use the term ``estimates" to stay consistent with FE, and reserve the term ``predictions" when referring to predictions for the outcome $(Y_{g[i]})$.} Note that FE obtains $\hat{\gamma}_{\mathrm{FE}}$ at the same time as it obtains $\hat{\beta}_{\mathrm{FE}}$ due to the joint maximization of $L_{\mathrm{FE}}$ in (\ref{eq:likelihood_fe}) over $(\beta, \gamma, \theta)$. However, this is not the case with MLM, which integrates out $\gamma$ in its likelihood function in (\ref{eq:likelihood_mlm}). While there are several options for estimating $\gamma$ in the MLM framework (see \citealp{jiang2021linear}), we focus on estimates $\hat{\gamma}_{\mathrm{MLM}}$ from maximizing the posterior probability distribution $p(\gamma \ | \ Y, X,Z,\hat\beta_{\mathrm{MLM}},\hat\theta_{\mathrm{MLM}}, \hat\Omega_{\mathrm{MLM}})$. Note that this is equivalent to maximizing:
    \begin{align}
        p(Y, \gamma \ | \ X, Z, \hat{\beta}_{\mathrm{MLM}}, \hat{\theta}_{\mathrm{MLM}}, \hat{\Omega}_{\mathrm{MLM}}) =  p_{\mathrm{GLM}} (Y | X, Z, \hat{\beta}_{\mathrm{MLM}}, \gamma, h, \hat{\theta}_{\mathrm{MLM}}) p(\gamma | X, Z, \hat{\Omega}_{\mathrm{MLM}})
    \end{align}
which is the inner term of the integral in (\ref{eq:likelihood_mlm}), after substituting in the estimate $(\hat\beta_{\mathrm{MLM}}, \hat{\theta}_{\mathrm{MLM}}, \hat\Omega_{\mathrm{MLM}})$. We focus on these specific estimates of $\gamma$ due to their connection to regularization, which we cover in Section~\ref{subsec:insight1}. Further, in the MLE-based framework, estimating $\gamma$ in MLM occurs after estimating $\beta$, which is our main concern, so our lessons below would change little were one to consider a different estimation approach for $\gamma$.

Finally, we turn our attention to the traditional variance estimator for $\hat{\beta}_{\mathrm{MLM}}$ that is obtained through MLE-based estimation. Per MLE generally, an estimated variance can be retrieved by evaluating the negative inverse Hessian of MLM's log likelihood, evaluated at MLM's estimated parameters (see \citealp{pawitan2001all} for more details),
\begin{align}\label{eq:se_mlm}
    \widehat{\var}_{\mathrm{MLE}} \biggr( (\hat{\beta}_{\mathrm{MLM}}, \hat{\theta}_{\mathrm{MLM}}, \hat{\Omega}_{\mathrm{MLM}} ) \biggr) &= - \biggr(\ell''_{\mathrm{MLM}} (\hat{\beta}_{\mathrm{MLM}}, \hat{\theta}_{\mathrm{MLM}}, \hat{\Omega}_{\mathrm{MLM}})\biggr)^{-1} \nonumber \\ \text{where} \ \ \ell_{\mathrm{MLM}} (\cdot) &= \mathrm{log} \ L_{\mathrm{MLM}} (\cdot).
\end{align}
However, proper specification of the model determines the validity of the resulting standard errors. As we demonstrate in Section~\ref{subsec:insight3}, misspecifying the intragroup dependence structure can result in standard errors for MLM from $(\ref{eq:se_mlm})$ that are too small, and confidence intervals that are too narrow.

\subsection{Identification to specification}\label{subsec:identification}

As do H\&W in the linear setting, we explain why, from a causal inference perspective, we would expect the random effects assumption in MLMs to yield meaningfully biased estimates of $\beta$ in the GLM setting. We also discuss the relationship between $\beta$ and causal quantities of interest.

To illustrate MLM's bias concern, consider a simplified setting where no within-group confounding is present. In the students-within-schools example, this would mean that only school context determines the treatment status of students. This assumption guarantees the identifiability of any causal quantity of interest related to the treatment within each group, meaning one would only need to account for group-level confounding. Given enough data, one could account for group structure by estimating this causal quantity of interest within each group, and then averaging these estimates across the groups (if desired). However, it is rare to have enough data in each group to feasibly do this. Thus, researchers often use models to account for group structure, which require additional model specification-related assumptions to hold for consistent estimation. For example, Group-FE and RI attempt to account for group structure through the inclusion of the $\gamma_g$ in the GLM in (\ref{eq:group_glm}). In particular, group-level confounding is represented by adding a value to the (transformed) conditional mean of $Y_{g[i]}$ that is constant within each group. 

The concern with the RI model is that even if this modeling assumption is correct, RI does not account for group-level confounding as desired. This introduces bias into the coefficient estimate for the treatment in $\beta$ even beyond the finite sample bias from MLE, as we demonstrate in Section~\ref{subsec:insight1}. In brief, bias arises because of the random effects assumption in MLMs that $\cor(\gamma_g, X_{g[i]}) = 0$. In the students-within-schools example, this means that the RI model assumes that school context does \textit{not} influence whether or not a student takes the treatment. This assumption contradicts one of the primary reasons for including the $\gamma_g$ in the model in the first place, which was to account for group-level confounding (that is, $\cor(\gamma_g, X_{g[i]}) \neq 0$). H\&W show this bias concern has long been ignored in practice despite being well-chronicled (e.g., \citealp{hausman1978specification, clark2015should}). 

So far, this discussion of bias has been essentially the same as that from H\&W for the linear case. However, there is a key difference in the GLM case: Group-FE is no longer assured to be unbiased for $\beta$ if the GLM in (\ref{eq:group_glm}) is correctly specified due to potential finite sample bias of MLE estimates.\footnote{In the linear case, the MLE estimates of $\beta$ (and $\gamma$) from Group-FE are unbiased because they are also the OLS estimates for the linear model in (\ref{eq:linear}), which are unbiased as long as $\E(\epsilon_{g[i]} \ | \ X, Z) = 0$. } As we demonstrate in Section~\ref{subsec:insight1}, this bias is often non-negligible due to large the number of parameters that FE estimates and its incidental parameters problem. This fact informs our recommendations (in Section~\ref{sec:conclusion}) for \textit{non-linear} GLMs, which differ from those from H\&W for linear models.

There is one final difference between linear models and GLMs that is worth highlighting: the two model types provide differing mappings of causal quantities of interest to model parameters. Assuming a correctly specified \textit{linear} model, the treatment coefficient maps to the ``average treatment effect", which is often the target estimand. However, this does not necessarily hold in the GLM case. For example, let $X_{g[i]}$ be a binary treatment in the varying intercept logistic regression model in (\ref{eq:logit_int}). Then $e^{\beta}$ is interpretable as the ratio in the odds ($\mathrm{odds}(t) = \frac{t}{1-t}$) of success ($Y_{g[i]} = 1$) after receiving the treatment. Estimating the average treatment effect would require using the model to calculate predicted probabilities for each observation in the data with and without the treatment, calculating the difference between these predictions, and then averaging the differences. This estimate involves not only $\hat{\beta}$, but also the $\hat{\gamma}_g$, and so bias in the estimate is affected by more than just bias in $\hat{\beta}$. Nevertheless, we focus on bias in estimates of $\beta$, as this is typically the most influential factor for bias in an estimated average treatment effect. How bias in estimates of other parameters influences estimates of the average treatment effect is left to future work.

\section{Analytical Insights}\label{sec:insights}

\subsection{Insight 1: Random effects as regularization and bias for MLM}\label{subsec:insight1}

In this section, we explore the connection in the GLM setting between MLM and a generalized regularized fixed effects (RegFE) class of models, which fits an FE model with shrinkage applied to the $\gamma$. In contrast to the linear models setting, we find that there is no longer necessarily an \textit{exact} equivalence in finite samples between MLM fit through MLE, and RegFE in the GLM case. Nevertheless, we also show that the models can produce similar parameter estimates, so MLM can still be understood as regularizing its random effects. As in the linear setting, this leads to ``incomplete conditioning" and thus meaningful bias in $\beta$.

In the linear setting, H\&W showed that MLMs fit using MLE can also be obtained by fitting FE models that include Tikhonov ($L^2$) regularization on the group-varying coefficients in $\gamma$. To demonstrate this equivalence, they introduce the RegFE class of models, which minimizes a similar objective function as FE, but with an additional penalty term that scales with the squared norm of $\gamma$. In the specific case where the model includes only group-varying intercepts (i.e., $Z_{g[i]} = 1$), RegFE obtains its coefficients by minimizing the penalty-included objective function:
    \begin{align}\label{eq:RegFE_linear}
(\hat{\beta}_{\text{RegFE}},\hat{\gamma}_{\text{RegFE}})=\argmin_{\beta,\gamma}\left(\sum_{g=1}^G\sum_{i=1}^{n_g} [ Y_{g[i]}-(X_{g[i]}^\top\beta+\gamma_g) ]^2+ \lambda_{\mathrm{Lin}} \sum_{g=1}^G\gamma_g^2\right)
    \end{align}
The objective function in (\ref{eq:RegFE_linear}) above  penalizes larger magnitude estimates of $\gamma$, and thus \textit{regularizes} the $\gamma$ estimates. H\&W then show an \textit{exact} equivalence between the parameter estimates from fitting RegFE using (\ref{eq:RegFE_linear}) above and the estimates from fitting a linear RI model to (\ref{eq:linear}) with spherical errors, $\epsilon_{g[i]} \ | \ X, Z \overset{iid}{\sim} N(0, \sigma^2)$: that is, when $\lambda_{\mathrm{Lin}} = \hat\sigma_{\text{RI}}^2/ \hat\omega_\text{RI}^2$, then the estimated $\beta$ and $\gamma$ from RegFE and RI are exactly equal. This apparent regularization by MLM on $\gamma$ leads to what H\&W call ``incomplete conditioning" --- the shrinkage applied to the $\gamma_g$ prevents them from fully ``soaking up" group-level confounding, which leads to bias in the estimate for $\beta$. 

We now introduce a more general class of RegFE models that apply to the GLM setting. For the same class of models specified by GLM MLM, we define GLM RegFE parameter estimates of $(\beta, \gamma)$ as:
    \begin{align}\label{eq:RegFE_glm}
        (\hat\beta_{\text{RegFE}},\hat\gamma_{\text{RegFE}}) = \underset{\beta, \gamma}{\argmax} \biggr( p_{\mathrm{GLM}} (Y \ | \ X, Z, \beta, \gamma, h, \theta) \cdot p(\gamma \ | \ X, Z, \Omega) \biggr) 
    \end{align}
where $\theta$ and $\Omega$ are fixed values. 
We show in Appendix~\ref{app:regfe_motivation} why the estimates from (\ref{eq:RegFE_glm}) are the natural extension of the estimates from (\ref{eq:RegFE_linear}) in the GLM case. Additionally, while we apply the above formula in our demonstrations below with MLE estimates of $\theta$ and $\Omega$, they could be obtained through other means, such as cross-validation.\footnote{Recall that in the linear RI model, $\theta = \sigma^2$ and $\Omega = \omega^2$. Then, as we demonstrate concretely in Appendix~\ref{app:regfe_motivation}, setting $\lambda_{\mathrm{Lin}} = \sigma^2/\omega^2$ recovers linear RegFE. Thus, allowing $\theta$ and $\Omega$ to be chosen by cross-validation in general RegFE is analogous to letting cross-validation choose the level of shrinkage in linear RegFE.} Note that this formulation allows for multiple group-varying coefficients (i.e., models where $Z_{g[i]}$ is not identically $1$), not just group-varying intercepts. Note also that $p_{\mathrm{GLM}} (Y \ | \ X, Z, \beta, \gamma, h, \theta)$ in (\ref{eq:RegFE_glm}) above is exactly the likelihood function that is maximized under MLE in FE (i.e., $L_{\mathrm{FE}}$ in (\ref{eq:likelihood_fe})). For RegFE, the specification of $p(\gamma \ | \ X, Z, \Omega)$ then determines the regularization of $\gamma$.\footnote{Note also that when $p_{\mathrm{GLM}} (\cdot)$ comes from the Exponential Family and $h(\cdot)$ is the associated canonical link function, exchanging $p(\gamma \ | \ X, Z, \Omega)$ in (\ref{eq:RegFE_glm}) with Jeffrey's invariant prior on both $\beta$ and $\gamma$ yields Firth's bias correction (\citealp{firth1993}), which induces regularization on $\gamma$ \textit{and} $\beta$.} In the context of MLMs, $\gamma_g$ is often specified to be normally distributed, which induces $L^2$ regularization, and yields a special case of the estimator studied by \cite{wood2011fast}. We demonstrate this in Appendix~\ref{app:regfe_motivation}, as well as motivate our definition of GLM RegFE by showing how, under linear RI model specifications, the estimates in (\ref{eq:RegFE_glm}) can be obtained from linear RegFE estimates in (\ref{eq:RegFE_linear}).

Comparing the preceding optimization problem for RegFE in (\ref{eq:RegFE_glm}) to that arising from estimating parameters in MLM through MLE reveals the connection between GLM MLM and RegFE. Recall from Section~\ref{subsec:estimation} that MLE-fit MLM estimates $\beta$ and $\gamma$ in a two-step process. Given $(\hat\theta_{\text{MLM}}, \hat\Omega_{\text{MLM}})$, the function $p(Y|X,Z,\beta,\hat\theta_{\text{MLM}}, \hat\Omega_{\text{MLM}})$ is maximized over $\beta$:
    \begin{align}\label{eq:mlm_step1}
        \hat{\beta}_{\mathrm{MLM}} = \underset{\beta}{\argmax} \biggr( p(Y \ | \ X,Z,\beta,\hat\theta_{\text{MLM}}, \hat\Omega_{\text{MLM}}) \biggr)
    \end{align}
Second, the function $p(\gamma | Y, X,Z, \hat\beta_{\text{MLM}},\hat\theta_{\text{MLM}},\hat\Omega_{\text{MLM}})$ is maximized over $\gamma$ to obtain $\hat{\gamma}_{\mathrm{MLM}}$:
    \begin{align}\label{eq:mlm_step2}
        \hat{\gamma}_{\mathrm{MLM}} = \underset{\gamma}{\argmax} \biggr( p(\gamma \ | \ Y, X,Z,\hat\beta_{\text{MLM}},\hat\theta_{\text{MLM}}, \hat\Omega_{\text{MLM}}) \biggr)
    \end{align}
Note then that the product of the objective functions in these two steps yields the objective function for RegFE in (\ref{eq:RegFE_glm}), with $(\theta, \Omega)$ set to $(\hat\theta_{\text{MLM}}, \hat\Omega_{\text{MLM}})$:
    \begin{align}\label{eq:mlmtoregfe}
        \underbrace{p(Y, \gamma \ | \ X, Z, \beta, \hat\theta_{\text{MLM}}, \hat\Omega_{\text{MLM}})}_{\text{from (\ref{eq:RegFE_glm})}} = \underbrace{ p(Y \ | \ X,Z,\beta,\hat\theta_{\text{MLM}}, \hat\Omega_{\text{MLM}})}_{\text{from (\ref{eq:mlm_step1})}} \cdot \underbrace{p(\gamma \ | \ Y, X,Z, \beta,\hat\theta_{\text{MLM}}, \hat\Omega_{\text{MLM}})}_{\text{from (\ref{eq:mlm_step2})}}
    \end{align}
That is, for fixed estimates $(\hat\theta_{\text{MLM}}, \hat\Omega_{\text{MLM}})$, both MLM and RegFE estimate parameters by maximizing (\ref{eq:mlmtoregfe}) over $\beta$ and $\gamma$, although they do so through different processes: RegFE maximizes jointly over $\beta$ and $\gamma$, while MLM maximizes separately the individual terms of the product on the right side of (\ref{eq:mlmtoregfe}). It is not evident \textit{a priori} that these two procedures yield the same coefficient estimates. However, H\&W show that this does occur in the linear case.\footnote{Generally speaking, the maximizer of the product of two functions is not necessarily the same as the individual maximizers of the two functions. For example, let $g(x) = 2 - (1 - x)^2$ and $h(x, y) = \frac{5(1-y^2)}{1  + 5x^2}$. Then, consider maximizing $f(x, y) = g(x) h(x, y)$. First consider maximizing $g$ and $h$ separately to find $(x_{\mathrm{sep}}, y_{\mathrm{sep}})$. Maximizing $g$ yields $x_{\mathrm{sep}} = \underset{x}{\mathrm{argmax}} \ g(x) = 1$, and then maximizing $h$ after setting $x = x_{\mathrm{sep}}$ yields $y_{\mathrm{sep}} = \underset{y}{\mathrm{argmax}} \ h(x_{\mathrm{sep}}, y) = 0$. Thus, $(x_{\mathrm{sep}}, y_{\mathrm{sep}}) = (1, 0)$ and  $f(x_{\mathrm{sep}}, y_{\mathrm{sep}}) = \frac{5}{3}$. However, $f(0, 0) = 5 > \frac{5}{3} = f(x_{\mathrm{sep}}, y_{\mathrm{sep}})$, meaning that $(x_{\mathrm{sep}}, y_{\mathrm{sep}}) \neq \underset{(x, y)}{\mathrm{argmax}} \ f(x, y).$} Appendix \ref{app:equivalence in linear case} elaborates on the special conditions that yield this equivalence.

In the GLM setting, however, even with $(\theta, \Omega)$ set to $(\hat\theta_{\text{MLM}}, \hat\Omega_{\text{MLM}})$ for RegFE, there is no guarantee that $\hat{\beta}_{\mathrm{MLM}} = \hat{\beta}_{\mathrm{RegFE}}$ or $\hat{\gamma}_{\mathrm{MLM}} = \hat{\gamma}_{\mathrm{RegFE}}$.
\begin{figure}[!h]
    \caption{Estimates of $\beta_1$ in (\ref{DGP:logistic}) from RI, RegFE, Group-FE, and a GLM without fixed or random effects  
    for logistic regression}\label{fig:logistic_noneq}
    
    \vspace{-0.1in}
    
    \begin{center}
    
 \begin{subfigure}{0.48\textwidth}
   
     \begin{center}
     \vspace{-.5in}
     \includegraphics[scale=.45]{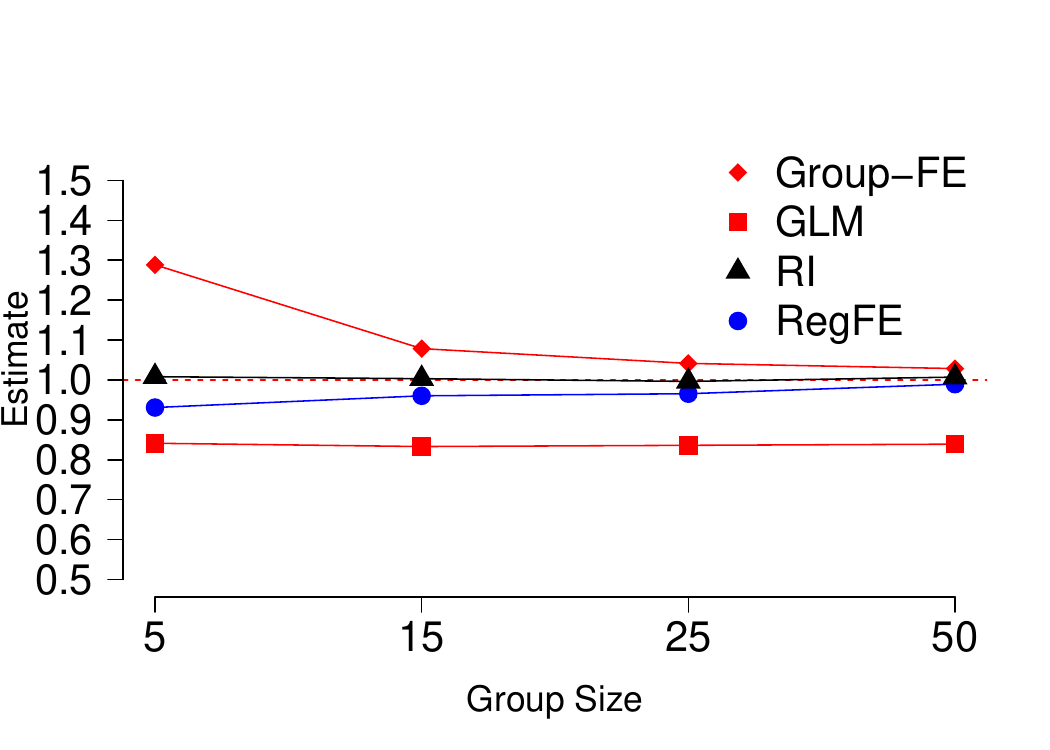}
     \subcaption{Median estimates for $G=50$}\label{fig:logistic_noneq_a}
     \end{center}
     \end{subfigure}
 \begin{subfigure}{0.50\textwidth}
     \begin{center}
     \vspace{-.5in}
     \includegraphics[scale=.45]{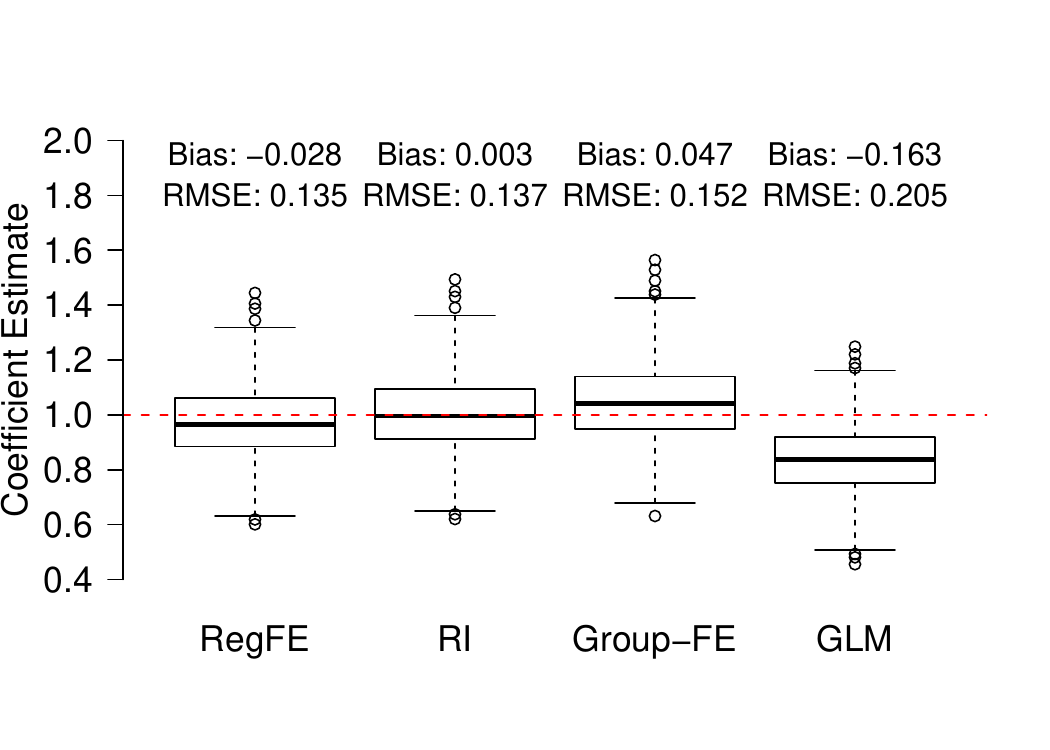}
     \subcaption{Estimate distributions for $G=50$ and $n_g=25$}\label{fig:logistic_noneq_b}
     \end{center}
     \end{subfigure}
          
    \caption*{\textit{Note:} Results across 1000 iterations at each sample size of the DGP in (\ref{DGP:logistic}). The RI model is a logistic regression RI model. The RegFE model is a logistic regression RegFE model with only varying intercepts, setting $\omega = \hat\omega_{\text{RI}}$. The Group-FE model is a logistic regression Group-FE model. The GLM model is a logistic regression that only includes $X_{g[i]}$ as a regressor, and omits fixed and random effects. \textit{(a)} Median estimates when $G=50$ for $\beta_1$. The red dashed line indicates the true value of $\beta_1 = 1$. \textit{(b)} Distribution of estimates when $G=50$ and $n_g=25$. The red dashed line indicates the true value of $\beta_1 = 1$.
    }
    \vspace{-.3in}
    
    \end{center}
    \end{figure}
Figure \ref{fig:logistic_noneq} demonstrates this nonequivalence in the case of logistic regression, showing the distribution of estimates of $\beta_1$ from RegFE and RI logistic regression models over 1000 iterations of the following data-generating process (DGP):
\begin{equation}\label{DGP:logistic}
    Y_{g[i]}\overset{}{\sim}\text{Bernoulli}(\operatorname{logit}^{-1}(\beta_0+X_{g[i]}\beta_1+\gamma_g)),\text{ where } X_{g[i]}\overset{iid}{\sim}N(0,0.5), ~ \gamma_{g}\overset{iid}{\sim}N(0,1)    
\end{equation}
We estimate bias and root mean square error (RMSE) of $\beta$ using
\begin{align}
    \mathrm{Bias}(\beta) = \frac{1}{M} \sum_{m=1}^M (\hat{\beta}^{(m)} - \beta), \qquad \mathrm{RMSE}(\beta) = \sqrt{\frac{1}{M} \sum_{m=1}^M (\hat{\beta}^{(m)} - \beta)^2}
\end{align}
where $m$ indexes the iteration number among the $M$ simulations, and $\hat{\beta}^{(m)}$ is the estimate of $\beta$ from the $m$th iteration. In Figure~\ref{fig:logistic_noneq}, the differences between the RI and RegFE estimates are largest when the group sizes are low.\footnote{Without formal mathematical proof, we cannot verify that these differences are not due to differences in the numerical optimization procedures for MLM (implemented with the \texttt{lme4} package in \texttt{R}) and RegFE (optimization done with the \texttt{optim()} function in \texttt{R}). However, we are confident that these are true differences for several reasons. First, particularly for smaller group sizes, the RegFE and RI estimates in Figure~\ref{fig:logistic_noneq} are meaningfully different -- for $G=50$ and $n_g=5$, the difference between the median RegFE and RI estimates is as large as the difference between the median RegFE and GLM estimates. Second, these are relatively simple models and we have set \texttt{nAGQ=100} (the maximum allowed, at time of writing) for \texttt{glmer()} in \texttt{R} for the best possible approximation of RI's MLE by Gauss-Hermite Quadrature. Finally, our simulation shows that RegFE's objective function in (\ref{eq:mlmtoregfe}) was higher when evaluated at $(\hat{\beta}_{\mathrm{RegFE}}, \hat{\gamma}_{\mathrm{RegFE}})$ than when evaluated at $(\hat{\beta}_{\mathrm{MLM}}, \hat{\gamma}_{\mathrm{MLM}})$ in over 99\% of iterations tried.}
Further, RI is effectively unbiased at all sample sizes, while RegFE shows consistent, slight negative bias even when $n_g=25$ (see Figure~\ref{fig:logistic_noneq_b}).

\begin{figure}[!h]
    \caption{Estimates of $\gamma$ in (\ref{DGP:logistic}) from RI, RegFE, and Group-FE for logistic regression}\label{fig:logistic_re}
    
    \vspace{-0.1in}
    
    \begin{center}
    
     \begin{subfigure}{0.48\textwidth}  
         \begin{center}
         \vspace{-.4in}
         \includegraphics[scale=.3]{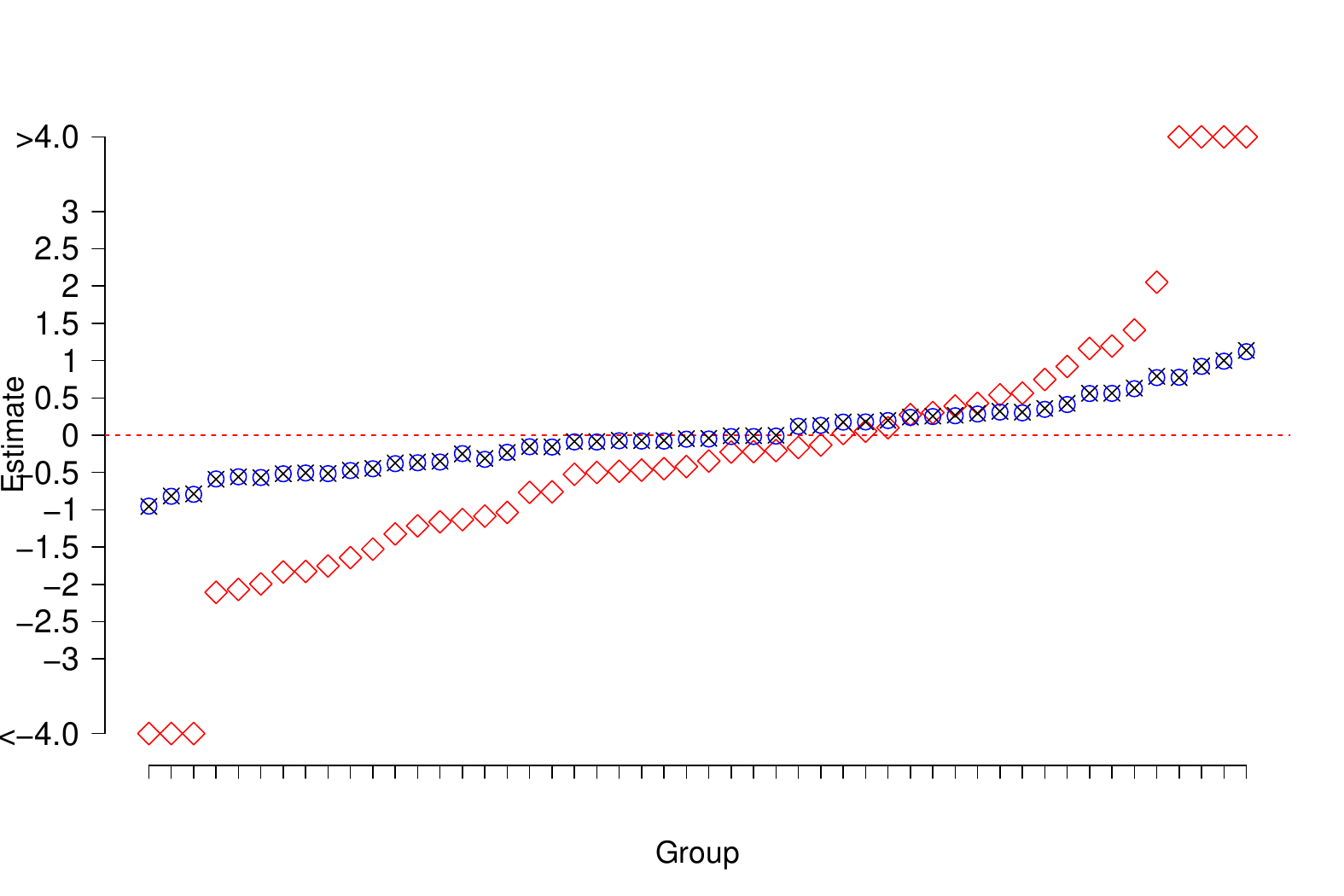}
         \subcaption{$n_g=5$}
         \end{center}
         \end{subfigure}
     \begin{subfigure}{0.48\textwidth}  
         \begin{center}
         \vspace{-.4in}
         \includegraphics[scale=.3]{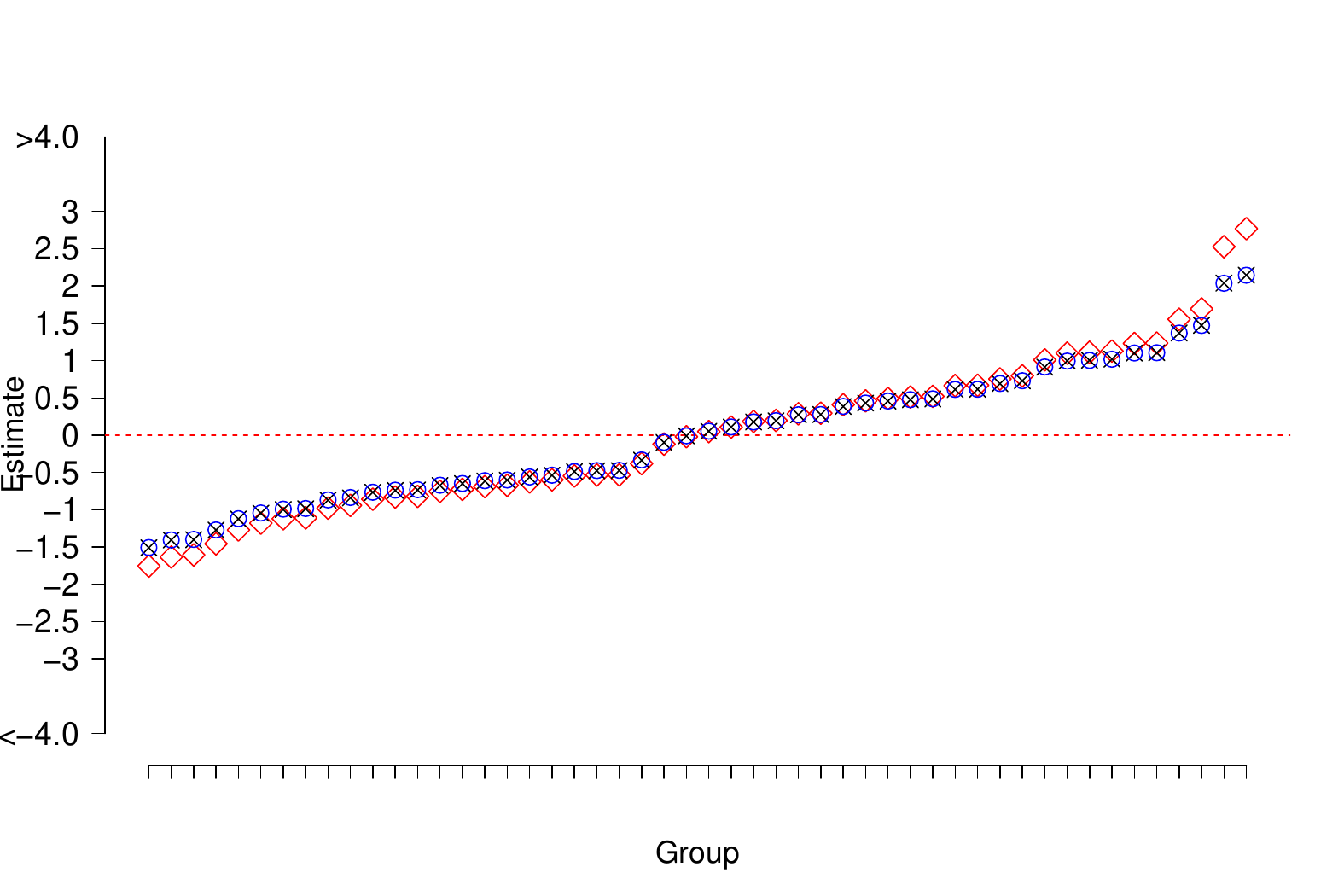}
         \subcaption{$n_g=50$}
         \end{center}
         \end{subfigure}
    
         \vspace{-0.5in}
         \begin{center}
         \includegraphics[scale=.4]{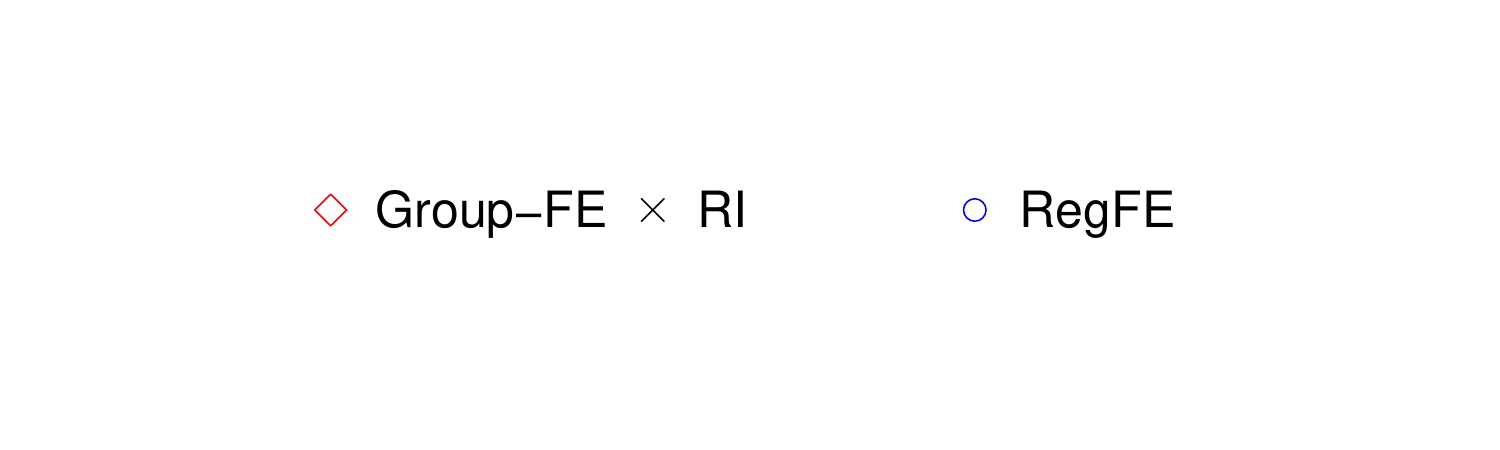}
         \end{center}
          \vspace{-0.5in}
          
    \caption*{\textit{Note:} Results across one iteration at each sample size of the DGP in (\ref{DGP:logistic}) with $G=50$. The RI model is a logistic regression RI model. The RegFE model is a logistic regression RegFE model with only varying intercepts, setting $\omega =  \hat\omega_{\text{RI}}$. The Group-FE model is a logistic regression Group-FE model. Estimates of each $\gamma_g$ from Group-FE are found by omitting the intercept term in $X_{g[i]}$ and retaining all group indicators in $Z_{g[i]}$, which yields a different intercept term for each group. The estimates of $\gamma_g$ for Group-FE plotted above are then the difference of each of these estimated intercepts from their overall average. The red dashed line represents an estimate of 0.}
    \vspace{-.3in}
    
    \end{center}
    \end{figure}

However, as the number of observations per group increases in Figure \ref{fig:logistic_noneq}, the coefficient estimates converge to one another, and at 50 observations per group the differences are slight. Further, both the RI and RegFE estimates act as shrinkage estimators, with median estimates of $\beta_1$ from simulation falling between Group-FE (which does not impose shrinkage; i.e., $\lambda_{\mathrm{GLM}} = 0$ in RegFE) and a GLM that only includes $X_{g[i]}$ as a regressor and omits fixed and random effects (which can be thought of complete shrinkage; i.e., $\lambda_{\mathrm{GLM}} = \infty$ in RegFE). Figure~\ref{fig:logistic_re} further demonstrates that the estimates of $\gamma$ from RegFE and RI are, in the large majority of cases, contracted towards 0 compared to Group-FE's estimates of $\gamma$. Shrinkage is most pronounced when group sizes are small ($n_g=5$), and very slight when group sizes are large ($n_g=50$). Additionally, the RegFE and RI estimates of $\gamma$ are approximately equal. Appendix~\ref{app:pois_gamma_sims} shows similar results for an application of MLM and RegFE to Poisson regression, with estimates that are even closer than in the logistic regression case considered here. 

In summary, MLE-fit MLM methods yield similar estimates of $\beta$ and $\gamma$ to those from RegFE when $\lambda_{\mathrm{GLM}}$ is fixed to the MLM estimates, which is unsurprising given the connection between the objective functions the two models optimize. Therefore, as in the linear setting, we continue to think of GLM MLM as regularizing its random effect coefficients, $\gamma_g$. This regularization on the random effects explains why MLMs can include group-level variables as regressors while FE cannot. In MLM, the regularization of the $\gamma$ prevents group-level variables from creating collinearities between the columns of $Z$ and $X$; these variables cannot be included for FE because the collinearity leads to the loss of a unique solution. The apparent regularization in MLMs also explains MLM's superior out-of-sample prediction error---the shrinkage on $\gamma$ prevents MLMs from overfitting to sample data. 

As discussed above, shrinking the estimates for $\gamma$ towards 0 also leads to ``incomplete conditioning" in GLM MLMs, biasing the estimate $\hat\beta_\text{MLM}$ when group-level confounding is present. Consider the data generation process (\ref{DGP1}) described below, which includes unobserved group-level variables $W_g^{(1)}$ and $W_g^{(2)}$ which influence the outcome, and where $W_g^{(1)}$ is a confounder: 
     \begin{align}\label{DGP1}
        & Y_{g[i]}\overset{}{\sim}\text{Bernoulli}(\operatorname{logit}^{-1}(\beta_0+X_{g[i]}\beta_1+W_g^{(1)}+W_g^{(2)})) \tag{DGP 1} \\ 
        \text{ where } \ \  & [W_g^{(1)} \ W_g^{(2)}]^{\top} \overset{iid}{\sim}\mathcal{N}(\Vec{0},I_2) \ \ \text{and} \ \ X_{g[i]}{\sim}N(W_g^{(1)},0.5) \nonumber 
    \end{align}
Here, MLM estimates of the $\gamma_g$ are shrunken towards 0, and cannot fully absorb the effect of the random intercept $(W_g^{(1)}+W_g^{(2)})$; therefore, some of the effect of the confounding is left unaccounted for. But because $W_g^{(1)}$ covaries with $X_{g[i]}$, some of this remaining effect can be captured through the bias of MLM's estimate of $\beta_1$. 

\begin{figure}
    \caption{Estimates of $\beta_1$ in \ref{DGP1} from RI, RegFE, Group-FE, and GLM without fixed or random effects}

    \vspace{-0.15in}
    \label{fig:bias_MLM_FE}

\begin{center}

 \begin{subfigure}{0.48\textwidth}  
     \begin{center}
     \vspace{-.5in}
     \includegraphics[scale=.45]{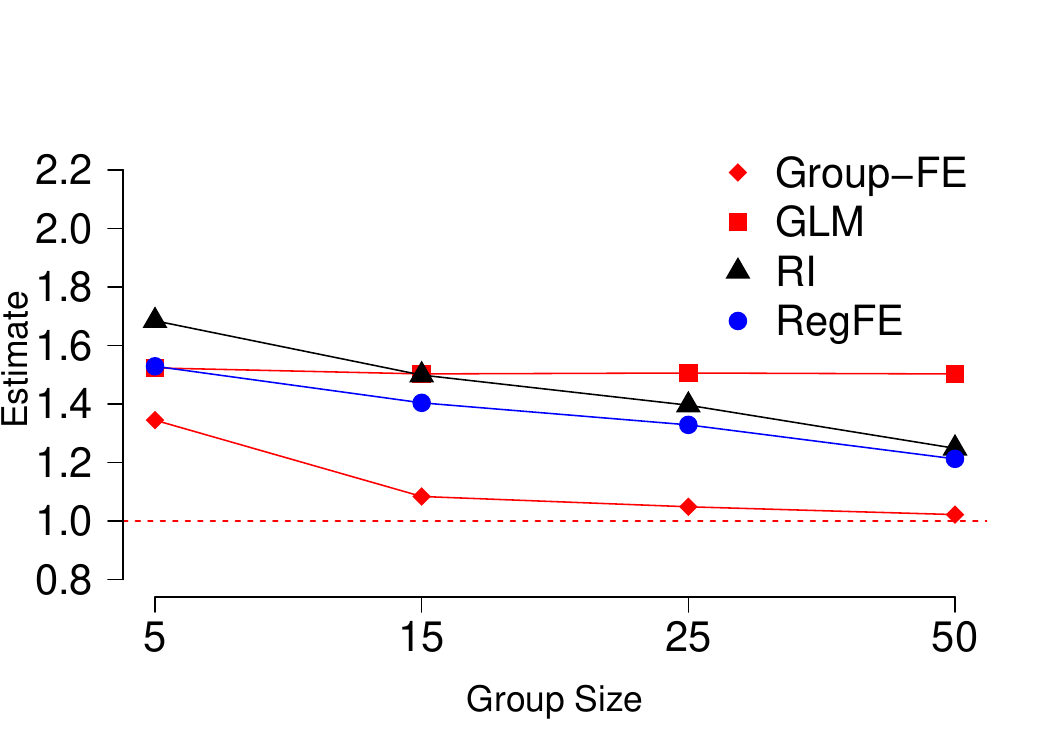}
     \subcaption{Mean estimates for $G=50$}
     \end{center}
     \end{subfigure}
 \begin{subfigure}{0.50\textwidth}  
     \begin{center}
     \vspace{-.5in}
     \includegraphics[scale=.45]{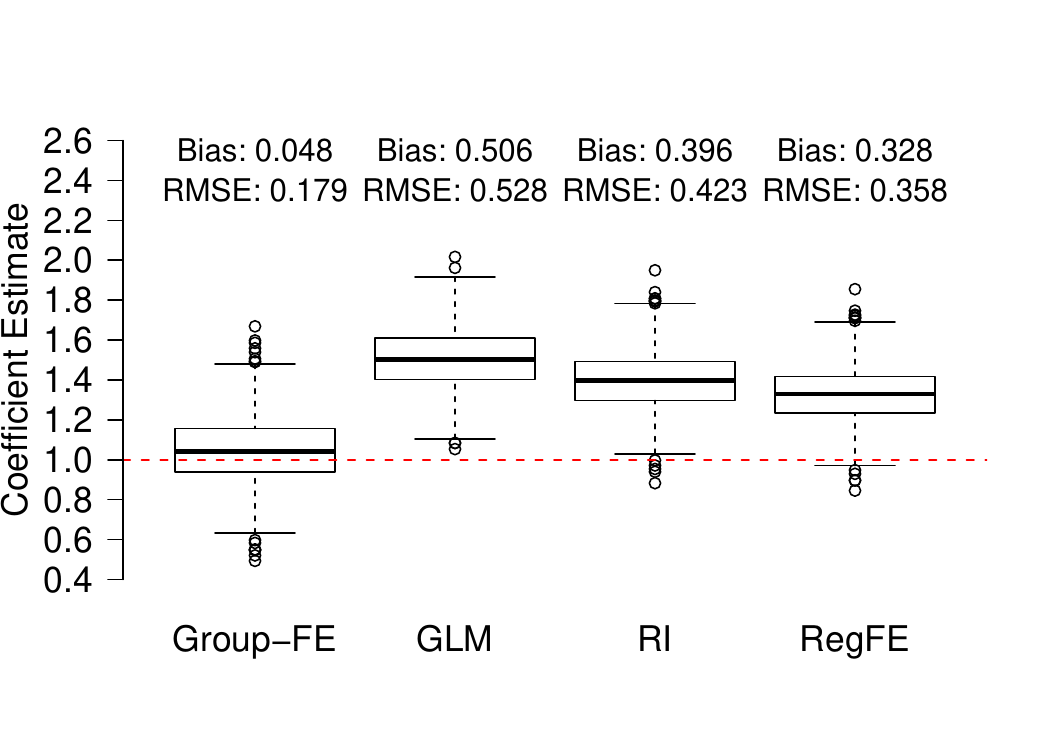}
     \subcaption{Estimate distributions for $G=50$ and $n_g=25$}
     \end{center}
     \end{subfigure}
      
\caption*{\textit{Note:} Results across 1000 iterations at each sample size of \ref{DGP1}. The RI model is a logistic regression RI model. The RegFE model is a logistic regression RegFE model with only varying intercepts, setting $\omega = \hat\omega_{\text{RI}}$. The Group-FE model is a logistic regression Group-FE model. The GLM model is a logistic regression that only includes $X_{g[i]}$ as a regressor, and omits fixed and random effects. \textit{(a)} Mean estimates when $G=50$ for $\beta_1$. The red dashed line indicates the true value of $\beta_1 = 1$. \textit{(b)} Distribution of estimates of $\beta_1$ when $G=50$ and $n_g=25$. The red dashed line indicates the true value of $\beta_1 = 1$.}
\vspace{-.3in}

\end{center}
\end{figure}

Figure~\ref{fig:bias_MLM_FE} demonstrates RI's large bias in \ref{DGP1} at every sample size tried. When $n_g$ is larger than 15, we observe that RI has lower bias than does a logistic regression model that does not include group-varying intercepts. However, RI  has higher bias than does Group-FE among all sample sizes simulated. Further, we observe that while RI has similar, or slightly lower variance to Group-FE when $n_g = 25$, it has far greater RMSE. This is not surprising, given the substantial bias at this group size. RegFE performs similarly to RI, given that its regularization also implies incomplete conditioning, though it has slightly lower bias at all sample sizes tried here.

H\&W demonstrated that a similar result occurs in the linear setting with a comparable DGP. However, the key difference in the GLM case is that, although Group-FE tends to have the least bias in Figure~\ref{fig:bias_MLM_FE}, estimates nevertheless are still noticeably biased. This is in contrast to the linear setting, in which Group-FE estimates are unbiased.\footnote{This is true as long as $X_{g[i]}$ is uncorrelated with the errors $\epsilon_{g[i]}$ in the model in (\ref{eq:linear}).} As previously mentioned, the bias from GLM Group-FE is a consequence of the well-documented finite-sample bias in MLE estimates for GLMs generally (e.g., \citealp{Gauss1991Bias}), and FE's incidental parameters problem (\citealp{neyman1948consistent, lancaster2000incidental}). The issue is particularly acute with grouped data when $n_g$ is small, as is demonstrated in Figure~\ref{fig:bias_MLM_FE}, because Group-FE then has little data in each group on which to base its $G$ intercept estimates. This bias ultimately influences our recommendations for non-linear GLMs, which differ from H\&W's recommendations in the linear setting. When group sizes are small, we instead recommend bias-corrected versions of MLM and RegFE, which will be explored in the next section.

Finally, while we focus on MLE-based estimation of MLMs, note that the lessons from this section, and from Sections~\ref{subsec:insight2} and~\ref{subsec:insight3} below, hold even when applying MLMs fit through PQL. We show this concretely in Appendix~\ref{app:pql}, where we replicate all of the demonstrations from Section~\ref{sec:insights} using PQL to fit MLMs. In fact, we find that the estimated coefficients from PQL-fit MLM are even more similar to those from RegFE than are coefficients from MLE-fit MLM. 

\subsection{Insight 2: Bias-correction for MLM and RegFE for GLMs}\label{subsec:insight2}

H\&W showed that in linear MLMs, it is possible to correct the bias in the estimate of $\beta$ resulting from correlated random effects. For RI, the correction, which originates from \cite{mundlak1978pooling}, is the inclusion of the group-level means of $X_{g[i]}$ as additional regressors. In the students-within-schools example, this involves adding the school-level averages of \textit{all} student-level covariates in $X_{g[i]}$ as regressors. H\&W refer to this modeling approach as ``bias-corrected MLM" (bcMLM).  This correction enables linear MLMs to obtain unbiased estimates for coefficients of individual-level covariates, providing estimates which are exactly equal to those from FE. Furthermore, bcMLM retains the ability from MLMs to estimate coefficients for group-level covariates, while also showing superior predictive accuracy. 

In this section, we examine the extension of bcMLM to the GLM setting, and a corresponding bias-corrected RegFE. We first consider the extension of bcMLM to GLMs with random intercepts, which we refer to as ``bias-corrected RI" (bcRI). As in the linear setting, we include group-level means $\bar{X}_g = \frac{1}{n_g} \sum_{i=1}^{n_g} X_{g[i]}$ as additional regressors to the RI model:
    \begin{align}\label{eq:MLMlm2}
         	\mu_{g[i]} &= h^{-1}(X^{\top}_{g[i]} \beta + \bar{X}_g^{\top} \alpha + \gamma_{g}), \ \ \ \gamma_g \ | \ X, Z \overset{iid} \sim N(0, \omega^2)
     \end{align}
Unlike the linear setting, this model does not necessarily produce equivalent estimates to those from Group-FE. Its estimates are also not necessarily unbiased, as MLE estimates can have finite-sample bias even when the model is correctly specified. Further, the magnitude of the model's bias depends on the true form of the conditional expectation (given $X$) of the random intercepts $\gamma_g$ from (\ref{eq:group_glm}). The bcRI model specifies that, in the general varying intercepts model in (\ref{eq:group_glm}),
    \begin{align}\label{eq:linear_re_assumption}
         	\E ( \gamma_g | X_g ) = d_1\bar{X}_{g} + d_2
     \end{align}
\noindent for $d_1, d_2 \in \mathbb{R}$. \cite{goetgeluk2008conditional} show that if the conditional expectation of $\gamma_g$ is not linear in $\bar{X}_g$, then bcRI may produce asymptotically inconsistent estimates of $\beta$. \cite{brumback2013adjusting} also presents a DGP where bcMLM's bias is substantial. 

\begin{figure}[!h]
    \caption{
    Estimates of $\beta_1$ in \ref{DGP1} from bcRI, Group-FE, uncorrected RI, and bcRegFE}
    \label{fig:estimates}
    \vspace{-.35in}
    \begin{subfigure}{.50\textwidth}
        \includegraphics[scale=0.48]{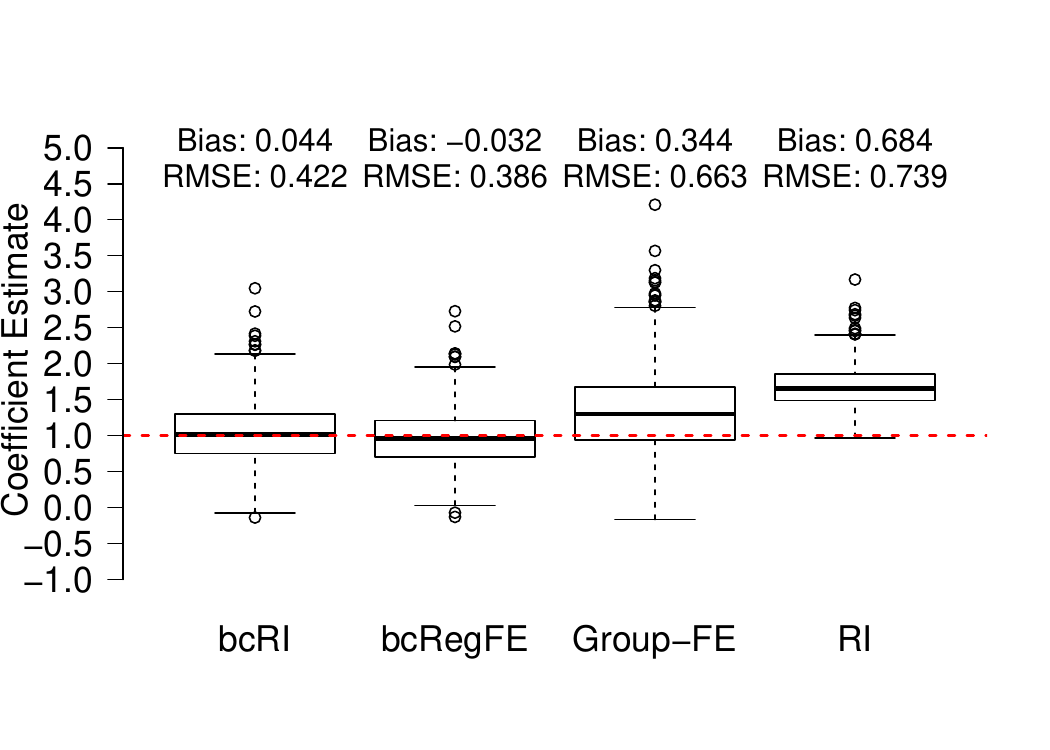}
        \vspace{-.20in}
        \subcaption{$G=50$ and $n_g=5$}
    \end{subfigure}
    \begin{subfigure}{.50\textwidth}
        \includegraphics[scale=0.48]{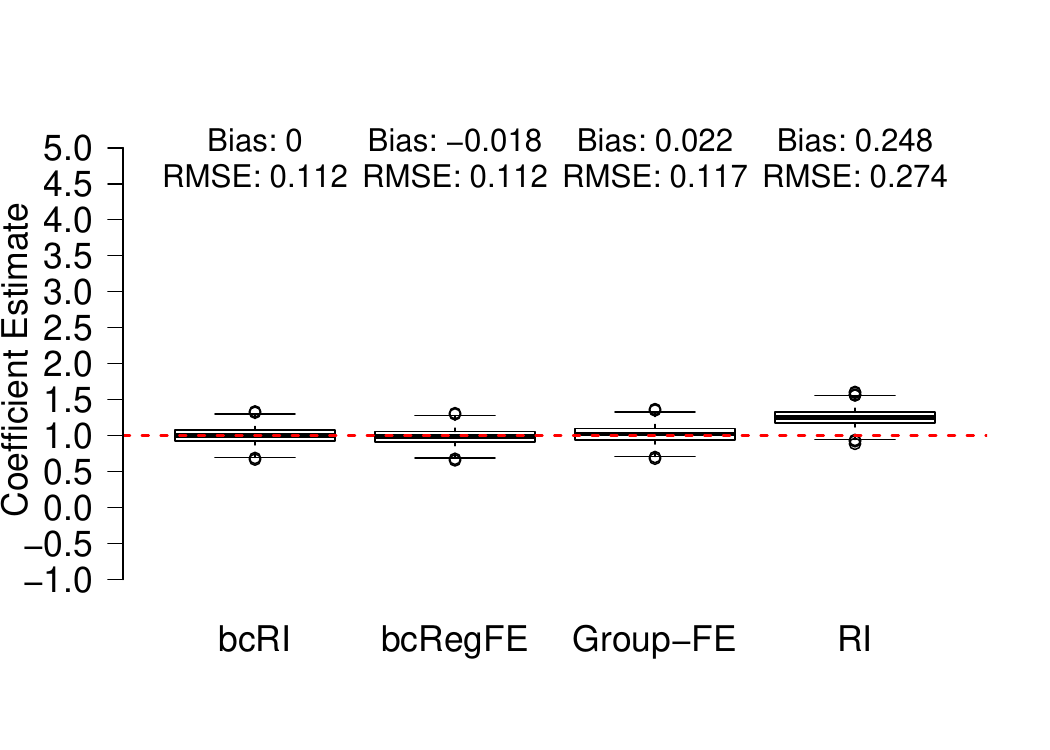}
        \vspace{-.20in}
        \subcaption{$G=50$ and $n_g=50$}
    \end{subfigure}

    \vspace{.25in}
    \subcaption*{\textit{Note:} Results across 1000 iterations at each sample size of DGP 1. Distributions of estimates for $\beta_1$ by logistic regression applications of Group-FE, bcRI, a varying intercepts bcRegFE model with $(\theta, \Omega) = (\hat{\theta}_{\mathrm{MLM}}, \hat{\Omega}_{\mathrm{MLM}})$, and uncorrected RI. The dashed horizontal line represents the true parameter value, $\beta_1 = 1$.}
\end{figure}

However, \cite{goetgeluk2008conditional} argue that bcMLM's bias is usually slight in more realistic scenarios. To illustrate, consider again DGP 1 from Section 3.1---Figure~\ref{fig:estimates} plots the distributions of the logistic regression estimates of $\beta_1$ from bcRI, Group-FE, and uncorrected RI. bcRI produces minimally-biased estimates of $\beta_1$, and has considerably less bias than does Group-FE at smaller sample sizes, and than does uncorrected RI at both large and small sample sizes. Further, in smaller samples, bcRI has lower variance than does Group-FE, which combined with lower bias yields much lower RMSE. bcRI and Group-FE perform similarly when group sizes are large. While in simulations with $n_g = 5$, RI tends to have the lowest variance among the estimators analyzed, it nevertheless has RMSE greater than that of bcRI and Group-FE, due to its substantial bias.
 
\begin{figure}[!h]
    \vspace{.05in}
    \caption{Average test error rates of bcRI, bcRegFE, and Group-FE in DGP 1}
    \label{fig:prediction}
        \vspace{-.35in}
    \begin{subfigure}{.49\textwidth}
        \includegraphics[scale=0.50]{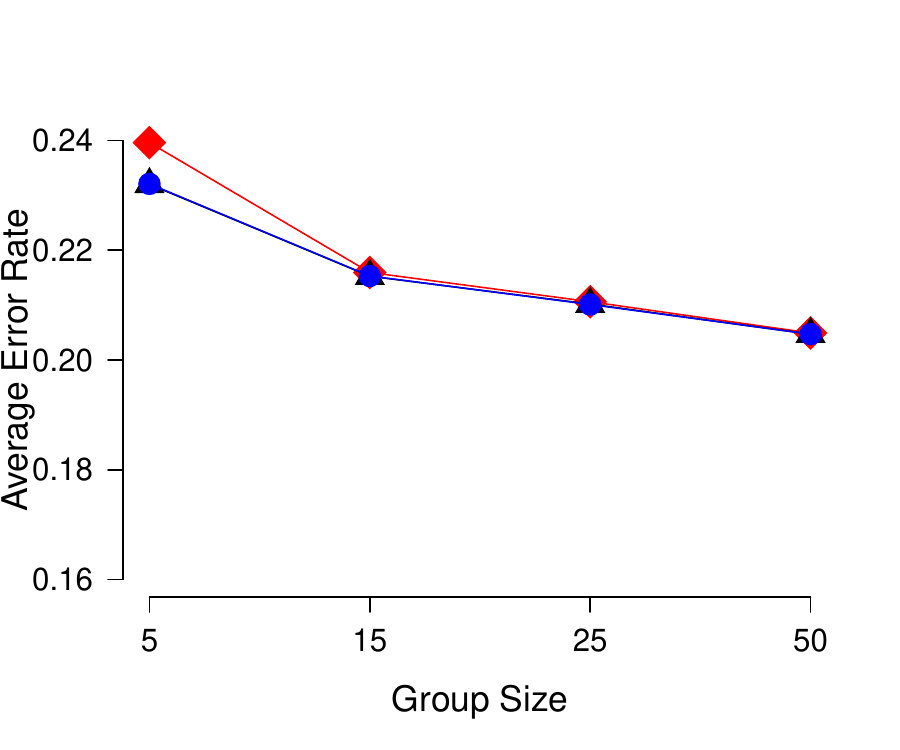}
        \subcaption{$G=15$}
    \end{subfigure}
    \begin{subfigure}{.49\textwidth}
        \includegraphics[scale=0.50]{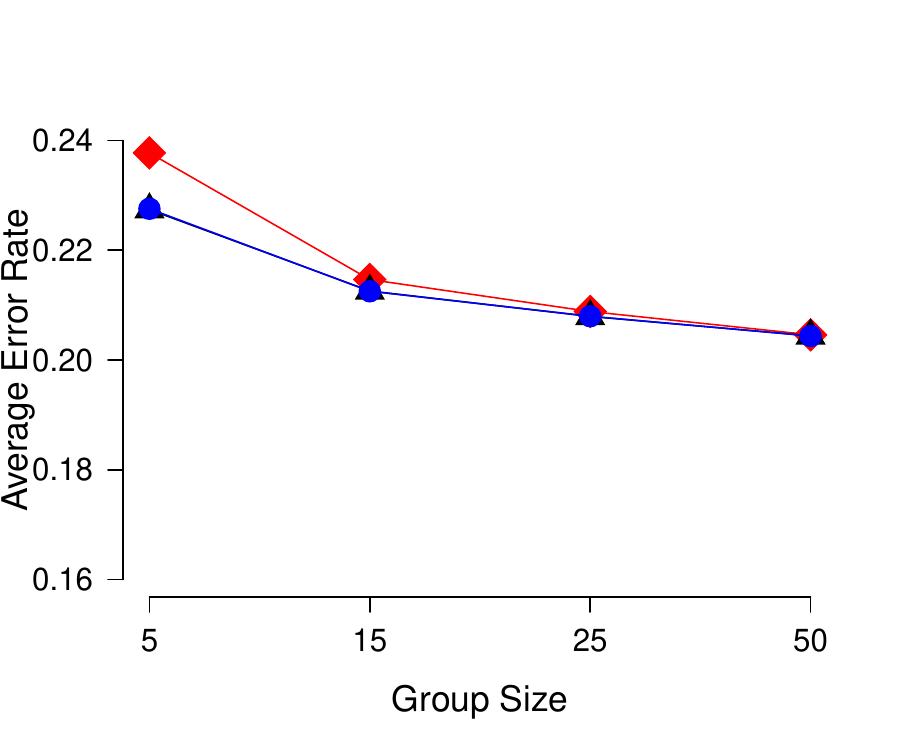}
        \subcaption{$G=50$}
    \end{subfigure}
    
         \vspace{-0.6in}
     \begin{center}
     \includegraphics[scale=.4]{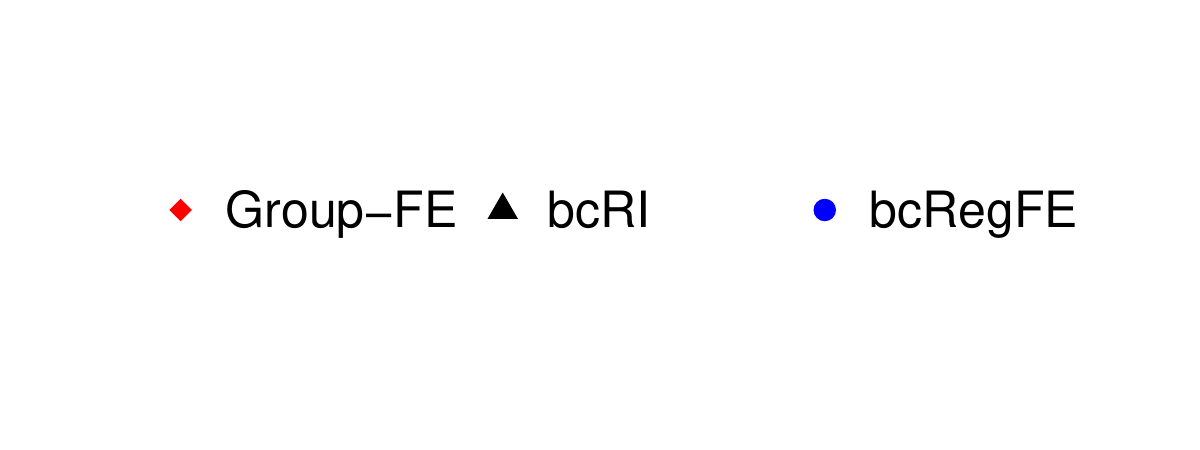}
     \end{center}
      \vspace{-0.6in}
      
    \subcaption*{\textit{Note:} Results across 1000 iterations at each sample size of DGP 1. Comparison of the average test error rates of logistic regression applications of Group-FE, bcRI, and a varying intercepts bcRegFE model with $(\theta, \Omega) = (\hat{\theta}_{\mathrm{MLM}}, \hat{\Omega}_{\mathrm{MLM}})$. The training and testing datasets were of the same size.}
    \vspace{-0.15in}
    \end{figure}

bcRI also retains its superior predictive accuracy over Group-FE in the GLM setting. To see this, we evaluate each model's classification accuracy, as measured by the proportion of incorrect predictions on a test data set with a binary response variable.\footnote{To make a prediction for a given test point, we first calculate predicted probabilities from each model based on the point's covariate values $X_{\mathrm{test}}$ and group $g_{\mathrm{test}}$. Then, if the predicted probability from a model is under 0.50, the model predicts $\hat{Y}_{\mathrm{test}} = 0$. If the predicted probability is over 0.50, then the model predicts $\hat{Y}_{\mathrm{test}} = 1$.} Figure~\ref{fig:prediction} shows the test error rates from test data generated according to DGP 1. Because of the regularization induced by random effects, bcRI has lower test error rates than does Group-FE, most notably when group sizes are small.
Given the lower bias and improved prediction in comparison to Group-FE, the prevailing recommendation in the literature is to use bcMLM to analyze multilevel data in real-world settings (e.g., \citealp{raudenbush2009adaptive}; \citealp{bell2019fixed}; \citealp{schunck2017within}).\footnote{Alternatives include the Conditional Likelihood approach (\citealp{neuhaus2006separating}), which produces consistent estimates of $\beta$, but it is only applicable for linear and logit link functions. See also \cite{goetgeluk2008conditional} for the conditional generalized estimating equations (CGEE) approach. Although, \cite{brumback2010adjusting} note that CGEE does not estimate $\gamma$ or admit link functions other than the identity and the exponential functions.} Less importantly, unlike FE, bcMLM can estimate coefficients for group-level covariates, though the estimates may be unreliable if the group-level covariate is not independent of the random effects (\citealp{hazlett2022understanding}).

Next, we consider the general extension of bcMLM to GLMs that can include random coefficients beyond just random intercepts. For linear MLMs, H\&W describe a procedure analogous to bcRI that debiases estimates of $\beta$ and produces the same estimates as general FE. The approach projects the fixed effect variables ($X_g$), excluding the intercept, onto the random effect variables ($Z_g$) within each group to obtain $\tilde{X}_{g[i]} = Z_{g[i]}^T (Z_g^T Z_g)^{-1}Z_g^T X_g$, and includes $\tilde{X}_{g[i]}$ as ``fixed effect" regressors in the model. In the students-within-schools example, this involves first performing linear regressions \textit{within each school} that predict \text{each} student-level covariate in $X_{g[i]}$ with the covariates in $Z_{g[i]}$. Then, the vectors $\tilde{X}_{g[i]}$ are the predicted values from these school-specific regressions, and are added as regressors. The extension of bcMLM to GLMs does the same:
    \begin{align}\label{eq:projections}
         	\mu_{g[i]} = h^{-1}(X_{g[i]}^T \beta + \tilde{X}_{g[i]}\alpha + Z_{g[i]}^{\top} \gamma_g), \ \ \ \gamma_g | X, Z \overset {iid} \sim N(0, \Omega) \tag{bcMLM}
     \end{align}
As in the RI case, this bias correction procedure for GLM bcMLM no longer guarantees the same estimates for $\beta$ as those from FE, and does not guarantee unbiasedness. However, it tends to show far less bias than FE in simulated examples (see Appendix \ref{app:rs}). 

Finally, we consider what we call ``bias-corrected RegFE" (bcRegFE). As with moving from uncorrected MLM to bcMLM, this involves simply adding the $\tilde{X}_{g[i]}$ variables from bcMLM as regressors to an uncorrected RegFE model. Figures~\ref{fig:estimates} and~\ref{fig:prediction} include estimates of $\beta_1$ and test error rates from a varying intercepts (i.e., $Z_{g[i]} = 1$) bcRegFE model, in which the group means $\tilde{X}_{g[i]} = \bar{X}_g$ are included regressors, as with bcRI. In Figure~\ref{fig:estimates}, bcRegFE's estimates of $\beta_1$ are very similar to bcRI's estimates, showing little to no bias, and much lower bias than Group-FE when $n_g=5$. Additionally, in Figure~\ref{fig:prediction}, bcRegFE has very similar test error rates to those of bcRI. In summary, although RegFE and MLM do not produce identical estimates in all GLMs, bcRegFE may be a similarly good option to bcMLM to correct for bias from correlated random effects and to avoid FE's bias from its incidental parameters problem, while retaining strong predictive accuracy on new data.

\subsection{Insight 3: Variance estimation}\label{subsec:insight3}

From their review of journal articles, H\&W note that one reason commonly provided by researchers for employing MLM is that it correctly estimates standard errors in grouped data. This statement, as H\&W point out, is incorrect. Instead, MLM makes stringent assumptions on the intragroup dependence structure that are often violated in practice. For example, the mostly commonly used linear RI model assumes that, for $i \neq i'$,
    \begin{align}\label{eq:ri_cov}
     	\text{Cov}(Y_{g[i]}, Y_{g[i']} \ | \ X, Z) = \omega^2
     \end{align}
In other words, linear RI relaxes the assumption of independence between observations, even those within the same group, of the traditional OLS standard error, and instead models the covariance between observations in the same group. However, the model assumes the covariance is \textit{constant}. This assumption is easily violated, which can lead to standard error estimates that are too small. A prime example is when the data are longitudinal, and observations are auto-correlated. To illustrate, the data may describe high school students (the groups here) over multiple grades in school (the observations). Here, it is very plausible that (after accounting for the covariates) a student's outcomes in 9th grade are more similar to their outcomes in 10th grade than they are to their outcomes in 12th grade. However, the dependence structure in (\ref{eq:ri_cov}) does not allow for this, and instead assumes that (after accounting for the covariates) each grade's outcomes are \textit{equally} correlated with each other. This problem extends to MLMs with a large number of random coefficients, as they also assume a dependence structure that could be misspecified.

To solve this  problem in the linear setting, H\&W recommend applying cluster robust standard errors (CRSEs) to linear MLMs. CRSEs require fewer assumptions than do default MLM standard errors---they only assume independence between groups, but impose no assumption on the intragroup dependence structure, instead learning the structure from the data (see \citealp{cameron2015practitioner} and H\&W for more detail). H\&W also show an equivalence between CRSEs from FE and bcMLM, and demonstrate that applying CRSEs to linear MLMs, if provided enough data, essentially eliminates undercoverage of confidence intervals formed using MLM's default standard errors. 

In this section, we discuss variance estimation in the GLM setting. We first consider the dependence structure implied by MLMs in GLMs. For any MLM, the conditional covariance between outcomes in the same group is
  \begin{align}\label{eq:mlm_cov1}
        \text{cov}(Y_{g[i]}, Y_{g[i']} | X, Z) =  \E \biggr( \cov(Y_{g[i]}, Y_{g[i']} | X, Z, \gamma) \biggr| X, Z \biggr) + \cov ( \mu_{g[i]}, \mu_{g[i']} | X, Z )
    \end{align}
by the Law of Total Covariance. The MLMs studied here specify that the only dependence between outcomes from the same group arises from the random effect $\gamma_g$, which means that $\cov(Y_{g[i]}, Y_{g[i']} | X, Z, \gamma) = 0$ for $i \neq i'$ in (\ref{eq:mlm_cov1}). Thus, we first focus on the right-most covariance term in (\ref{eq:mlm_cov1}). For a general link function $h$, this covariance does not necessarily have a closed form. 
However, in the special case of Poisson RI regression with the canonical log link, where
    \begin{align}
        Y_{g[i]} \sim \mathrm{Pois} (\lambda = \mu_{g[i]}) \ \ \text{with} \ \ \mu_{g[i]} = \mathrm{exp} (X_{g[i]}^{\top} \beta + \gamma_g)
    \end{align}
the  right-most covariance term in (\ref{eq:mlm_cov1}) is:
    \begin{align}\label{eq:cov_pois}
        \cov ( \mu_{g[i]}, \mu_{g[i']} | X, Z ) &= \exp(X_{g[i]}^{\top} \beta + X_{g[i']}^{\top} \beta) \var ( e^{\gamma_g} | X, Z) \nonumber \\
        &= \exp(X_{g[i]}^{\top} \beta + X_{g[i']}^{\top} \beta) (e^{2\omega^2} - e^{\omega^2} )
    \end{align}
Thus, unlike the linear setting in (\ref{eq:ri_cov}), the covariance between units in the same group is not necessarily constant in the RI model---instead, it is an increasing function in $X_{g[i]}^{\top} \beta$ and $X_{g[i']}^{\top} \beta$, and scaled by a function of the variance of the random effect (i.e.,  $e^{2\omega^2} - e^{\omega^2}$). See Appendix~\ref{app:dependence_structure} for a more extensive analysis of the right-most covariance term in (\ref{eq:mlm_cov1}). 

\begin{figure}[!h]
        \caption{Coverage rates of 95\% confidence intervals for $\beta_1$ in DGP 2}
    \label{fig:coverage}
        \vspace{-.35in}
    \begin{subfigure}{.24\textwidth}
        \includegraphics[scale=0.39]{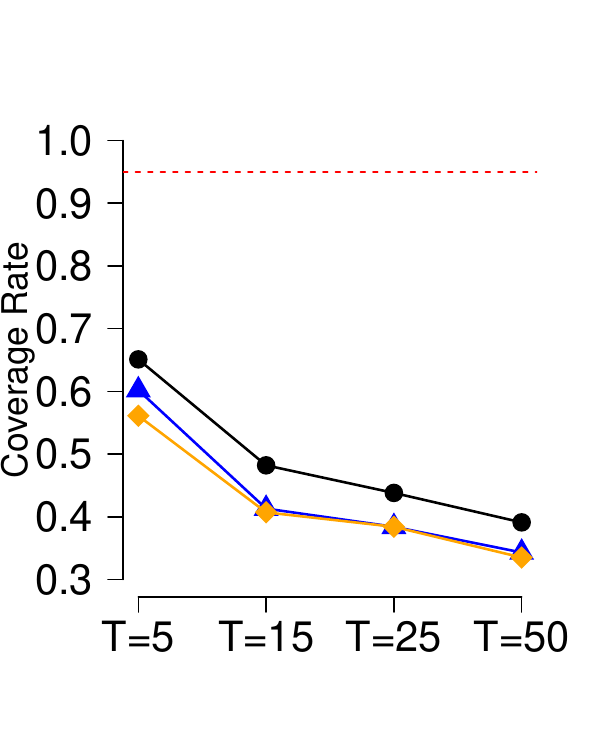}
        \subcaption{RI}\label{fig:coverage_a}
    \end{subfigure}
    \begin{subfigure}{.24\textwidth}
        \includegraphics[scale=0.39]{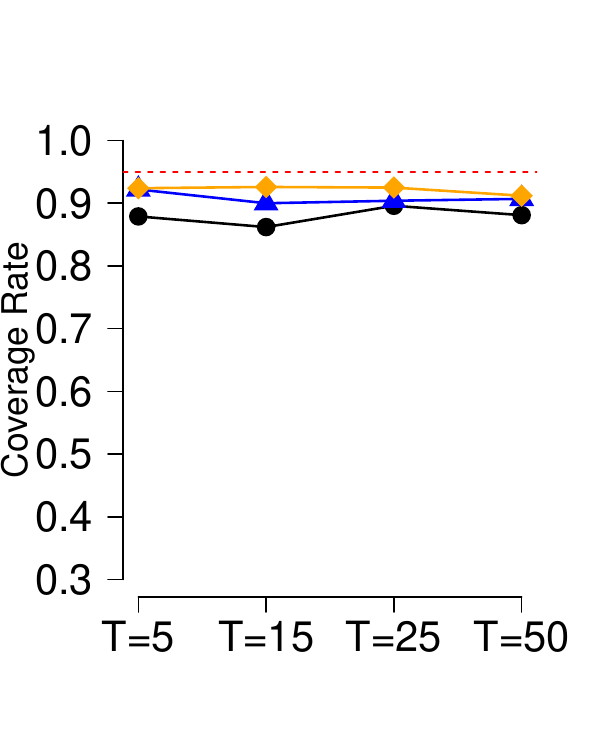}
        \subcaption{RI, Bootstrap}\label{fig:coverage_b}
    \end{subfigure}
    \begin{subfigure}{.24\textwidth}
        \includegraphics[scale=0.39]{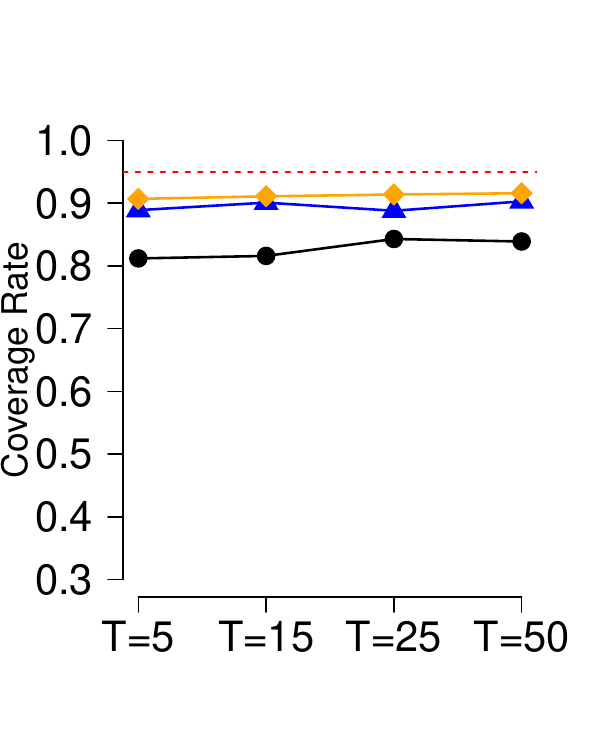}
        \subcaption{Group-FE, CRSE}\label{fig:coverage_c}
    \end{subfigure}
     \begin{subfigure}{.24\textwidth}
        \includegraphics[scale=0.39]{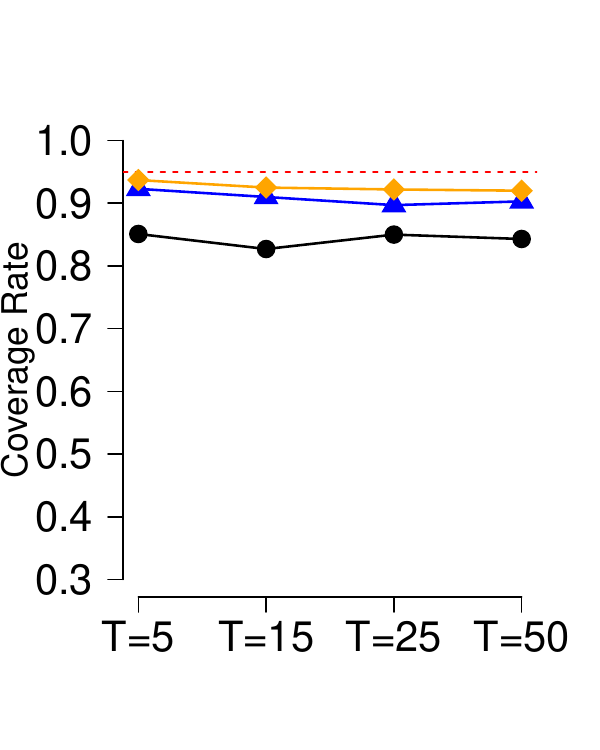}
        \subcaption{RegFE, CRSE}\label{fig:coverage_d}
    \end{subfigure}   
         \vspace{-0.6in}
     \begin{center}
     \includegraphics[scale=.4]{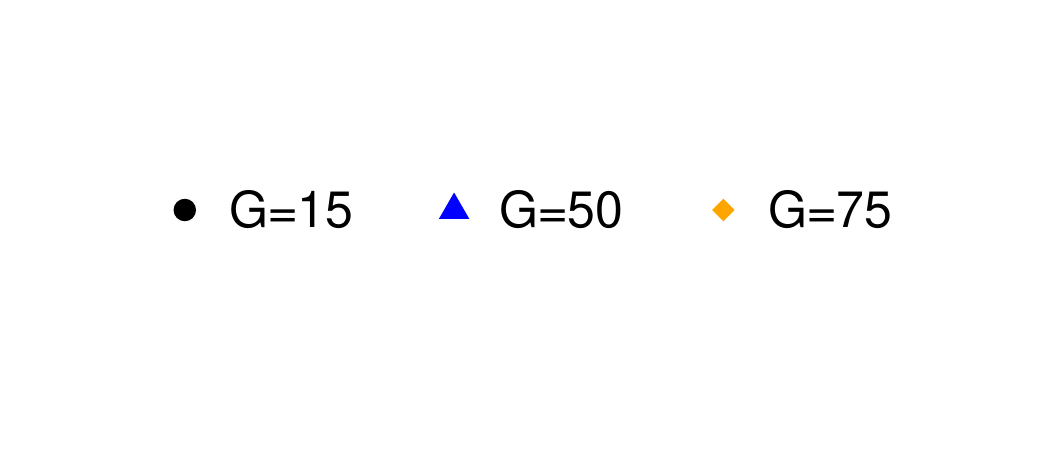}
     \end{center}
      \vspace{-0.6in}
    \subcaption*{\textit{Note:} Results across 1000 iterations at each sample size of DGP 2. Comparison of the coverage rates of 95\% confidence intervals constructed using \textit{(a)} 
    Poisson RI with its default standard errors; \textit{(b)} Poisson RI with a percentile  cluster-bootstrap (using 200 bootstrap samples); \textit{(c)} Poisson Group-FE with CRSEs; and \textit{(d)} Poisson RegFE with CRSEs. The dashed horizontal line shows the target 95\% coverage rate.}
    \end{figure}

Of course, it is still possible that MLM model specifications are violated by the true data generation process. For example, it is possible that intragroup dependence does not only arise from $\gamma_g$---it may be that $\cov(Y_{g[i]}, Y_{g[i']} | X, Z, \gamma) \neq 0$ for $i \neq i'$ in (\ref{eq:mlm_cov1}). Consider the following longitudinal data generation process (\ref{eq:dgp2}), where $g$ indexes an individual and $t=1, \dots, T$ indexes the time-point:
\begin{align*}\label{eq:dgp2}
        & Y_{g[t]} \sim \mathrm{Pois} (\lambda = \mathrm{exp} (\beta_0 + \beta_1 X_{g[t]} + W_g + \epsilon_{g[t]})) \tag{DGP 2} \\
        \text{where} \ \ & W_g \overset{iid}{\sim} N(0, 1), \ \  X_{g[t]} \overset{iid}{\sim} N(0, 0.5) \\
        & \epsilon_{g[t]} \sim N(0, 0.5) \ \ \text{and} \ \ \cor(\epsilon_{g[t]}, \epsilon_{g[t+k]}) = (0.75)^k
\end{align*}
For example, $g$ might index students in the data who are measured over multiple time points $t$ (e.g., grade years in school). Despite misspecifying the true model, Poisson RI and Group-FE both show negligible bias for $\beta_1$ in this DGP.\footnote{This finding is aligned with the result in \cite{davis2000autocorrelation} that MLE coefficient estimates from a Poisson GLM are consistent in a similarly autocorrelated Poisson DGP, albeit one where the data is not clustered.} However, the left-most covariance term on the right hand side of (\ref{eq:mlm_cov1}) is non-zero for \ref{eq:dgp2} because of the inclusion of the unobserved $\epsilon_{g[t]}$, which are autocorrelated. Thus, Poisson RI substantially misspecifies the intragroup dependence structure, and one should expect its traditional standard errors to be biased. This is evident in Figure~\ref{fig:coverage_a}, where 95\% confidence intervals for Poisson RI using default, MLE-based standard errors show coverage rates for $\beta_1$ well below the target rate of 95\%.

CRSEs would be useful in practice for MLMs in the GLM setting, just as CRSEs are in the linear setting. Though while CRSEs have been generalized to FE in GLMs (e.g., \citealp{angrist2009mostly}; \citealp{cameron2008bootstrap}), at present, the authors are unaware of a comprehensive extension to MLE-fit MLMs in the GLM framework.\footnote{
CRSEs come naturally for estimates of $\beta$ from the Generalized Estimating Equations (GEE; \citealp{liang1986longitudinal}) approach, a general estimation framework that can accommodate MLM's assumptions in (\ref{eq:ri}) or (\ref{eq:mlm_general}) (e.g., see Section 7.9 of \citealp{demidenko2013mixed}). In fact, in the linear case, the MLE estimate of $\beta$ from MLM is exactly a GEE estimator for a specific choice of ``working" covariance structure, which is one reason why CRSEs are available for linear MLMs. However, we are unaware of extensions to non-linear GLM MLMs estimated through an approximate MLE-based approach, such as Laplace Approximation or  Gauss-Hermite Quadrature.
} However, one potential remedy is the use of a cluster bootstrap method. A cluster bootstrap is obtained in a similar manner to the traditional bootstrap. However, instead of sampling $N$ observations from the entire data set with replacement, a cluster bootstrap samples $G$ \textit{groups} with replacement. Figures~\ref{fig:coverage_b} and~\ref{fig:coverage_c} report the coverage rates of 95\% confidence intervals for $\beta_1$ in \ref{eq:dgp2} from Poisson RI using a cluster-bootstrap and Poisson Group-FE with CRSEs, respectively. When $G=50$ and $G=75$, confidence intervals from RI with the cluster-bootstrap show very  slight undercoverage, with coverage rates in the 90-95\% range. Group-FE with CRSEs also shows slight undercoverage, hovering around 90\% when $G=50$, and in the 90-95\% range when $G=75$. When $G=15$, both RI with the cluster bootstrap and Group-FE show consistent undercoverage, although RI's cluster boostrap (just below 90\%) is consistently superior to Group-FE (around 80-85\%). This undercoverage when $G=15$ is not surprising---\cite{cameron2015practitioner} suggest that 20 to 50 groups may be required for stable CRSEs and a cluster-bootstrap. Further, the asymptotic validity of cluster-robust inference relies on $G \rightarrow \infty$, which may not be realistic in a given setting. Nevertheless, even with smaller $G$ the additional permissiveness of cluster-robust inference to model misspecifications may still be preferable over an incorrect dependence structure specified by MLM---the coverage rates at $G=15$ for Group-FE with CRSEs and RI with a cluster bootstrap are much closer to the target rate of 95\% than are those from RI with its default standard errors (in Figure~\ref{fig:coverage_a}). 
    
There is also a natural extension of CRSEs to RegFE. We leave the details of this extension to Appendix~\ref{app:crses_regfe}, but implement these CRSEs with RegFE in \ref{eq:dgp2} here. Figure~\ref{fig:coverage_d} reports the associated coverage rates. As with the cluster-bootstrap for RI, and CRSEs for Group-FE, these CRSEs for RegFE struggle when the number of groups is smaller ($G=15$), but achieve near nominal coverage rates when $G$ is large ($G=50$ or $75$). Thus, (bc)RegFE with CRSEs may be a good alternative to (bc)MLM with a cluster-bootstrap if the boostrap is too computationally intensive, and to FE in order to avoid FE's bias concerns in certain GLMs.

\section{Application 1: Twelfth grade math in L.A. Unified}\label{sec:application}

We now demonstrate our lessons from Section~\ref{sec:insights} in a real-data setting: estimating the effects of taking a mathematics course in the twelfth grade for a student in the Los Angeles Unified School District (L.A. Unified) of Los Angeles, California, United States of America (USA). This application uses data from the \href{https://laeri.luskin.ucla.edu/}{Los Angeles Education Research Institute} (LAERI) data archive—a longitudinal archive of L.A. Unified student, staff, and school-level administrative and survey data. 

L.A. Unified is the second-largest school district in the USA, serving more than 500,000 students, of which over 100,000 are in high school (\citealp{lausd2025finger}). 
L.A. Unified requires that students complete three years of high school (grades 9 through 12) mathematics in order to graduate high school (\citealp{lausd2025pathways}). Normatively, the corresponding three courses are Algebra 1, Geometry, and Algebra 2, occurring in grades 9, 10, and 11, respectively.\footnote{L.A. Unified allows other mathematics pathways, such as accelerated pathways that could involve taking one or more of these courses before ninth grade (\citealp{lausd2017middle}), or taking two or more of the courses in the same year (\citealp{lausd2025pathways}). L.A. Unified also allows alternative sequences, such as taking Financial Algebra 1 and/or 2 instead of traditional Algebra 1 and/or 2 (\citealp{lausd2025pathways}).} This leaves students the option to choose whether or not to take a mathematics course (or a course in a related field, such as statistics) in twelfth grade. Along with student-level factors such as interest in mathematics and grades in mathematics courses, schools can influence whether or not a student chooses to take a twelfth grade math through course offerings, course availability, and advising. Thus, analyzing the effects of taking a twelfth grade math class in this district requires accounting for both student- and school-level factors. 

\cite{wainstein2023access, wainstein2023success} originally analyzed the effects of taking math in twelfth grade in L.A. Unified on college enrollment, persistence, and success outcomes in two research reports, which were published by LAERI. The authors found evidence that, even after accounting for student characteristics and school, taking math in twelfth grade increased the likelihood of students enrolling in college in the years they examined, despite small decreases to their cumulative high school GPA. They also found that taking twelfth grade math, after accounting for student characteristics and school, increased the number of credits accumulated for \textit{some} groups of community college-going students, though these effects were less pronounced.

In their analyses, \cite{wainstein2023access, wainstein2023success} combined matching and weighting methods with linear, weighted least squares models.\footnote{The authors' primary estimation strategy uses ``preferential within-cluster matching"  (\citealp{arpino2016propensity}) to reduce the dependence on the linear model, per \cite{ho2007matching}. The authors then employ CRSEs for inference.} We re-analyze a subset of the data from their reports with the multilevel and fixed effects GLMs we have studied thus far. Due to data confidentiality, we cannot make the data and code from this analysis publicly available. However, in Appendix C, we demonstrate a similar analysis on mock, simulated data that was generated to mimic aspects of the real data (these aspects are publicly available in \citealp{wainstein2023access}). Appendix C also demonstrates our \href{https://github.com/lwainstein/mlmfe}{\texttt{mlmfe}} package, includes \texttt{R} code, and the simulated data is publicly available through \href{https://github.com/lwainstein/mlmfe}{\texttt{mlmfe}}.

\subsection{Data and methods}\label{subsec:application_data}

We focus on a subset of the students studied in the LAERI reports: students who, by the end of eleventh grade, had completed Algebra 1, Geometry, and Algebra 2; earned at least a C in these math courses, and thus had completed the math requirements for eligibility for admission to a public four-year university in California (e.g., California State University or University of California campuses); and had not taken any math courses beyond Algebra 2 (e.g., Precalculus). Per \cite{wainstein2023access}, the most common courses in mathematics (or a related field) that these students took in twelfth grade were Precalculus (62.6\%) and Statistics (17.4\%). The data we use describes 9,993 students in 110 high schools who were in twelfth grade in academic years 2016-17 and 2017-18. 

We consider three outcomes: (1) cumulative weighted GPA at the end of twelfth grade; (2) enrolling in a two-year or four-year college within one year of high school graduation; and (3) for students who eventually enrolled in a California community college, the number of college credits accumulated in the first two years.\footnote{We analyze \textit{whole} credits for this third outcome, so it can be modeled by Poisson regression. For example, if a student earned 20.5 credits in the first two years of community college, we treat that as earning 20 \textit{whole} credits.} We model these outcomes with linear, logistic, and Poisson regressions, respectively. To analyze community college credit accumulation, we restrict our sample to students who enrolled in a California community college within one year of graduating high school (3,541 students in 100 high schools). 

We estimate the effects of taking a mathematics course (e.g., Precalculus), or a course in a related field (e.g., Statistics), in twelfth grade on these three outcomes.\footnote{As do  \cite{wainstein2023access, wainstein2023success}, we treat taking a mathematics course, such as Precalculus, and a course in a related field, such as Statistics, as being the same ``treatment" for simplicity. However, note that \cite{wainstein2023access, wainstein2023success} do compare the effects of taking different types of courses (e.g., Precalculus versus Statistics).} When modeling the outcomes as a function of twelfth grade math course-taking, we control for both student-level factors and school. For student-level factors, the data includes various demographics variables (e.g., gender and race/ethnicity), academics-related variables (e.g., eleventh grade standardized test scores and GPAs), behavioral variables (e.g., behavioral grades and attendance rate), measures of self-perception (e.g., self-reported educational expectations), and the academic year in which the student was in twelfth grade.\footnote{We utilize exactly the same student-level covariates as do \cite{wainstein2023access, wainstein2023success}, and point readers to their LAERI reports for a complete list of these variables and how they are defined. One difference here is that the LAERI researchers include the squared terms of the continuous academics-related variables in their models, but we omit these terms because including them yielded instability in model convergence.}

\begin{figure}[!h]
    \caption{Visualization of school-level confounding in L.A. Unified}
    \label{fig:lausd}
    \vspace{-.35in}
    \begin{subfigure}{.50\textwidth}
        \includegraphics[scale=0.45]{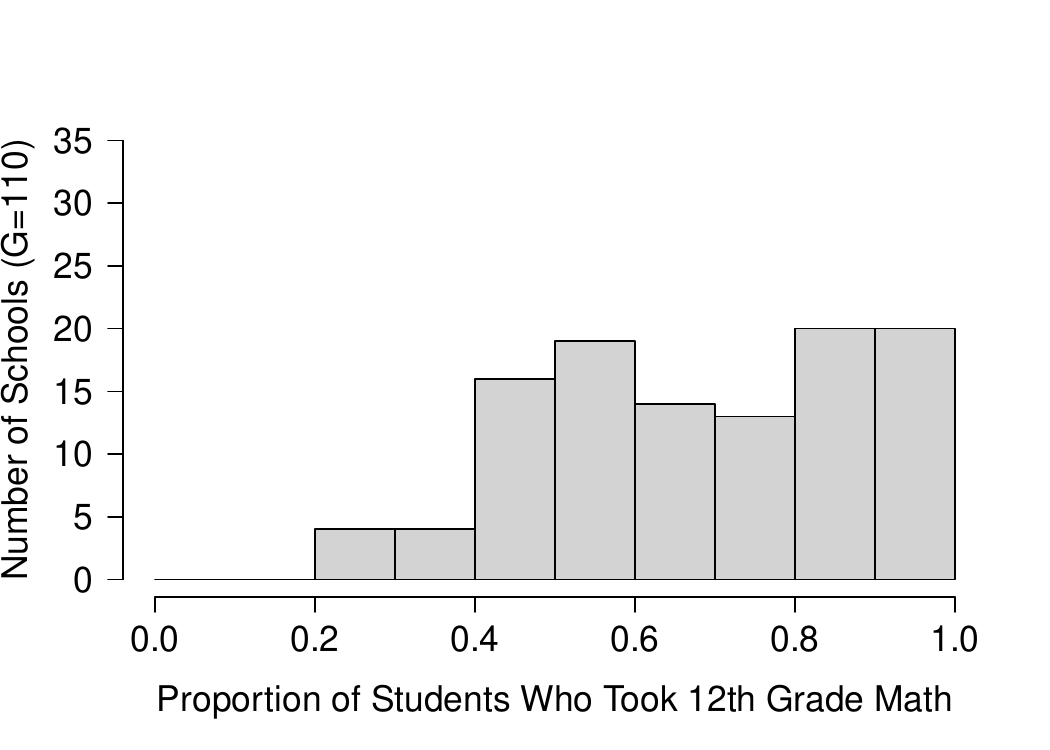}
        \vspace{-.05in}
        \subcaption{Treatment: Twelfth grade math course-taking}\label{fig:lausd_math}
    \end{subfigure}
    \begin{subfigure}{.50\textwidth}
        \includegraphics[scale=0.45]{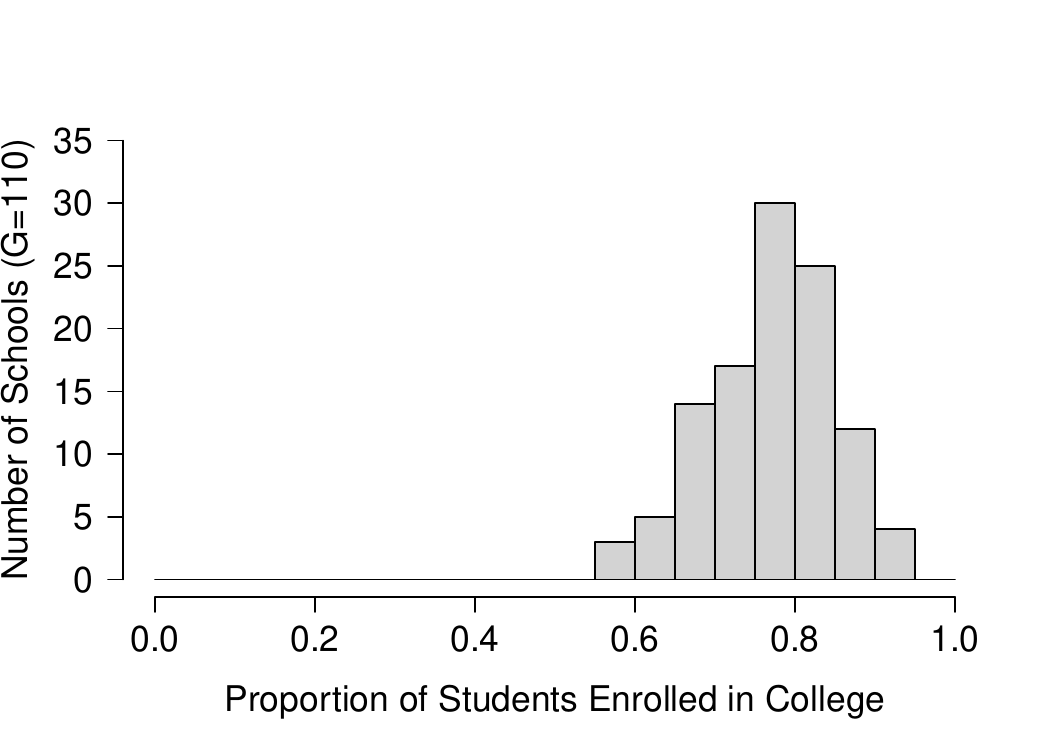}
        \vspace{-.05in}
        \subcaption{Outcome: College enrollment}\label{fig:lausd_enroll}
    \end{subfigure}
    \subcaption*{\textit{Note:} Distributions of the proportions of students in each school who \textit{(a)} took math in twelfth grade (the treatment) and \textit{(b)} enrolled in a two-year or four-year college within one year of graduating high school (an outcome). The data describes $G=110$ schools. }
\end{figure}

We control for school using the models we have focused on throughout this paper: multilevel, fixed effects, and regularized fixed effect models with school-varying intercepts. Controlling for school is required because school-level factors undoubtedly influence both twelfth grade math course-taking, and the student outcomes we examine. For example, schools may or may not offer enough classes for all students to take math in twelfth grade, or certain schools may have limited course offerings (e.g., only offer Precalculus, and \textit{not} offer Statistics). Figure~\ref{fig:lausd} visualizes the school-level confounding in our data. In Figure~\ref{fig:lausd_math}, the proportion of students in our data who took a twelfth grade math course varies greatly by school.  The average college enrollment outcome also noticeably varies by school (Figure~\ref{fig:lausd_enroll}). 

\subsection{Results}\label{subsec:application_results}

Table~\ref{tab.lausd} reports the estimated coefficients on twelfth grade math course-taking, along with the associated standard errors and levels of statistical significance, from all of the models considered in this paper for the three outcomes we examine. For comparison, we also include results from standard GLMs that include only twelfth grade math course-taking as a regressor (``Unadjusted"), and models that include the student-level covariates, but do not control for school at all (``Only Student-Level Covariates"). In summary, our recommendations, to (i) favor bcMLM, bcRegFE, and FE over uncorrected MLM and RegFE when group-level confounding is plausible and to (ii) apply CRSEs or a cluster bootstrap for inference, yield more conservative conclusions in this dataset, which are also more consistent with the original findings from \cite{wainstein2023access, wainstein2023success}. 

\begin{table}[!h]
\caption{Model coefficients for twelfth grade math course-taking}
\label{tab.lausd}
\vspace{-0.2in}
\scriptsize
\begin{center}
     \begin{tabular}{ |m{5.1cm} | m{1.38cm} m{1.38cm} | m{1.38cm} m{1.38cm} | m{1.38cm} m{1.38cm} |}
     \hline
     \multicolumn{1}{ |c| }{ \textbf{Model} } & \multicolumn{2}{ c| }{\textbf{High School GPA}} & \multicolumn{2}{  c | }{ \textbf{College Enrollment} } & \multicolumn{2}{ c| }{ \textbf{College Credits} } \\
      & Coefficient & (SE) & Coefficient & (SE) & Coefficient & (SE)  \\
     \hline
     Unadjusted (CRSE) & 0.2320*** &  (0.0220) & 0.5529*** &  (0.0551) & 0.1351*** &  (0.0301) \\
     Only Student-Level Covariates (CRSE) & -0.0453*** &  (0.0047) & 0.2365*** &  (0.0574) & 0.0235 &  (0.0268) \\
     RI (Default) & -0.0424*** &  (0.0028) & 0.2839*** &  (0.0586) & 0.0506*** &  (0.0075) \\
     RegFE (CRSE) & -0.0424*** &  (0.0040) & 0.2822*** &  (0.0567) & 0.0506 &  (0.0264) \\
     bcRI (Default) & -0.0419*** &  (0.0029) & 0.3368*** &  (0.0625) & 0.0510*** &  (0.0076) \\
     bcRI (CRSE/Boot) & -0.0419*** &  (0.0040) & 0.3368*** &  (0.0629) & 0.0510* &  (0.0244) \\
     bcRegFE (CRSE) & -0.0419*** &  (0.0040) & 0.3362*** &  (0.0631) & 0.0510 &  (0.0265) \\
     Group-FE (CRSE) & -0.0419*** &  (0.0040) & 0.3460*** &  (0.0636) & 0.0509* &  (0.0259) \\
     \hline
     $N$ (Students) & 9993 &  & 9993 &  & 3541 &  \\
     $G$ (Schools) & 110 &  & 110 &  & 100 &  \\
     \hline
     \end{tabular}

\end{center}
\normalsize
    \subcaption*{\textit{Note:} Coefficients on twelfth grade math course-taking (the treatment) in models for cumulative high school weighted GPA (linear regression); enrolling in a two-year or four-year college within one year of high school graduation (logistic regression); and credits accumulated in the first two years of community college (Poisson regression). In the ``Models" column, the method for inference is given in parentheses. ``Default" inference is pulled straight from the \texttt{lme4} output for ``RI (Default)" and ``bcRI (Default)". For ``bcRI (CRSE/Boot)", CRSEs are used for the high school GPA model (linear regression), and a cluster bootstrap (based on 200 bootstrap samples) is used for the college enrollment (logistic regression) and college credits (Poisson regression) models. The ``Unadjusted" model only includes twelfth grade math course-taking as a regressor. The ``Only Student-Level Covariates" model includes twelfth grade math course-taking and all student-level covariates as regressors, but does not control for school. *$p<0.05$, **$p<0.01$, ***$p<0.001$.}
\end{table}

Starting with cumulative high school weighted GPA, uncorrected RI and RegFE produce coefficients for twelfth grade math course-taking (-0.0424) that are very similar to the coefficients from the models that are more robust to school-level confounding: bcRI, bcRegFE, and Group-FE (-0.0419). All of these models imply that taking math in twelfth grade decreases one's GPA by approximately 0.04 GPA points on average, and that this result is highly statistically significant. This finding is consistent with \cite{wainstein2023access}. However, the default standard errors for RI and bcRI ($\approx 0.0029$) are over 25\% smaller than the standard errors obtained through CRSE ($\approx 0.0040$). Thus, default MLM standard errors would yield confidence intervals that are potentially too narrow for this outcome.

For college enrollment (within one year), uncorrected RI and RegFE produce coefficients ($\approx 0.28$) that are notably smaller than the coefficients from bcRI, bcRegFE, and Group-FE ($\approx 0.34$), potentially underestimating the effects of twelfth grade math on this outcome.  Nevertheless, all of the coefficient estimates from these models are positive and highly statistically significant, leading to the same overall conclusion: taking math in twelfth grade increases the likelihood of enrolling in college for these students. This conclusion is consistent with \cite{wainstein2023access}.

Finally, for community college credit accumulation, all of the models that account for school produce very similar coefficient estimates ($\approx 0.05$), which are also all positive. However, the default standard errors for RI and bcRI ($\approx 0.008$) are much smaller (by about 70\%) than the standard errors obtained through CRSE or a cluster-bootstrap for RegFE, bcRI, bcRegFE, and Group-FE ($\approx 0.025$). Were they trusted, the smaller standard errors would meaningfully change the overall conclusion from the analysis of this outcome: the smaller, default standard errors (for RI and bcRI) imply that the estimated effect of twelfth grade math on this outcome is highly statistically significant, while the CRSEs and bootstrapped standard errors imply that the estimated effect is on the cusp of statistical significance at the 0.05 level (statistically significant for bcRI and Group-FE, but not for RegFE or bcRegFE). The latter story is more consistent with \cite{wainstein2023success}, who also estimate a positive effect of twelfth grade math on community college credit accumulation for the same types of students, but the estimate is \textit{not} statistically significant at the 0.05 level. 

\section{Conclusion}\label{sec:conclusion}

Two commonly used approaches for analyzing grouped data are FE with specialized standard errors, and MLMs, which employ random effects. H\&W identified misunderstandings about these approaches in applied works, and explicated their similarities and differences in the linear setting using three analytical insights. We investigated if these insights, and H\&W's ultimate recommendations, carry over to GLMs, finding: (i) MLM can still be thought of as a regularized form of FE, which explains MLM's bias problem, but there is no longer an exact equivalence between RegFE and MLE-fit MLM like in the linear setting; (ii) none of FE, bcMLM, or bcRegFE entirely solves MLM's bias problem in GLMs, but bcMLM and bcRegFE tend to show little bias, and FE's bias lessens as group sizes increase; and (iii) like in the linear setting, MLM's assumptions can misspecify the true intragroup dependence structure, leading to default standard errors that are too small. MLM with a cluster bootstrap, FE with cluster robust standard errors, or RegFE with cluster robust standard errors are more agnostic alternatives to MLM's default standard errors, and can perform well given enough data. Additionally, while we focused on MLMs fit through (approximate) MLE in the text,  Appendix~\ref{app:pql} shows that these lessons carry over to PQL-fit MLMs. 

This brings us to our recommendations. For \textit{non-linear} GLMs, we recommend bcMLM for estimating the treatment coefficient, and a cluster-bootstrap for standard errors and confidence intervals. For models that only allow group-varying intercepts, bcMLM involves simply including the group-level averages of the covariates as fixed-effect regressors. We note that this differs from H\&W's recommendations in the linear setting, which were to use FE \textit{or} bcMLM for coefficient estimation, which yield equivalent estimates, and cluster robust standard errors, which are also equal for FE and bcMLM. The difference in our recommendations comes from the fact that FE and bcMLM are not necessarily equivalent in non-linear GLMs, and in fact FE shows non-negligible finite sample bias due to its incidental parameters problem that tends to be higher than that from bcMLM, particularly when group sizes are small. As for variance estimation, at the time of writing, we are unaware of a comprehensive extension of cluster robust standard errors to MLE-fit MLMs in the GLM framework.

However, in larger samples, a cluster bootstrap may be infeasible due to computation time. In these settings, we recommend bcRegFE with cluster robust standard errors, or FE with cluster robust standard errors when group sizes are larger. bcRegFE assumes the same model as does bcMLM, but fits the model with explicit regularization on the group-varying coefficients. 
In demonstrations here, we find that bcRegFE with cluster-robust standard errors produces similarly low bias coefficient estimates to those of bcMLM, and similar coverage rates to a cluster bootstrap. Further, cluster-robust standard errors are far less computationally intensive than a bootstrap. As for FE, when group sizes are larger, its bias is less of a concern, and the ability to feasibly apply cluster robust standard errors can avoid MLM's strict assumptions on the intragroup dependence structure that can lead to greatly biased default standard errors and incorrect inference.

We make our recommendations easily implementable in the \texttt{R} programming language with our package \href{https://github.com/lwainstein/mlmfe}{\texttt{mlmfe}}. In Appendix C, we demonstrate this package through an analysis of mock, simulated data, and include the analysis code. The analysis and data in Appendix C are purely for learning purposes, but are meant to mimic the analysis and data used in our investigation of twelfth grade math in L.A. Unified in Section~\ref{sec:application}.

Finally, we note alternatives to FE, bcMLM, and bcRegFE that reduce FE's  bias in smaller samples: Conditional Logistic Regression (\citealp{breslow1978estimation}) for logistic regression settings, and Firth's correction (\citealp{firth1993}) and its extensions (e.g., \citealp{kosmidis2009bias, kenne2017median, kosmidis2020mean}). Whether these methods are preferable to bcMLM or bcRegFE is beyond the scope of this paper. However, we do note that these methods perform remarkably similarly to bcMLM and bcRegFE in the simple setting with group-level confounding considered here (see Appendix~\ref{app:firth_condlr}). 
If these alternatives are preferred by the reader, we maintain the importance of using accompanying standard errors that are robust to a wide variety of intragroup dependence structures, for example applying a cluster bootstrap, or an extension of cluster robust standard errors to these methods.

\bibliographystyle{apalike}
\nocite{*}
\bibliography{MLMFE_GLM}

\begin{thebibliography}{}

\bibitem[Angrist and Pischke, 2009]{angrist2009mostly}
Angrist, J.~D. and Pischke, J.-S. (2009).
\newblock {\em Mostly harmless econometrics: An empiricist's companion}.
\newblock Princeton university press.

\bibitem[Arpino and Cannas, 2016]{arpino2016propensity}
Arpino, B. and Cannas, M. (2016).
\newblock Propensity score matching with clustered data. an application to the
  estimation of the impact of caesarean section on the apgar score.
\newblock {\em Statistics in medicine}, 35(12):2074--2091.

\bibitem[Bell et~al., 2019]{bell2019fixed}
Bell, A., Fairbrother, M., and Jones, K. (2019).
\newblock Fixed and random effects models: making an informed choice.
\newblock {\em Quality \& quantity}, 53:1051--1074.

\bibitem[Bell and Jones, 2015]{bell2015explaining}
Bell, A. and Jones, K. (2015).
\newblock Explaining fixed effects: Random effects modeling of time-series
  cross-sectional and panel data.
\newblock {\em Political Science Research and Methods}, 3(1):133--153.

\bibitem[Breslow et~al., 1978]{breslow1978estimation}
Breslow, N., Day, N., Halvorsen, K., Prentice, R., and Sabai, C. (1978).
\newblock Estimation of multiple relative risk functions in matched
  case-control studies.
\newblock {\em American Journal of Epidemiology}, 108(4):299--307.

\bibitem[Brumback et~al., 2010]{brumback2010adjusting}
Brumback, B.~A., Dailey, A.~B., Brumback, L.~C., Livingston, M.~D., and He, Z.
  (2010).
\newblock Adjusting for confounding by cluster using generalized linear mixed
  models.
\newblock {\em Statistics \& probability letters}, 80(21-22):1650--1654.

\bibitem[Brumback et~al., 2013]{brumback2013adjusting}
Brumback, B.~A., Zheng, H.~W., and Dailey, A.~B. (2013).
\newblock Adjusting for confounding by neighborhood using generalized linear
  mixed models and complex survey data.
\newblock {\em Statistics in medicine}, 32(8):1313--1324.

\bibitem[Cameron et~al., 2008]{cameron2008bootstrap}
Cameron, A.~C., Gelbach, J.~B., and Miller, D.~L. (2008).
\newblock Bootstrap-based improvements for inference with clustered errors.
\newblock {\em The review of economics and statistics}, 90(3):414--427.

\bibitem[Cameron and Miller, 2015]{cameron2015practitioner}
Cameron, A.~C. and Miller, D.~L. (2015).
\newblock A practitioner’s guide to cluster-robust inference.
\newblock {\em Journal of human resources}, 50(2):317--372.

\bibitem[Chang and Goplerud, 2024]{chang2024generalized}
Chang, Q. and Goplerud, M. (2024).
\newblock Generalized kernel regularized least squares.
\newblock {\em Political Analysis}, 32(2):157--171.

\bibitem[Clark and Linzer, 2015]{clark2015should}
Clark, T.~S. and Linzer, D.~A. (2015).
\newblock Should i use fixed or random effects?
\newblock {\em Political Science Research and Methods}, 3(2):399--408.

\bibitem[Cordeiro and McCullagh, 1991]{Gauss1991Bias}
Cordeiro, G.~M. and McCullagh, P. (1991).
\newblock Bias correction in generalized linear models.
\newblock {\em Journal of the Royal Statistical Society. Series B
  (Methodological)}, 53(3):629--643.

\bibitem[Czado, 2017]{czadonotes}
Czado, C. (2017).
\newblock Lecture 10: Linear mixed models (linear models with random effects).

\bibitem[Davis et~al., 2000]{davis2000autocorrelation}
Davis, R.~A., Dunsmuir, W.~T., and Wang, Y. (2000).
\newblock On autocorrelation in a poisson regression model.
\newblock {\em Biometrika}, 87(3):491--505.

\bibitem[Demidenko, 2013]{demidenko2013mixed}
Demidenko, E. (2013).
\newblock {\em Mixed models: theory and applications with R}.
\newblock John Wiley \& Sons.

\bibitem[Firth, 1993]{firth1993}
Firth, D. (1993).
\newblock Bias reduction of maximum likelihood estimates.
\newblock {\em Biometrika}, 80(1):27--38.

\bibitem[Gelman and Hill, 2006]{gelman2006data}
Gelman, A. and Hill, J. (2006).
\newblock {\em Data analysis using regression and multilevel/hierarchical
  models}.
\newblock Cambridge university press.

\bibitem[Goetgeluk and Vansteelandt, 2008]{goetgeluk2008conditional}
Goetgeluk, S. and Vansteelandt, S. (2008).
\newblock Conditional generalized estimating equations for the analysis of
  clustered and longitudinal data.
\newblock {\em Biometrics}, 64(3):772--780.

\bibitem[Green, 1987]{green1987penalized}
Green, P.~J. (1987).
\newblock Penalized likelihood for general semi-parametric regression models.
\newblock {\em International Statistical Review/Revue Internationale de
  Statistique}, pages 245--259.

\bibitem[Hausman, 1978]{hausman1978specification}
Hausman, J.~A. (1978).
\newblock Specification tests in econometrics.
\newblock {\em Econometrica: Journal of the econometric society}, pages
  1251--1271.

\bibitem[Hazlett and Wainstein, 2022]{hazlett2022understanding}
Hazlett, C. and Wainstein, L. (2022).
\newblock Understanding, choosing, and unifying multilevel and fixed effect
  approaches.
\newblock {\em Political Analysis}, 30(1):46--65.

\bibitem[Ho et~al., 2007]{ho2007matching}
Ho, D.~E., Imai, K., King, G., and Stuart, E.~A. (2007).
\newblock Matching as nonparametric preprocessing for reducing model dependence
  in parametric causal inference.
\newblock {\em Political analysis}, 15(3):199--236.

\bibitem[Jiang and Nguyen, 2021]{jiang2021linear}
Jiang, J. and Nguyen, T. (2021).
\newblock {\em Linear and generalized linear mixed models and their
  applications (second edition)}.
\newblock Springer.

\bibitem[Kabaila and Ranathunga, 2019]{kabaila2019adaptive}
Kabaila, P. and Ranathunga, N. (2019).
\newblock On adaptive gauss-hermite quadrature for estimation in glmm’s.
\newblock In {\em Statistics and Data Science: Research School on Statistics
  and Data Science, RSSDS 2019, Melbourne, VIC, Australia, July 24--26, 2019,
  Proceedings 1}, pages 130--139. Springer.

\bibitem[Kenne~Pagui et~al., 2017]{kenne2017median}
Kenne~Pagui, E.~C., Salvan, A., and Sartori, N. (2017).
\newblock Median bias reduction of maximum likelihood estimates.
\newblock {\em Biometrika}, 104(4):923--938.

\bibitem[Kim and Steiner, 2019]{kim2019causal}
Kim, Y. and Steiner, P. (2019).
\newblock Causal graphical views of fixed effects and random effects models.

\bibitem[Kosmidis and Firth, 2009]{kosmidis2009bias}
Kosmidis, I. and Firth, D. (2009).
\newblock Bias reduction in exponential family nonlinear models.
\newblock {\em Biometrika}, 96(4):793--804.

\bibitem[Kosmidis et~al., 2020]{kosmidis2020mean}
Kosmidis, I., Kenne~Pagui, E.~C., and Sartori, N. (2020).
\newblock Mean and median bias reduction in generalized linear models.
\newblock {\em Statistics and Computing}, 30(1):43--59.

\bibitem[{L}.{A}.{ }Unified, 2017]{lausd2017middle}
{L}.{A}.{ }Unified (2017).
\newblock {M}iddle {S}chool {M}ath {P}lacement and {P}athways.
\newblock Viewed 3/10/2026 from
  \href{https://lausd-p12.edlioschool.com/apps/pages/index.jsp?uREC_ID=4397387&type=d&pREC_ID=2620106}{here}.

\bibitem[{L}.{A}.{ }Unified, 2025a]{lausd2025finger}
{L}.{A}.{ }Unified (2025a).
\newblock {F}ingertip {F}acts 2025-26.
\newblock Accessed 3/10/2026 from
  \href{https://4.files.edl.io/aef4/08/14/25/170225-cc3e26de-f440-4470-9e85-486022db3fa6.pdf}{here}.

\bibitem[{L}.{A}.{ }Unified, 2025b]{lausd2025pathways}
{L}.{A}.{ }Unified (2025b).
\newblock {REG}-069702.3 {H}igh {S}chool {M}athematics {P}lacement and
  {P}athways.
\newblock Accessed 3/10/2026 from
  \href{https://my.lausd.net/eLibrary/DocView.aspx?id=144076&dbid=0&repo=eLibrary&cr=1}{here}.

\bibitem[Lancaster, 2000]{lancaster2000incidental}
Lancaster, T. (2000).
\newblock The incidental parameter problem since 1948.
\newblock {\em Journal of econometrics}, 95(2):391--413.

\bibitem[Lehmann and Casella, 1996]{lehmanncasella1998}
Lehmann, E. and Casella, G. (1996).
\newblock {\em Theory of Point Estimation}.
\newblock Springer New York.

\bibitem[Liang and Zeger, 1986]{liang1986longitudinal}
Liang, K.-Y. and Zeger, S.~L. (1986).
\newblock Longitudinal data analysis using generalized linear models.
\newblock {\em Biometrika}, 73(1):13--22.

\bibitem[McCullagh and Nelder, 1989]{mccullaghnelder1989}
McCullagh, P. and Nelder, J. (1989).
\newblock {\em Generalized Linear Models}.
\newblock CRC Press.

\bibitem[Mundlak, 1978]{mundlak1978pooling}
Mundlak, Y. (1978).
\newblock On the pooling of time series and cross section data.
\newblock {\em Econometrica: journal of the Econometric Society}, pages 69--85.

\bibitem[Neuhaus and McCulloch, 2006]{neuhaus2006separating}
Neuhaus, J.~M. and McCulloch, C.~E. (2006).
\newblock Separating between-and within-cluster covariate effects by using
  conditional and partitioning methods.
\newblock {\em Journal of the Royal Statistical Society Series B: Statistical
  Methodology}, 68(5):859--872.

\bibitem[Neyman and Scott, 1948]{neyman1948consistent}
Neyman, J. and Scott, E.~L. (1948).
\newblock Consistent estimates based on partially consistent observations.
\newblock {\em Econometrica: journal of the Econometric Society}, pages 1--32.

\bibitem[Pawitan, 2001]{pawitan2001all}
Pawitan, Y. (2001).
\newblock {\em In all likelihood: statistical modelling and inference using
  likelihood}.
\newblock Oxford University Press.

\bibitem[Raudenbush, 2009]{raudenbush2009adaptive}
Raudenbush, S.~W. (2009).
\newblock Adaptive centering with random effects: An alternative to the fixed
  effects model for studying time-varying treatments in school settings.
\newblock {\em Education Finance and Policy}, 4(4):468--491.

\bibitem[Schunck and Perales, 2017]{schunck2017within}
Schunck, R. and Perales, F. (2017).
\newblock Within-and between-cluster effects in generalized linear mixed
  models: A discussion of approaches and the xthybrid command.
\newblock {\em The Stata Journal}, 17(1):89--115.

\bibitem[Wainstein et~al., 2023a]{wainstein2023access}
Wainstein, L., Miller, C.~E., Phillips, M., Yamashiro, K., and Melguizo, T.
  (2023a).
\newblock Twelfth grade math and college access.
\newblock {\em Los Angeles Education Research Institute}.
\newblock Available at
  https://laeri.luskin.ucla.edu/12thgrademathandcollegeaccess/.

\bibitem[Wainstein et~al., 2023b]{wainstein2023success}
Wainstein, L., Miller, C.~E., Phillips, M., Yamashiro, K., and Melguizo, T.
  (2023b).
\newblock Twelfth grade math and college success.
\newblock {\em Los Angeles Education Research Institute}.
\newblock Available at
  https://laeri.luskin.ucla.edu/12thgrademathandcollegesuccess/.

\bibitem[White, 1984]{white1984asymptotic}
White, H. (1984).
\newblock Asymptotic theory for econometricians.
\newblock Technical report.

\bibitem[Wood, 2011]{wood2011fast}
Wood, S.~N. (2011).
\newblock Fast stable restricted maximum likelihood and marginal likelihood
  estimation of semiparametric generalized linear models.
\newblock {\em Journal of the Royal Statistical Society Series B: Statistical
  Methodology}, 73(1):3--36.

\end{thebibliography}

\newpage

\appendix

\section{Mathematical derivations}\label{app}

\subsection{Symbols and abbreviations}\label{app:tables}


\begin{table}[!h]
\caption{Abbreviations for model-related terms}
\label{tab.abbreviations}
\scriptsize
\begin{center}
\vspace{-0.1in}
\begin{tabular}{| l | l | l |}
	\hline
		\textbf{Abbreviation} & \textbf{Full name} & \textbf{Location} \\
	\hline
		bcMLM & Bias-corrected multilevel model & Section~\ref{subsec:insight2} \\
	\hline
		bcRegFE & Bias-corrected regularized fixed effects model & Section~\ref{subsec:insight2} \\
	\hline
		bcRI & Bias-corrected random intercepts model & Section~\ref{subsec:insight2} \\
        \hline 
		CRSE & Cluster-robust standard error & Section~\ref{subsec:insight3}  \\	
	\hline 
		FE & Fixed effects model & Section~\ref{subsec:mlmfe} \\
	\hline
		Group-FE & Group fixed effects model & Section~\ref{subsec:mlmfe} \\
	\hline
		GLM & Generalized linear model & Section~\ref{subsec:mlmfe} \\
	\hline
		MLE & Maximum likelihood estimation & Section~\ref{subsec:estimation}  \\
	\hline 
		MLM & Multilevel model & Section~\ref{subsec:mlmfe}  \\
	\hline 
		PQL & Penalized quasi-likelihood & Section~\ref{subsec:estimation}  \\
    \hline 
		RegFE & Regularized fixed effects model & Section~\ref{subsec:insight1} \\
	\hline
		RI & Random intercepts model & Section~\ref{subsec:mlmfe} \\
	\hline 
\end{tabular}
\normalsize
\end{center}
\end{table}

\vspace{-0.2in}

\begin{table}[!h]
\caption{Symbols}
\label{tab.symbols}
\tiny
\vspace{-0.1in}
\begin{center}
\begin{tabular}{| l | l | l | l |}
	\hline
		\textbf{Symbol} & \textbf{Description} & \textbf{Relevant model(s)} & \textbf{Location} \\
	\hline
		$\alpha$ & Coefficient vector & bcMLM, bcRegFE & Section~\ref{subsec:insight2} \\
	\hline 
		$\beta$ & Coefficient vector &  FE, Group-FE, MLM, RI, RegFE, bcMLM, bcRI, bcRegFE & Section~\ref{subsec:notation} \\
	\hline
		$c$ & Scalar for CRSEs & FE, Group-FE, MLM, RI, RegFE, bcMLM, bcRI, bcRegFE & Appendix~\ref{app:crses_regfe} \\
	\hline
		$\gamma_g$ and $\gamma$ & Coefficient vector &  FE, Group-FE, MLM, RI, RegFE, bcMLM, bcRI, bcRegFE & Section~\ref{subsec:notation} \\
	\hline 
		$\omega^2$ & Scalar variance &  RI, bcRI & Section~\ref{subsec:mlmfe} \\
	\hline
		$\Omega$ & Covariance matrix &  MLM, bcMLM, RegFE, bcRegFE & Section~\ref{subsec:mlmfe} \\
	\hline
		$\lambda$ & Scalar tuning parameter & RegFE, bcRegFE & Section~\ref{subsec:insight1} \\
	\hline
		$\theta$ & Parameter vector for GLM & FE, Group-FE, MLM, RI, RegFE, bcMLM, bcRI, bcRegFE & Section~\ref{subsec:mlmfe} \\
	\hline
		$\Theta$ & Vector of fixed parameters for GLM & FE, Group-FE, MLM, RI, bcMLM, bcRI & Section~\ref{subsec:estimation} \\
	\hline
		$\sigma^2$ & Scalar variance for linear model &  FE, Group-FE, MLM, RI, RegFE, bcMLM, bcRI, bcRegFE & Section~\ref{subsec:mlmfe} \\
	\hline
 		$h$ & Link function for GLM & FE, Group-FE, MLM, RI, RegFE, bcMLM, bcRI, bcRegFE & Section~\ref{subsec:mlmfe} \\
	\hline
 		$\mu_{g[i]}$ & Conditional mean for GLM & FE, Group-FE, MLM, RI, RegFE, bcMLM, bcRI, bcRegFE & Section~\ref{subsec:mlmfe} \\
	\hline
 		$p_{\mathrm{GLM}}$ & Conditional distribution for GLM & FE, Group-FE, MLM, RI, RegFE, bcMLM, bcRI, bcRegFE & Section~\ref{subsec:mlmfe} \\
	\hline
 		$L_{\mathrm{FE}}$ & Likelihood for FE & FE, Group-FE & Section~\ref{subsec:estimation} \\
	\hline
 		$L_{\mathrm{MLM}}$ & Likelihood for MLM & MLM, RI & Section~\ref{subsec:estimation} \\
	\hline
		$s$ & Scale function for exponential family & RegFE, bcRegFE & Appendix~\ref{app:crses_regfe} \\
	\hline
		$v$ & Variance function for exponential family & RegFE, bcRegFE & Appendix~\ref{app:crses_regfe} \\
	\hline
		$\hat{W}$ & Weight matrix & RegFE, bcRegFE & Appendix~\ref{app:crses_regfe} \\
	\hline
		$S$ & Regularization matrix & RegFE, bcRegFE & Appendix~\ref{app:crses_regfe} \\
	\hline
		$\hat{M}$ & Matrix for RegFE CRSEs & RegFE, bcRegFE & Appendix~\ref{app:crses_regfe} \\
	\hline
		$X_{g[i]}$ & Random (covariate) vector & FE, Group-FE, MLM, RI, RegFE, bcMLM, bcRI, bcRegFE & Section~\ref{subsec:notation} \\
	\hline 
		$X_{g}$ and $X$ & Random (covariate) matrix & FE, Group-FE, MLM, RI, RegFE, bcMLM, bcRI, bcRegFE & Section~\ref{subsec:notation} \\
	\hline $\bar{X}_g$ & Random (covariate) vector & bcMLM, bcRI, bcRegFE & Section~\ref{subsec:insight2} \\
	\hline
		$\tilde{X}_{g[i]}$ & Random (covariate) vector & bcMLM, bcRegFE & Section~\ref{subsec:insight2}  \\
	\hline
		$Y_{g[i]}$ & Random (outcome) variable  &  FE, Group-FE, MLM, RI, RegFE, bcMLM, bcRI, bcRegFE & Section~\ref{subsec:notation} \\
	\hline
		$Y_{g}$ and $Y$ & Random (outcome) vector  & FE, Group-FE, MLM, RI, RegFE, bcMLM, bcRI, bcRegFE & Section~\ref{subsec:notation} \\
	\hline
		$Z_{g[i]}$ & Random (covariate) vector & FE, MLM, RegFE, bcMLM, bcRegFE & Section~\ref{subsec:notation} \\
	\hline 
		$Z_{g}$ and $Z$ & Random (covariate) matrix & FE, MLM, RegFE, bcMLM, bcRegFE & Section~\ref{subsec:notation} \\
	\hline
		$\hat{A}_{g[i]}$ & Transformed outcome variable & RegFE, bcRegFE & Appendix~\ref{app:crses_regfe} \\
	\hline
		$\hat{A}$ & Transformed outcome vector & RegFE, bcRegFE & Appendix~\ref{app:crses_regfe} \\
	\hline
		$\hat{e}_{g[i]}$ & Transformed residual & RegFE, bcRegFE & Appendix~\ref{app:crses_regfe} \\
	\hline
		$\hat{e}_g$ & Transformed residual vector & RegFE, bcRegFE & Appendix~\ref{app:crses_regfe} \\
	\hline
\end{tabular}
\end{center}
\normalsize
\end{table}

\newpage

\subsection{Random effects assumption}\label{app:re_assumption}

The assumptions on $\gamma_g$ in RI, given in (\ref{eq:ri}), and in general MLMs, given in (\ref{eq:mlm_general}), imply what has been called the ``random effects assumption'' (\citealp{bell2015explaining, kim2019causal}). In the solely group-varying intercepts models in (\ref{eq:group_glm}), this assumption is that $\cor(\gamma_g, X_{g[i]}) =0$. This assumption follows from RI in (\ref{eq:ri}) because:
\begin{align}
	\cov(\gamma_g, X_{g[i]}) &= \E (\gamma_g X_{g[i]}) - \E (\gamma_g ) \E( X_{g[i]}) \nonumber \\
	&= \E \biggr( \underbrace{\E( \gamma_g \ | \ X, Z)}_{=0} X_{g[i]} \biggr) - \E \biggr( \underbrace{\E( \gamma_g \ | \ X, Z)}_{=0} \biggr) \E( X_{g[i]}) \nonumber \\
	&= 0
\end{align}
where $\E( \gamma_g \ | \ X, Z) = 0$ because $\gamma \ | \ X, Z \overset{iid}{\sim} N(0, \omega^2)$ in  (\ref{eq:ri}).

For general models (\ref{eq:link}) with potentially multiple group-varying coefficients, the random effects assumption has a  more general form: that $\cor(Z_{g[i]}^{\top} \gamma_g, X_{g[i]}) = 0$. In other words, the whole ``random effect contribution", $Z_{g[i]}^{\top} \gamma_g$, is uncorrelated with $X_{g[i]}$. This follows from general MLMs in (\ref{eq:mlm_general}) because:
\begin{align}
    \cov(Z_{g[i]}^{\top}  \gamma_g, X_{g[i]}) &= \E (Z_{g[i]}^{\top} \gamma_g X_{g[i]}) - \E (Z_{g[i]}^{\top} \gamma_g ) \E( X_{g[i]}) \nonumber \\
    &= \E \biggr( Z_{g[i]}^{\top}  \underbrace{\E( \gamma_g \ | \ X, Z)}_{=0} X_{g[i]} \biggr) - \E \biggr( Z_{g[i]}^{\top}  \underbrace{\E( \gamma_g \ | \ X, Z)}_{=0} \biggr) \E( X_{g[i]}) \nonumber \\
    &= 0
\end{align}
where $\E( \gamma_g \ | \ X, Z) = 0$ because $\gamma \ | \ X, Z \overset{iid}{\sim} N(0, \Omega)$ in (\ref{eq:mlm_general}).

\subsection{Motivation for GLM RegFE}\label{app:regfe_motivation}

In order to motivate the general class of RegFE models for the GLM setting given in (\ref{eq:RegFE_glm}), we provide an alternate method for obtaining the parameter estimates for the linear RegFE model in (\ref{eq:RegFE_linear}). Consider the linear model in (\ref{eq:linear}) with only group-varying intercepts. Using the specifications for linear RI (i.e., that $\gamma_g \ | \ X, Z \overset{iid}{\sim} N(0, \omega^2)$ and $\epsilon_{g[i]} \ | \ X, Z \overset{iid}{\sim} N(0, \sigma^2)$ with $\gamma_g$ conditionally independent of $\epsilon_{g'[i]}$ given $ X, Z$ for all $g, g'$, and $i$), consider parameter estimates obtained by maximizing, over $\beta$ and $\gamma$, the conditional joint density of $Y$ and $\gamma$, given $X$, $Z$, and the model parameters:
    \begin{align}\label{eq:RegFE_glm_start}
        p(Y, \gamma \ | \ X, Z, \beta, \sigma, \omega) &= p(Y \ | \ X, Z, \beta, \gamma, \sigma) \cdot p(\gamma \ | \ X, Z, \omega)
    \end{align} 
Note that due to the iid assumptions on $\epsilon_{g[i]}$ and $\gamma_g$, we can rewrite (\ref{eq:RegFE_glm_start}) as
    \begin{align}\label{eq:RegFE_glm_indep}
        p(Y, \gamma \ | \ X, Z, \beta, \sigma, \omega) &= \prod_{g=1}^G\prod_{i=1}^{n_g} p(Y_{g[i]}|X,Z,\beta,\gamma,\sigma)\cdot \prod_{g=1}^G p(\gamma_g|X,Z,\omega)
    \end{align}
Maximizing the above expression is equivalent to minimizing its negative natural logarithm. After substituting the specified model distribution, the negative natural logarithm of the conditional joint distribution is given by 
    \begin{align}\label{eq:RegFE_deriv}
        -\mathrm{log} \ p(Y, \gamma \ | \ X, Z, \beta, \sigma, \omega) &= c_0+\sum_{g=1}^G\sum_{i=1}^{n_g} \frac{[Y_{g[i]}-(X^\top_{g[i]}\beta+\gamma_g)]^2}{2\sigma^2}+\sum_{g=1}^G\frac{\gamma_g^2}{2\omega^2} \nonumber \\ 
        &\propto     c_1 + \sum_{g=1}^G\sum_{i=1}^{n_g} [Y_{g[i]}-(X^\top_{g[i]}\beta+\gamma_g)]^2+ \frac{\sigma^2}{\omega^2} \sum_{g=1}^G\gamma_g^2
    \end{align}
where $c_0$ and $c_1$ are constant with respect to $\beta$ and $\gamma$.\footnote{Here $c_0$ and $c_1$ are functions of $\sigma^2$ and $\omega^2$, which will ultimately be fixed.} Disregarding $c_1$ and letting $\lambda_{\mathrm{Lin}} = \frac{\sigma^2}{\omega^2}$ then yields the objective function for linear RegFE in (\ref{eq:RegFE_linear}). This shows that parameter estimates for linear RegFE can also be obtained from maximization over $\beta$ and $\gamma$ of the joint density of $Y$ and $\gamma$ given in (\ref{eq:RegFE_glm_start}) under linear RI specification. 

Now, in the GLM case, $p(Y \ | \ X, Z, \beta, \gamma, \sigma)$ in (\ref{eq:RegFE_glm_start}) is specified by $p_{\mathrm{GLM}}$, the conditional distribution of the response variable $Y$ given the model parameters. Applying this observation allows us to extend RegFE from linear models to any GLM, and potentially multiple random coefficients (i.e., models where $Z_{g[i]}$ is not identically $1$): 
    \begin{align}\label{eq:RegFE_glm_appendix}
        (\hat\beta_{\text{RegFE}},\hat\gamma_{\text{RegFE}}) &= \underset{\beta, \gamma}{\argmax} \biggr( \underbrace{ p (Y, \gamma \ | \ X, Z, \beta, \theta, \Omega)}_{\text{LHS of (\ref{eq:RegFE_glm_start})}} \biggr) \nonumber \\
        &= \underset{\beta, \gamma}{\argmax} \biggr( \underbrace{ p (Y \ | \ X, Z, \beta, \gamma, \theta) \cdot p(\gamma \ | \ X, Z, \Omega) }_{\text{RHS of (\ref{eq:RegFE_glm_start})}} \biggr) \nonumber \\
        &= \underset{\beta, \gamma}{\argmax} \biggr( p_{\mathrm{GLM}} (Y \ | \ X, Z, \beta, \gamma, h, \theta) \cdot p(\gamma \ | \ X, Z, \Omega) \biggr) 
    \end{align}
where $\theta$ and $\Omega$ are fixed values. The above expression is the exact same expression as the GLM RegFE parameter estimates in (\ref{eq:RegFE_glm}). 

As previously noted, we primarily assign $\theta$ and $\Omega$ in (\ref{eq:RegFE_glm_appendix}) (and (\ref{eq:RegFE_glm})) to be the MLM estimates through MLE. However, they could be assigned through other means, for example cross-validation.\footnote{Recall that in the linear RI model, $\theta = \sigma^2$ and $\Omega = \omega^2$. Then, setting $\lambda_{\mathrm{Lin}} = \sigma^2/\omega^2$ recovers Linear RegFE. Thus, allowing $\theta$ and $\Omega$ to be chosen by cross-validation in general RegFE is analogous to letting cross-validation choose the level of shrinkage in linear RegFE.} Note that $p_{\mathrm{GLM}} (Y \ | \ X, Z, \beta, \gamma, h, \theta)$ is exactly the likelihood function that is maximized under MLE in FE (i.e., $L_{\mathrm{FE}}$ in (\ref{eq:likelihood_fe})). For RegFE, the specification of $p(\gamma \ | \ X, Z, \Omega)$ then determines the regularization of $\gamma$.\footnote{Note also that when $p_{\mathrm{GLM}} (\cdot)$ comes from the Exponential Family and $h(\cdot)$ is the associated canonical link function, exchanging $p(\gamma \ | \ X, Z, \Omega)$ in (\ref{eq:RegFE_glm_appendix}) (and (\ref{eq:RegFE_glm})) with Jeffrey's invariant prior on both $\beta$ and $\gamma$ yields Firth's bias correction (\citealp{firth1993}), which induces regularization on $\gamma$ \textit{and} $\beta$.} In the context of MLMs, $\gamma_g$ is often specified to be normally distributed, which induces $L^2$ regularization, and yields a special case of the estimator studied by \cite{wood2011fast}. To see this, consider the RI context, where the $\gamma_g$ are varying intercepts and $\gamma_g \ | \ X, Z \overset{iid}{\sim} N(0, \omega^2)$. The RegFE maximization problem in (\ref{eq:RegFE_glm_appendix}) (and (\ref{eq:RegFE_glm})) is equivalent to the minimization problem:
    \begin{equation}\label{eq:RegFE_ri}
        (\hat\beta_{\text{RegFE}},\hat\gamma_{\text{RegFE}})=\argmin_{\beta,\gamma}\left(-\log\Big(p_{\text{GLM}}(Y|X,Z,\beta,\gamma,h,\theta) \Big)+ \lambda_{\mathrm{GLM}} \sum_{g=1}^G\gamma_g^2\right)
    \end{equation}
where $\lambda_{\mathrm{GLM}} = \frac{1}{2\omega^2}$. While this does not reduce to the form of the objective function in (\ref{eq:RegFE_linear}), except in the special case where a linear link function and normally distributed errors are used, we nevertheless see that $\lambda_{\mathrm{GLM}} = \frac{1}{2\omega^2}$ determines the extent of the regularization on the $\gamma_g$, analogous to how $\lambda_{\mathrm{Lin}}$ determines the extent of regularization in the linear version of RegFE from (\ref{eq:RegFE_linear}).

\subsection{The equivalence of MLM and RegFE in the case of linear regression}\label{app:equivalence in linear case}

Here we show that \textit{in the linear regression setting with homoscedastic errors}, jointly maximizing $p(Y, \gamma \ | \ X, Z, \beta, \hat\theta_{\text{MLM}}, \hat\Omega_{\text{MLM}})$ over $\beta$ and $\gamma$, as in RegFE, is the same as first maximizing \\ 
$p(Y \ | \ X,Z,\beta,\hat\theta_{\text{MLM}}, \hat\Omega_{\text{MLM}})$ over $\beta$ to find $\hat{\beta}_{\mathrm{MLM}}$ and then maximizing \\ 
$p(\gamma \ | \ Y, X,Z, \hat{\beta}_{\text{MLM}},\hat\theta_{\text{MLM}}, \hat\Omega_{\text{MLM}})$ over $\gamma$ to find $\hat{\gamma}_{\mathrm{MLM}}$ in the case of MLM. To reduce notation, we define:
    \begin{align}
        L_Y (\beta) &= p(Y \ | \ X,Z,\beta,\hat\theta_{\text{MLM}}, \hat\Omega_{\text{MLM}}) \\
        L_{\gamma|Y}(\beta,\gamma) &= p(\gamma \ | \ Y, X,Z, \beta,\hat\theta_{\text{MLM}}, \hat\Omega_{\text{MLM}}) \\
        L_{Y, \gamma} (\beta, \gamma) &= L_Y (\beta) \times L_{\gamma|Y}(\beta,\gamma) = p(Y, \gamma \ | \ X, Z, \beta, \hat\theta_{\text{MLM}}, \hat\Omega_{\text{MLM}}) 
    \end{align}
So that
    \begin{align}
        \hat{\beta}_{\mathrm{MLM}} &= \underset{\beta}{\mathrm{argmax}} \  L_Y (\beta) \\
        \hat{\gamma}_{\mathrm{MLM}} &= \underset{\gamma}{\mathrm{argmax}} \  L_{\gamma|Y}(\hat{\beta}_{\mathrm{MLM}},\gamma) \\
        (\hat{\beta}_{\mathrm{RegFE}}, \hat{\gamma}_{\mathrm{RegFE}}) &= \underset{\beta, \gamma}{\mathrm{argmax}} \  L_{Y, \gamma} (\beta, \gamma)
    \end{align}

Consider the following condition: 
    \begin{align}\label{eq:lin_mlm_key_prop}
        \hat{\beta}_{\mathrm{MLM}} \in \Bigg\{\beta \ \bigg| \ \exists \gamma_0 \text{ such that } L_{\gamma|Y} (\beta,\gamma_0)=\underset{\beta, \gamma}{\textrm{max}} \ L_{\gamma|Y}(\beta, \gamma)\Bigg\}
    \end{align}
In words, (\ref{eq:lin_mlm_key_prop}) states that setting $\beta = \hat{\beta}_{\mathrm{MLM}}$ does not change the maximum possible value for $L_{\gamma|Y}(\beta, \gamma)$. For $(\hat{\beta}_{\mathrm{MLM}}, \hat{\gamma}_{\mathrm{MLM}}) = (\hat{\beta}_{\mathrm{RegFE}}, \hat{\gamma}_{\mathrm{RegFE}})$, it is sufficient to show that (\ref{eq:lin_mlm_key_prop}) holds. To see this, consider if (\ref{eq:lin_mlm_key_prop}) were to hold: then,
\begin{align}
    L_{Y, \gamma} ( \hat{\beta}_{\mathrm{MLM}}, \hat{\gamma}_{\mathrm{MLM}}) &= \biggr(\underset{\beta}{\max} \ L_Y (\beta) \biggr) \times \biggr(\underset{\beta, \gamma}{\max} \ L_{\gamma|Y} (\beta, \gamma) \biggr) \nonumber \\
    & \geq  \underset{\beta, \gamma}{\max} \ \biggr( L_Y (\beta) \times  \ L_{\gamma|Y} (\beta, \gamma) \biggr)
    = \underset{\beta, \gamma}{\max} \ L_{Y, \gamma} (\beta, \gamma) \nonumber \\
    &= L_{Y, \gamma} ( \hat{\beta}_{\mathrm{RegFE}}, \hat{\gamma}_{\mathrm{RegFE}})
\end{align}
Then using that $L_{Y, \gamma} ( \hat{\beta}_{\mathrm{RegFE}}, \hat{\gamma}_{\mathrm{RegFE}}) \geq L_{Y, \gamma} ( \hat{\beta}_{\mathrm{MLM}}, \hat{\gamma}_{\mathrm{MLM}})$ by definition implies that $(\hat{\beta}_{\mathrm{MLM}}, \hat{\gamma}_{\mathrm{MLM}}) = (\hat{\beta}_{\mathrm{RegFE}}, \hat{\gamma}_{\mathrm{RegFE}})$. 

We show that (\ref{eq:lin_mlm_key_prop}) holds in a Linear MLM with homoscedastic errors. Here, $\theta = \sigma^2$ and one assumes
    \begin{align}
        Y_{g[i]}&=X_{g[i]}^{\top} \beta+Z_{g[i]}^{\top} \gamma_g+\epsilon_{g[i]} \nonumber \\
        \text{where} \ \ \ \gamma_g&\overset{iid}{\sim}\mathcal{N}(\Vec{\mathbf{0}}, \Omega ) \ \  \text{and} \ \ \epsilon_{g[i]} \overset{iid}{\sim}N    (0,\sigma^2 )
    \end{align}
We rewrite
    \begin{align}
        \epsilon^{*}_{g[i]} = Z_{g[i]}^{\top} \gamma_g + \epsilon_{g[i]} 
    \end{align}
and letting $\epsilon^{*}$ be an $N \times 1$ vector that combines all of the  $\epsilon^{*}_{g[i]}$ for all units, the model can be rewritten as
    \begin{align*}
        Y&=X\beta+\epsilon^* \\
        \text{where} \ \ \ \epsilon^*&\sim\mathcal{N}(\Vec{0}, V ) \ \ \text{with} \ \ V = Z \Omega_{\operatorname{block}} Z^\top+ \sigma^2I_N
    \end{align*}
where
    \begin{align*}
       \Omega_{\mathrm{block}} = \begin{bmatrix} \Omega & \dots & 0 \\ \vdots & \ddots & \vdots \\ 0 & \dots & \Omega \end{bmatrix} \in \mathbbm{R}^{Gd \times Gd}
    \end{align*}
Because two normally distributed variables are jointly normal, it holds that
\begin{align}
\begin{bmatrix}
    Y\\
    \gamma
\end{bmatrix}\Biggr| X,Z \sim \mathcal{N}\left(\begin{bmatrix}
    X\beta\\
    \Vec{0}
\end{bmatrix},\begin{bmatrix}
    V&Z\Omega_{\operatorname{block}}\\
    \Omega_{\operatorname{block}}Z^\top& \Omega_{\operatorname{block}}
\end{bmatrix}\right),
\end{align}
Thus, using the closed form for the conditional distribution of a multivariate normal,
\begin{align}\label{eq:gamma_dist_norm}
    \gamma \ | \ Y, X,Z\sim\mathcal{N}\left(\Omega_{\operatorname{block}}Z^\top V^{-1}(Y-X\beta), \ \Omega_{\operatorname{block}}-\Omega_{\operatorname{block}}Z^\top V^{-1}Z\Omega_{\operatorname{block}} \right)
    \end{align}
The preceding derivation can be found in \cite{czadonotes}. Now, because $\gamma \ | \ Y, X,Z$ is normally distributed, given any $\beta$, $L_{\gamma|Y}(\beta,\gamma)$ can be maximized by predicting the conditional mean for $\gamma$ shown in (\ref{eq:gamma_dist_norm}),
    \begin{align}
        \gamma = \hat{\Omega}_{\operatorname{block}}Z^\top  \hat{V}_{\mathrm{MLM}}^{-1}(Y-X\beta)
    \end{align}
and the exact value of this maximum is independent of $\beta$, because the maximum value of a multivariate normal distribution depends only on its variance, and the variance in (\ref{eq:gamma_dist_norm}) is independent of $\beta$. In other words,
    \begin{align}
        \forall \beta^{*}, \ \ \underset{\gamma} {\mathrm{max}} \ L_{ \gamma | Y} (\beta^{*}, \gamma) = \underset{\beta, \gamma} {\mathrm{max}} \ L_{ \gamma | Y} (\beta, \gamma)
    \end{align}
Which means the above applies for when $\beta^{*} = \hat{\beta}_{\mathrm{MLM}}$, meaning that the condition in (\ref{eq:lin_mlm_key_prop}) holds by allowing
    \begin{align}
        \gamma_0 = \hat{\Omega}_{\operatorname{block}}Z^\top  \hat{V}_{\mathrm{MLM}}^{-1}(Y-X \hat{\beta}_{\mathrm{MLM}})
    \end{align}
Thus, $(\hat{\beta}_{\mathrm{MLM}}, \hat{\gamma}_{\mathrm{MLM}}) = (\hat{\beta}_{\mathrm{RegFE}}, \hat{\gamma}_{\mathrm{RegFE}})$ as previously shown.

\subsection{Analysis of $\cov ( \mu_{g[i]}, \mu_{g[i']} | X, Z )$ in (\ref{eq:mlm_cov1})}\label{app:dependence_structure}

In this appendix, we analyze the $\cov ( \mu_{g[i]}, \mu_{g[i']} | X, Z )$ term in (\ref{eq:mlm_cov1}) more extensively. For a general link function, consider this term's approximation through a second-order Taylor expansion:
    \begin{align}
      \cov( \mu_{g[i]}, \mu_{g[i']} | X, Z ) &\approx (h^{-1})'\biggr( X^{\top}_{g[i]} \beta \biggr) (h^{-1}) ' \biggr(X^{\top}_{g[i']} \beta \biggr) \cov ( Z_{g[i]}^{\top} \gamma_g, Z_{g[i']}^{\top} \gamma_g \ | \ X, Z) \nonumber \\
      & -  \frac{1}{4}  (h^{-1})'' \biggr( X^{\top}_{g[i]} \beta \biggr) (h^{-1})'' \biggr( X^{\top}_{g[i']} \beta \biggr) \var ( Z_{g[i]}^{\top} \gamma_g \ | \ X, Z) \var ( Z_{g[i']}^{\top} \gamma_g \ | \ X, Z)
    \end{align}
To see a specific case of this, consider the RI model for logistic regression (i.e., $h^{-1} (t) = \frac{\mathrm{exp}(t)}{1 + \mathrm{exp}(t)}$). The above covariance becomes:
    \begin{align}
      & \cov( \mu_{g[i]}, \mu_{g[i']} | X, Z ) \approx \nonumber \\
      & \ \ \ \ \ \ \ \ p_{g[i]}(1-p_{g[i]})p_{g[i']}(1-p_{g[i']})\omega^2 - \frac{1}{4}p_{g[i]}(1-p_{g[i]})(1-2p_{g[i]})p_{g[i']}(1-p_{g[i']})(1-2p_{g[i']})\omega^4
    \end{align}
where $p_{g[i]} = \frac{\exp(X^\top_{g[i]}\beta)}{1+\exp(X^\top_{g[i]}\beta)}$ denotes the probability of success for the $i^{\text{th}}$ unit in group $g$, before the influence of the random intercept, $\gamma_g$. The covariance between units in the same group is a function of $X_{g[i]}$ and $X_{g[i']}$. This expression is largest when $p_{g[i]} = p_{g[i']} = 0.5$, and decreases as either probability deviates from 0.5.

\subsection{CRSEs for RegFE}\label{app:crses_regfe}

We detail here an extension of CRSEs to RegFE. Per \cite{wood2011fast}, when $p_{\mathrm{GLM}} (\cdot)$ comes from the Exponential Family and $\gamma_g \overset{iid}{\sim} N(0, \Omega)$, then $(\hat{\beta}_{\mathrm{RegFE}}, \hat{\gamma}_{\mathrm{RegFE}})$ can be found by iterative (re)weighted least squares. Because $p_{\mathrm{GLM}} (\cdot)$ comes from the Exponential Family, one can express $\var(Y_{g[i]} \ | \ X, Z) = s (\theta) v (\mu_{g[i]})$
for some scale function $s(\cdot)$ and variance function $v(\cdot)$.\footnote{For example, in a Normal model, $s(\theta) = \sigma^2$ and $v(\mu_{g[i]}) = 1$. In a Poisson model, $s(\theta) = 1$ and $v(\mu_{g[i]}) = \mu_{g[i]}$. And in a Bernoulli model, $s(\theta) = 1$ and $v(\mu_{g[i]}) = \mu_{g[i]}(1-\mu_{g[i]})$.}
Then, letting the superscript $(k)$ denote parameter estimates from the $k$th step of the optimization process, the $(k+1)$th step estimates are:
\begin{align}\label{eq:regfe_iwls}
    (\hat{\beta}_{\mathrm{RegFE}}^{(k+1)}, \hat{\gamma}_{\mathrm{RegFE}}^{(k+1)}) = \biggr( [X \ Z]^{\top} \hat{W}^{(k)}  [X \ Z]  + S \biggr)^{-1}  [X \ Z]^{\top}  \hat{W}^{(k)} \hat{A}^{(k)}
\end{align}
where $\hat{W}^{(k)} = \mathrm{diag}( \ [h'(\hat{\mu}^{(k)}_{g[i]})]^{-2} * [v (\hat{\mu}^{(k)}_{g[i]})]^{-1} \ )$ is a diagonal matrix of weights,
\begin{align}
    S = \begin{bmatrix} 
        0_{p \times p} & 0_{p \times d} & \cdots & 0_{p \times d} \\
        0_{d \times p} & s(\theta) \Omega^{-1} & \cdots & 0_{d \times d} \\
        \vdots & \vdots & \ddots & \vdots \\
         0_{d \times p} & 0_{d \times d} & \cdots & s(\theta) \Omega^{-1}
    \end{bmatrix}
\end{align}
is a block-diagonal matrix that induces regularization in estimates of $\gamma$, and $\hat{A}^{(k)}$ is a $N \times 1$ vector of transformed responses, $\hat{A}^{(k)}_{g[i]} = (X_{g[i]}^{\top}\hat{\beta}^{(k)} + Z_{g[i]}^{\top}\hat{\gamma}_g^{(k)}) + (Y_{g[i]} - \hat{\mu}^{(k)}_{g[i]} ) h'(\hat{\mu}^{(k)}_{g[i]})$. In words, the $(k+1)$th step parameter estimates come from a regularized (by $S$) and weighted (by $\hat{W}^{(k)}$) linear regression of the transformed outcome, $\hat{A}^{(k)}_{g[i]}$, on $X_{g[i]}$ and $Z_{g[i]}$. After initiating starting parameter values, updating the parameter estimates as in (\ref{eq:regfe_iwls}) until convergence ultimately yields $(\hat{\beta}_{\mathrm{RegFE}}, \hat{\gamma}_{\mathrm{RegFE}})$. By taking the variance (conditional on $X$ and $Z$) of both sides of (\ref{eq:regfe_iwls}) and treating the weights as fixed, a natural cluster-robust variance estimator reveals itself as:
\begin{align}\label{eq:crse_regfe}
    \widehat{\var}_{\mathrm{CRSE}} \biggr( (\hat{\beta}_{\mathrm{RegFE}}, \hat{\gamma}_{\mathrm{RegFE}}) \biggr)  = c \times \hat{M}  \times 
    \begin{bmatrix} 
    \hat{e}_1 \hat{e}_1^{\top} & \cdots & 0 \\
    \vdots & \ddots & \vdots \\
    0 & \cdots & \hat{e}_G \hat{e}_G^{\top}
    \end{bmatrix} \times \hat{M}^{\top}
\end{align}
where $\hat{M}  = ( [X \ Z]^{\top} \hat{W}  [X \ Z]  + S )^{-1}  [X \ Z]^{\top}  \hat{W}$, $\hat{e}_g$ is the vector of transformed residuals $\hat{e}_{g[i]} = (Y_{g[i]} - \hat{\mu}_{g[i]} ) h'(\hat{\mu}_{g[i]})$ for group $g$, and $c$ is a finite sample correction scalar.\footnote{See \cite{cameron2015practitioner} for thorough discussion of the scalar $c$ for cluster-robust inference.} Note that in the linear case, with a Normal model for $p_{\mathrm{GLM}} (\cdot)$ and an identity link function, the CRSEs in (\ref{eq:crse_regfe}) for $\beta$ from bcRegFE are (given the same $c$) exactly equal to the CRSEs from bcMLM (see e.g., \citealp{cameron2015practitioner, hazlett2022understanding, chang2024generalized}), as are the estimates of $\beta$ --- this is easily verified using a strategy similar to the one used by H\&W when proving that the CRSEs from linear FE and bcMLM are exactly equal (see Appendix A.16 of H\&W). Additionally, without regularization (i.e., $S=0$) the CRSEs in (\ref{eq:crse_regfe}) from RegFE are --- at least for linear regression, logistic regression, and Poisson regression --- exactly equal (given the same $c$) to the CRSEs from FE, as are the estimates of $\beta$. We describe how to verify this fact in Appendix~\ref{app:crses_fe_regfe}.

\subsection{The equivalence of CRSEs from FE and CRSEs from RegFE without regularization}\label{app:crses_fe_regfe}

In this appendix, we describe how to verify that the CRSEs from FE and the CRSEs from RegFE (detailed in Appendix~\ref{app:crses_regfe}) without regularization ($S=0$) are exactly equal (given the same $c$). We do not provide the complete proofs here because they are highly algebraic, and thus are not particularly enlightening. However, we have verified mathematically and through simulation that the CRSEs are indeed equal for linear regression, logistic regression, and Poisson regression. We leave verifying the equivalence for other GLMs to the reader. 

To start, we define CRSEs for FE. Let
\begin{align}
    \ell_{g[i]} (\beta, \gamma) = \log \ p_{\mathrm{GLM}} (Y_{g[i]} \ | \ X, Z, \beta, \gamma, h, \theta)
\end{align}
be the log-likelihood for a single observation. Then,
\begin{align}\label{eq:crse_fe}
    \widehat{\mathrm{var}}_{\mathrm{CRSE}} \biggr( (\hat{\beta}_{\mathrm{FE}}, \hat{\gamma}_{\mathrm{FE}}) \biggr) = c &\times \biggr( - \sum_{g=1}^{G} \sum_{i=1}^{n_g} H_{g[i]} (\hat{\beta}_{\mathrm{FE}}, \hat{\gamma}_{\mathrm{FE}}) \biggr)^{-1} \nonumber \\
    & \times \biggr[ \sum_{g=1}^G \biggr( \sum_{i=1}^{n_g} S_{g[i]} (\hat{\beta}_{\mathrm{FE}}, \hat{\gamma}_{\mathrm{FE}}) \biggr)  \biggr( \sum_{i=1}^{n_g} S_{g[i]} (\hat{\beta}_{\mathrm{FE}}, \hat{\gamma}_{\mathrm{FE}})\biggr)^{\top} \biggr] \nonumber \\
    & \times \biggr( - \sum_{g=1}^{G} \sum_{i=1}^{n_g} H_{g[i]} (\hat{\beta}_{\mathrm{FE}}, \hat{\gamma}_{\mathrm{FE}}) \biggr)^{-1}
\end{align}
where
\begin{align}
    S_{g[i]} (\beta, \gamma) &= \frac{ \partial \ell_{g[i]} }{\partial (\beta, \gamma)} \\ 
    H_{g[i]} (\beta, \gamma) &= \frac{ \partial^2 \ell_{g[i]} }{\partial (\beta, \gamma) \partial (\beta, \gamma)^{\top}}
\end{align}
To prove the desired equivalence for a given GLM, using the corresponding $p_{\mathrm{GLM}}(\cdot)$, $h(\cdot)$, and $v(\cdot)$, one simply must:
\begin{enumerate}
    \item Evaluate (\ref{eq:crse_regfe}) with $S=0$.
    \item Evaluate (\ref{eq:crse_fe}).
    \item Confirm that the two expressions are equal.
\end{enumerate}

When doing this, it is useful to condense notation. For example, one could define $\eta = (\beta, \gamma)$, and $U = [X \ Z]$ with $U_{g[i]}$ being the $g[i]$th row of $U$, and $U_g$ being the matrix of $U_{g[i]}$ for group $g$:
\begin{align}
    U = \begin{bmatrix} U_1 \\ \vdots \\  U_G \end{bmatrix} \ \ \ \text{with} \ \ \ U_{g} = \begin{bmatrix} U_{g[1]}^{\top} \\ \vdots \\  U_{g[n_g]}^{\top} \end{bmatrix}
\end{align}
This way, one can express the linear predictor more compactly as
$$
    U_{g[i]}^{\top} \eta = X_{g[i]}^{\top} \beta + Z_{g[i]}^{\top} \gamma_g
$$
Also, note that $(\hat{\beta}_{\mathrm{RegFE}}, \hat{\gamma}_{\mathrm{RegFE}}) = (\hat{\beta}_{\mathrm{FE}}, \hat{\gamma}_{\mathrm{FE}})$ without regularization for RegFE. So it is useful to simply let a general $\hat{\eta} = (\hat{\beta}, \hat{\gamma})$ be the vector of coefficient estimates from either method, and $\hat{\mu}_{g[i]} = h^{-1} (U_{g[i]}^{\top} \hat{\eta})$. For example, with this notation, the logistic regression FE CRSEs take the form:
\begin{align}\label{eq:crse_fe_logistic}
    \widehat{\mathrm{var}}_{\mathrm{CRSE}} \biggr( (\hat{\beta}_{\mathrm{FE}}, \hat{\gamma}_{\mathrm{FE}}) \biggr) = c &\times \biggr( \sum_{g=1}^{G} \sum_{i=1}^{n_g} \hat{\mu}_{g[i]} ( 1 - \hat{\mu}_{g[i]}) U_{g[i]} U_{g[i]}^{\top} \biggr)^{-1} \nonumber \\
    & \times \biggr[ \sum_{g=1}^G \biggr( \sum_{i=1}^{n_g} (Y_{g[i]} - \hat{\mu}_{g[i]}) U_{g[i]} \biggr)  \biggr( \sum_{i=1}^{n_g}  (Y_{g[i]} - \hat{\mu}_{g[i]}) U_{g[i]} \biggr)^{\top} \biggr] \nonumber \\
    & \times \biggr( \sum_{g=1}^{G} \sum_{i=1}^{n_g} \hat{\mu}_{g[i]} ( 1 - \hat{\mu}_{g[i]}) U_{g[i]} U_{g[i]}^{\top} \biggr)^{-1}
\end{align}
and the logistic regression RegFE CRSEs with $S=0$ take the form:
\begin{align}\label{eq:crse_regfe_logistic}
    \widehat{\var}_{\mathrm{CRSE}} \biggr( (\hat{\beta}_{\mathrm{RegFE}}, \hat{\gamma}_{\mathrm{RegFE}}) \biggr)  = c \times (U^{\top} \hat{W} U)^{-1} U^{\top} \hat{W}
    \begin{bmatrix} 
    \hat{e}_1 \hat{e}_1^{\top} & \cdots & 0 \\
    \vdots & \ddots & \vdots \\
    0 & \cdots & \hat{e}_G \hat{e}_G^{\top}
    \end{bmatrix} \hat{W} U (U^{\top} \hat{W} U)^{-1}
\end{align}
where $\hat{e}_{g[i]} = \frac{1}{\hat{\mu}_{g[i]}(1-\hat{\mu}_{g[i]})}(Y_{g[i]} - \hat{\mu}_{g[i]} ) $ and $\hat{W} = \mathrm{diag}( \ \hat{\mu}_{g[i]}(1-\hat{\mu}_{g[i]}) \ )$. It is easy to verify that (\ref{eq:crse_fe_logistic}) and (\ref{eq:crse_regfe_logistic}) are equal.

\newpage 

\section{Simulations}\label{app_sim}

\subsection{Replication of results from Section~\ref{sec:insights} using PQL-fit MLMs}\label{app:pql}

In this appendix, we replicate the demonstrations from Section~\ref{sec:insights} using MLMs that are fit through PQL estimation, instead of through (approximate) MLE, which is the focus of the main text. In summary, our lessons and recommendations in the main text carry over to MLMs fit through PQL. Additionally, PQL-fit MLMs perform even more similarly to RegFE than do MLE-fit MLMs. 

Throughout this section, we implement PQL estimation of GLM (bc)MLMs in \texttt{R} with the \texttt{glmmPQL()} function from the \texttt{MASS} package. When applying (bc)RegFE, we then set $\theta$ and $\Omega$ to be the PQL-estimated values from the corresponding (bc)MLM. 

\subsubsection*{Results from Section~\ref{subsec:insight1} (Insight 1)}

\begin{figure}[!h]
    \caption{Replicating Figure~\ref{fig:logistic_noneq} -- Estimates of $\beta_1$ in (\ref{DGP:logistic}) from PQL-fit RI, RegFE, Group-FE, and a GLM without fixed or random effects  
    for logistic regression}\label{fig:logistic_noneq_pql}
    
    \vspace{-0.1in}
    
    \begin{center}
    
 \begin{subfigure}{0.48\textwidth}
   
     \begin{center}
     \vspace{-.5in}
     \includegraphics[scale=.45]{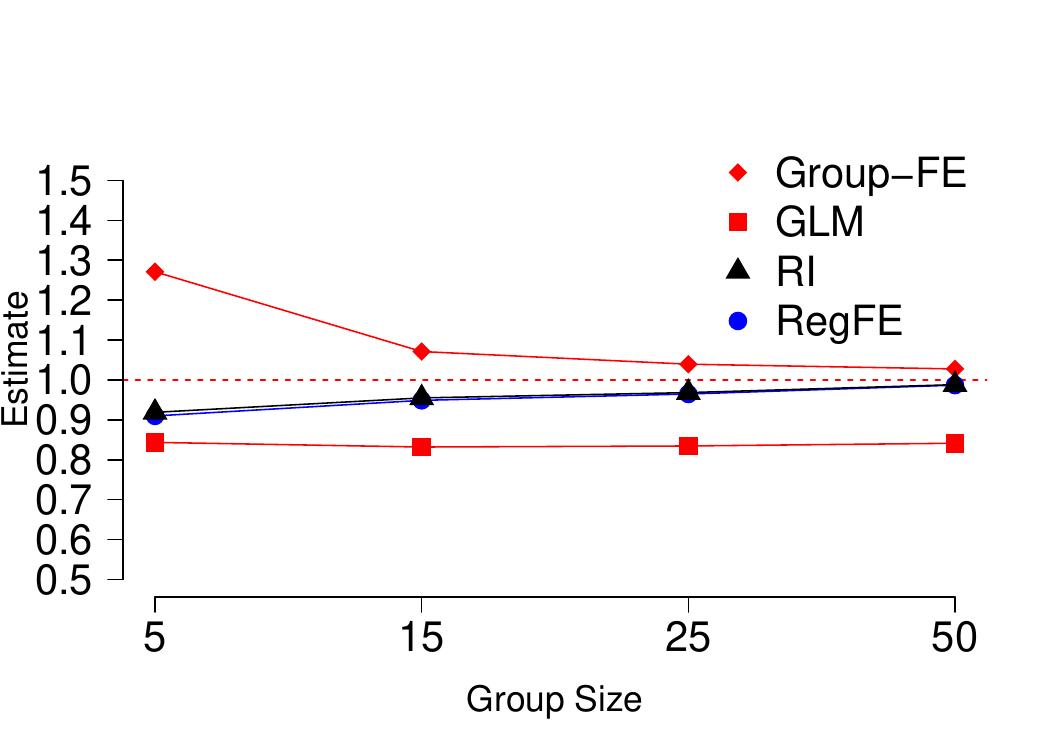}
     \subcaption{Median estimates for $G=50$}\label{fig:logistic_noneq_pql_a}
     \end{center}
     \end{subfigure}
 \begin{subfigure}{0.50\textwidth}
     \begin{center}
     \vspace{-.5in}
     \includegraphics[scale=.45]{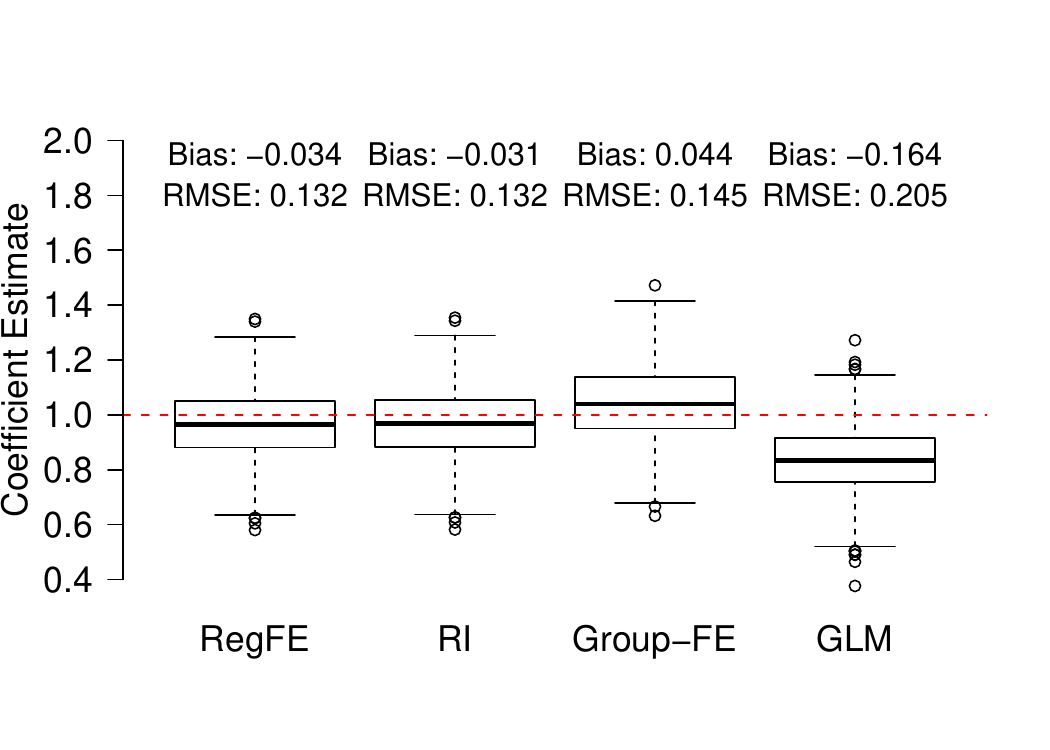}
     \subcaption{Estimate distributions for $G=50$ and $n_g=25$}\label{fig:logistic_noneq_pql_b}
     \end{center}
     \end{subfigure}
          
    \caption*{\textit{Note:} This figure replicates the results from Figure~\ref{fig:logistic_noneq} with MLMs fit through PQL. Results across 1000 iterations at each sample size of the DGP in (\ref{DGP:logistic}). The RI model is a logistic regression RI model, fit with PQL. The RegFE model is a logistic regression RegFE with only varying intercepts, setting $\omega$ to be the PQL-fit value. The Group-FE model is a logistic regression Group-FE model. The GLM model is a logistic regression that only includes $X_{g[i]}$ as a regressor, and omits fixed and random effects. \textit{(a)} Median estimates when $G=50$ for $\beta_1$. The red dashed line indicates the true value of $\beta_1 = 1$. \textit{(b)} Distribution of estimates when $G=50$ and $n_g=25$. The red dashed line indicates the true value of $\beta_1 = 1$.
    }
    \vspace{-.3in}
    
    \end{center}
    \end{figure}

\begin{figure}[!h]
    \caption{Replicating Figure~\ref{fig:logistic_re} --- Estimates of $\gamma$ in (\ref{DGP:logistic}) from PQL-fit RI, RegFE, and Group-FE for logistic regression}\label{fig:logistic_re_pql}
    
    \vspace{-0.1in}
    
    \begin{center}
    
     \begin{subfigure}{0.48\textwidth}  
         \begin{center}
         \vspace{-.4in}
         \includegraphics[scale=.3]{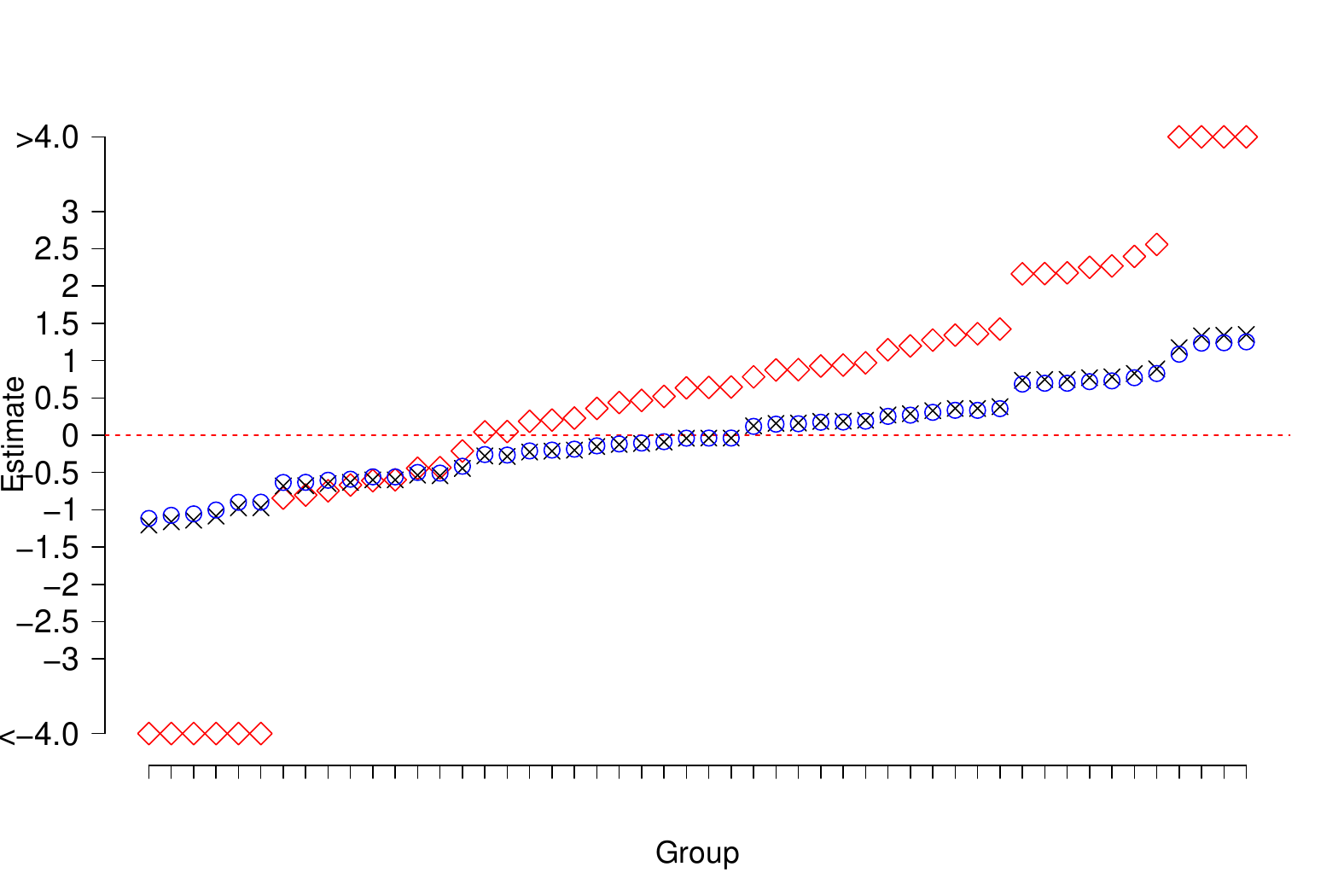}
         \subcaption{$n_g=5$}
         \end{center}
         \end{subfigure}
     \begin{subfigure}{0.48\textwidth}  
         \begin{center}
         \vspace{-.4in}
         \includegraphics[scale=.3]{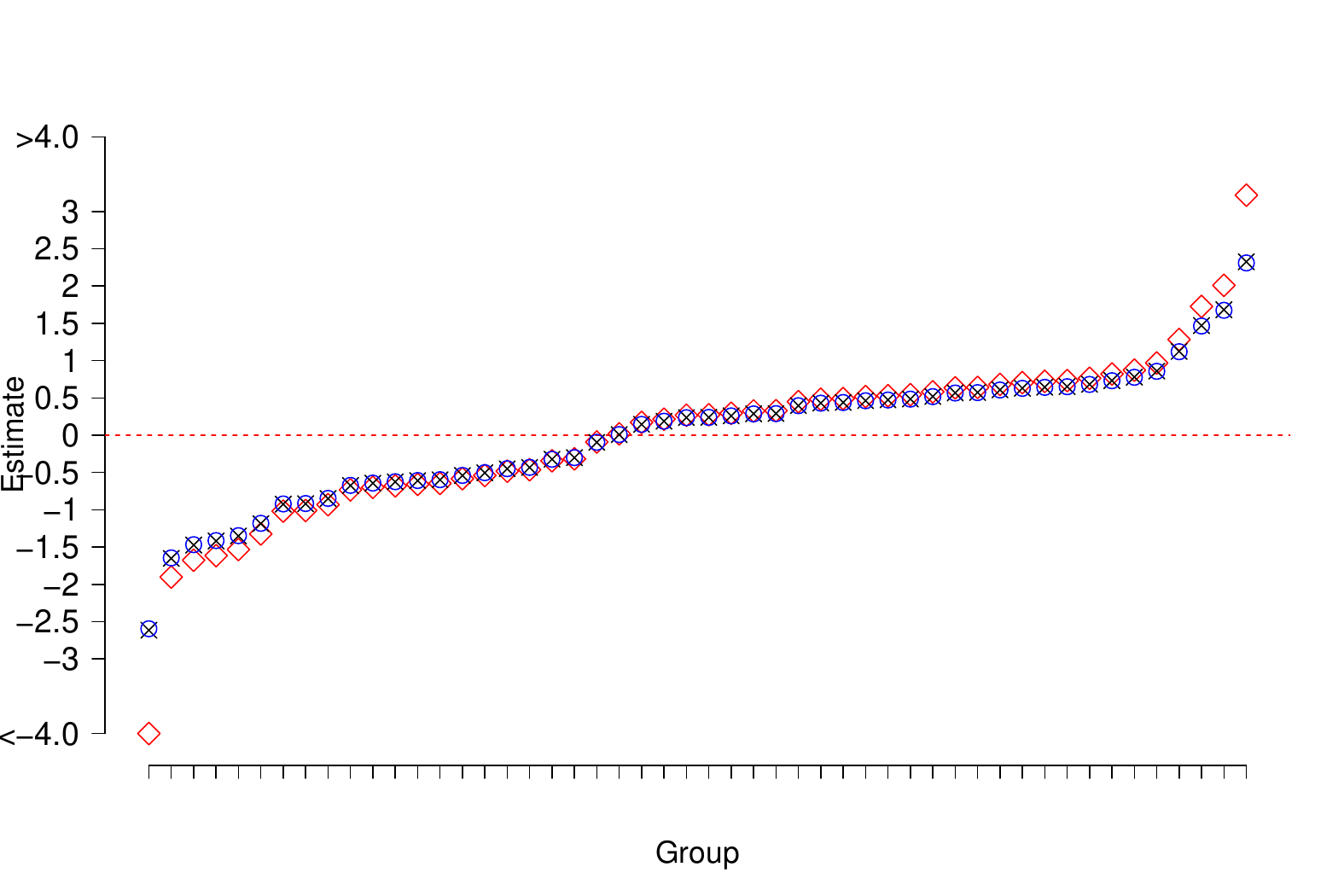}
         \subcaption{$n_g=50$}
         \end{center}
         \end{subfigure}
    
         \vspace{-0.6in}
         \begin{center}
         \includegraphics[scale=.4]{Figures_Tables/Figure02_Legend.pdf}
         \end{center}
          \vspace{-0.5in}
          
    \caption*{\textit{Note:} This figure replicates the results from Figure~\ref{fig:logistic_re} with MLMs fit through PQL. Results across one iteration at each sample size of the DGP in (\ref{DGP:logistic}) with $G=50$. The RI model is a logistic regression RI model, fit through PQL. The RegFE model is a logistic regression RegFE with only varying intercepts, setting $\omega$ to be the PQL-fit value. The Group-FE model is a logistic regression Group-FE model. Estimates of each $\gamma_g$ from Group-FE are found by omitting the intercept term in $X_{g[i]}$ and retaining all group indicators in $Z_{g[i]}$, which yields a different intercept term for each group. The estimates of $\gamma_g$ for Group-FE plotted above are then the difference of each of these estimated intercepts from their overall average. The red dashed line represents an estimate of 0.}
    \vspace{-.3in}
    
    \end{center}
    \end{figure}

Figures~\ref{fig:logistic_noneq_pql} and~\ref{fig:logistic_re_pql} replicate the results that underlie Figures~\ref{fig:logistic_noneq} and~\ref{fig:logistic_re}, respectively. We find that the PQL-fit RI model and RegFE (in Figures~\ref{fig:logistic_noneq_pql} and~\ref{fig:logistic_re_pql}) are even more similar than were MLE-fit RI and RegFE (in Figures~\ref{fig:logistic_noneq} and~\ref{fig:logistic_re}). 
In fact, the PQL-fit RI and RegFE models appear to produce effectively equivalent estimates of $\beta$ (Figure~\ref{fig:logistic_noneq_pql}) and $\gamma$ (Figure~\ref{fig:logistic_re_pql}). 
For example, while in Figure~\ref{fig:logistic_noneq_pql_a} the median estimates of $\beta_1$ (in (\ref{DGP:logistic})) between RegFE and RI when $G=50$ and $n_g=5$ appear to be slightly different, the estimates are in fact correlated at 0.999. Thus, when using PQL-based estimation, we can certainly still think of MLM as a regularized form of FE. 

\begin{figure}
    \caption{Replicating Figure~\ref{fig:bias_MLM_FE} --- Estimates of $\beta_1$ in \ref{DGP1} from PQL-fit RI, RegFE, Group-FE, and GLM without fixed or random effects}

    \vspace{-0.15in}
    \label{fig:bias_MLM_FE_pql}

\begin{center}

 \begin{subfigure}{0.48\textwidth}  
     \begin{center}
     \vspace{-.5in}
     \includegraphics[scale=.45]{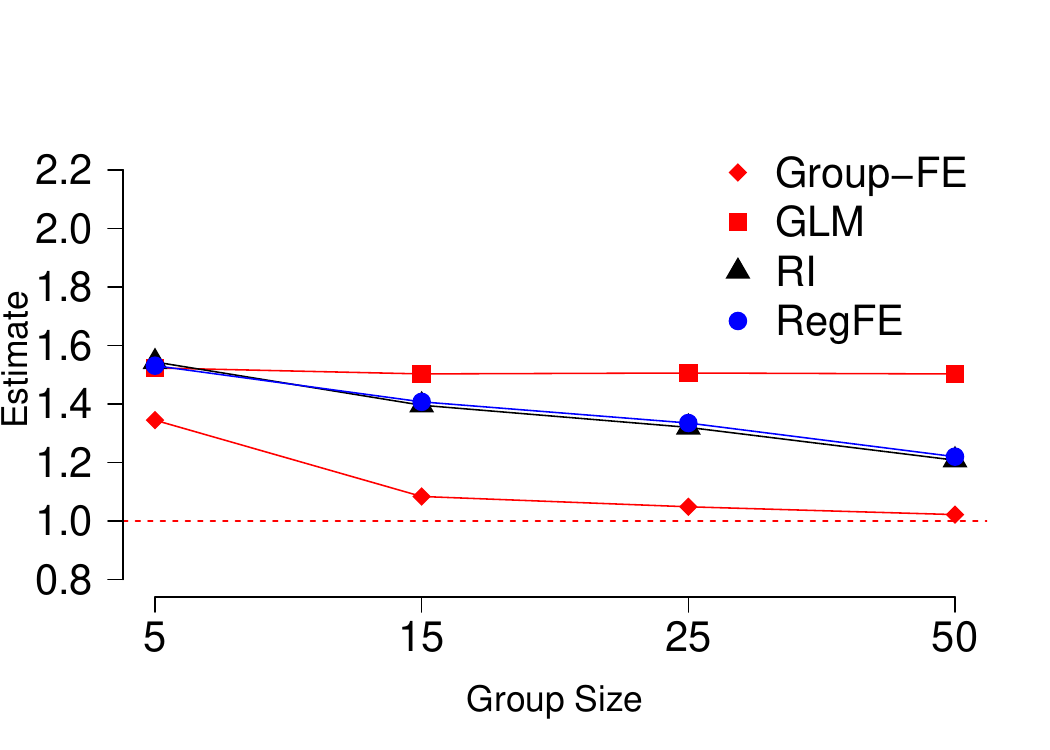}
     \subcaption{Mean estimates for $G=50$}
     \end{center}
     \end{subfigure}
 \begin{subfigure}{0.50\textwidth}  
     \begin{center}
     \vspace{-.5in}
     \includegraphics[scale=.45]{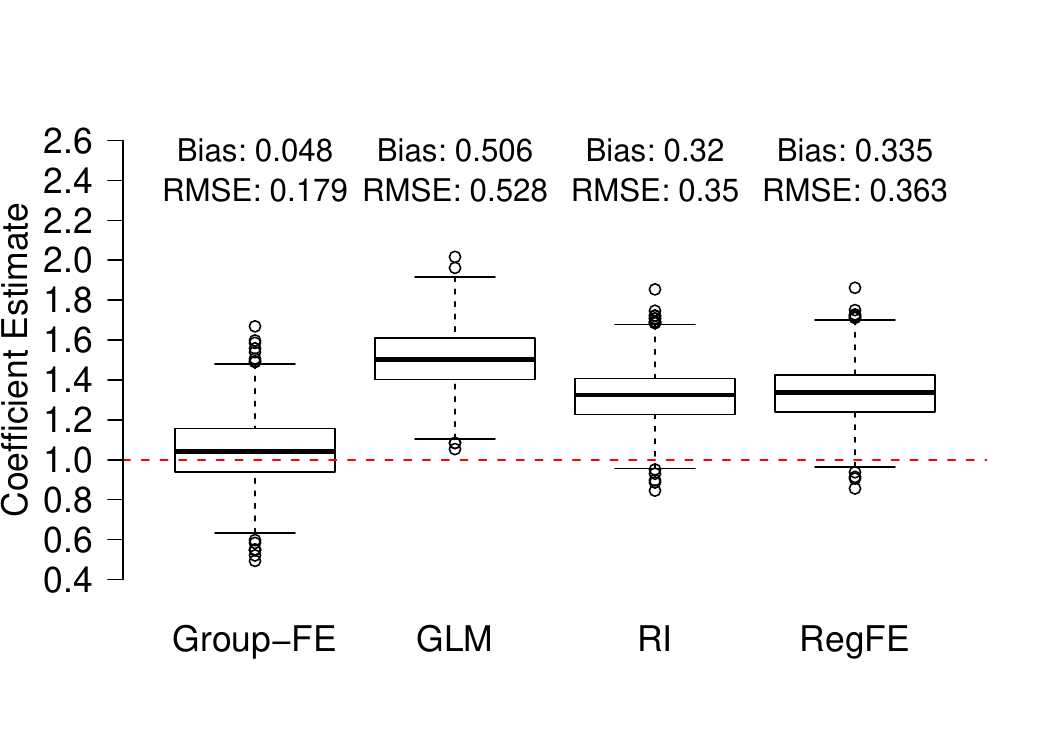}
     \subcaption{Estimate distributions for $G=50$ and $n_g=25$}
     \end{center}
     \end{subfigure}
      
\caption*{\textit{Note:} This figure replicates the results from Figure~\ref{fig:bias_MLM_FE} with MLMs fit through PQL. Results across 1000 iterations at each sample size of \ref{DGP1}. The RI model is a logistic regression RI model, fit through PQL. The RegFE model is a logistic regression RegFE with only varying intercepts, setting $\omega$ to be the PQL-fit estimate. The Group-FE model is a logistic regression Group-FE model. The GLM model is a logistic regression that only includes $X_{g[i]}$ as a regressor, and omits fixed and random effects. \textit{(a)} Mean estimates when $G=50$ for $\beta_1$. The red dashed line indicates the true value of $\beta_1 = 1$. \textit{(b)} Distribution of estimates of $\beta_1$ when $G=50$ and $n_g=25$. The red dashed line indicates the true value of $\beta_1 = 1$.}
\vspace{-.3in}

\end{center}
\end{figure}

Figure~\ref{fig:bias_MLM_FE_pql} then replicates Figure~\ref{fig:bias_MLM_FE}. We again see that PQL-fit RI and RegFE are 
effectively equivalent. And as with MLE-fit RI (in Figure~\ref{fig:bias_MLM_FE}), the PQL-fit RI (in Figure~\ref{fig:bias_MLM_FE_pql}) shows clear bias due to the group-level confounding in \ref{DGP1}, with consistently more bias than Group-FE. In summary, the lessons from Section~\ref{subsec:insight1} carry over to PQL-fit MLMs.

\subsubsection*{Results from Section~\ref{subsec:insight2} (Insight 2)}

\begin{figure}[!h]
    \caption{
    Replicating Figure~\ref{fig:estimates} --- Estimates of $\beta_1$ in \ref{DGP1} from PQL-fit bcRI, Group-FE, PQL-fit uncorrected RI, and bcRegFE}
    \label{fig:estimates_pql}
    \vspace{-.35in}
    \begin{subfigure}{.50\textwidth}
        \includegraphics[scale=0.48]{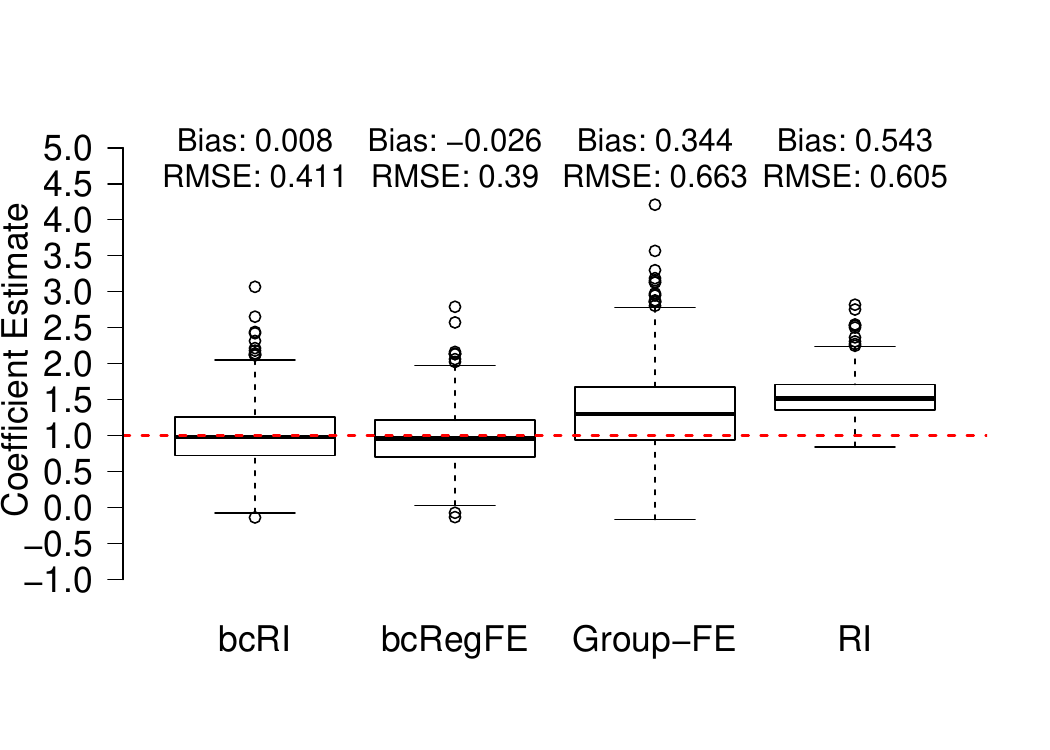}
        \vspace{-.20in}
        \subcaption{$G=50$ and $n_g=5$}
    \end{subfigure}
    \begin{subfigure}{.50\textwidth}
        \includegraphics[scale=0.48]{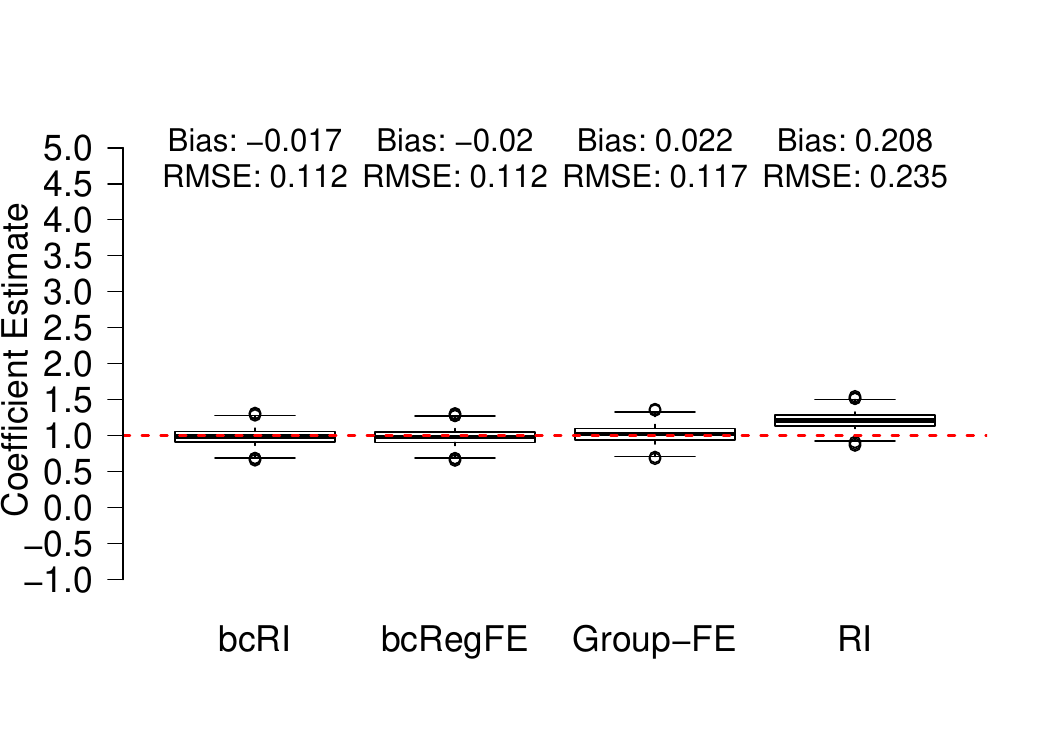}
        \vspace{-.20in}
        \subcaption{$G=50$ and $n_g=50$}
    \end{subfigure}

    \vspace{.25in}
    \subcaption*{\textit{Note:} This figure replicates the results from Figure~\ref{fig:estimates} with MLMs fit through PQL. Results across 1000 iterations at each sample size of DGP 1. Distributions of estimates for $\beta_1$ by logistic regression applications of Group-FE, PQL-fit bcRI, a varying intercepts bcRegFE model with $(\theta, \Omega)$ set to the PQL-fit values, and uncorrected PQL-fit RI. The dashed horizontal line represents the true parameter value, $\beta_1 = 1$.}
\end{figure}

\begin{figure}[!h]
    \vspace{.05in}
    \caption{Replicating Figure~\ref{fig:prediction} --- Average test error rates of PQL-fit bcRI, bcRegFE, and Group-FE in DGP 1}
    \label{fig:prediction_pql}
        \vspace{-.35in}
    \begin{subfigure}{.49\textwidth}
        \includegraphics[scale=0.50]{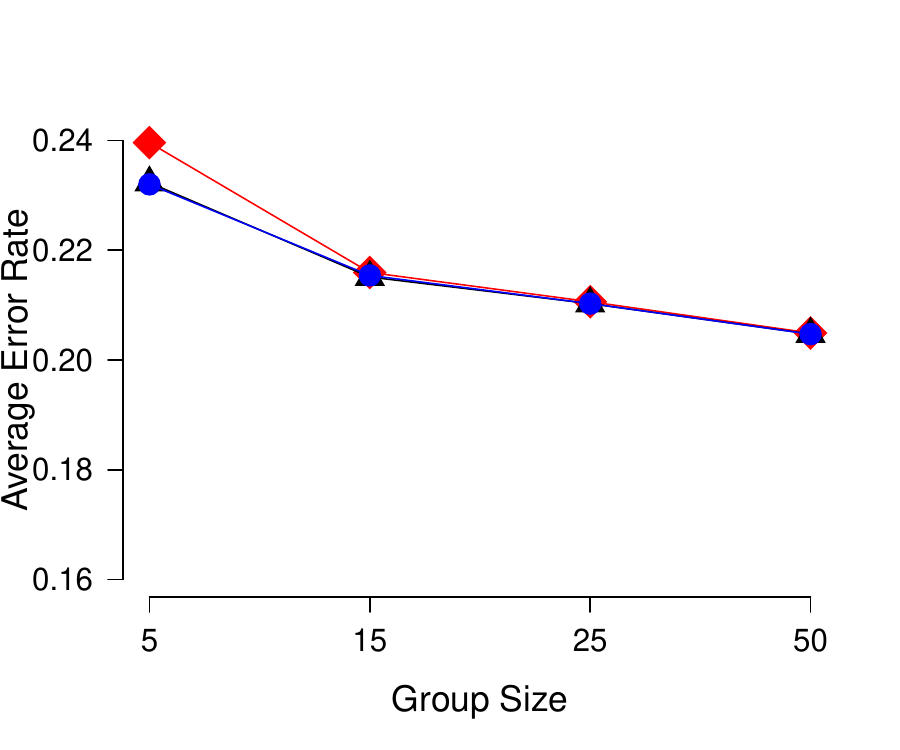}
        \subcaption{$G=15$}
    \end{subfigure}
    \begin{subfigure}{.49\textwidth}
        \includegraphics[scale=0.50]{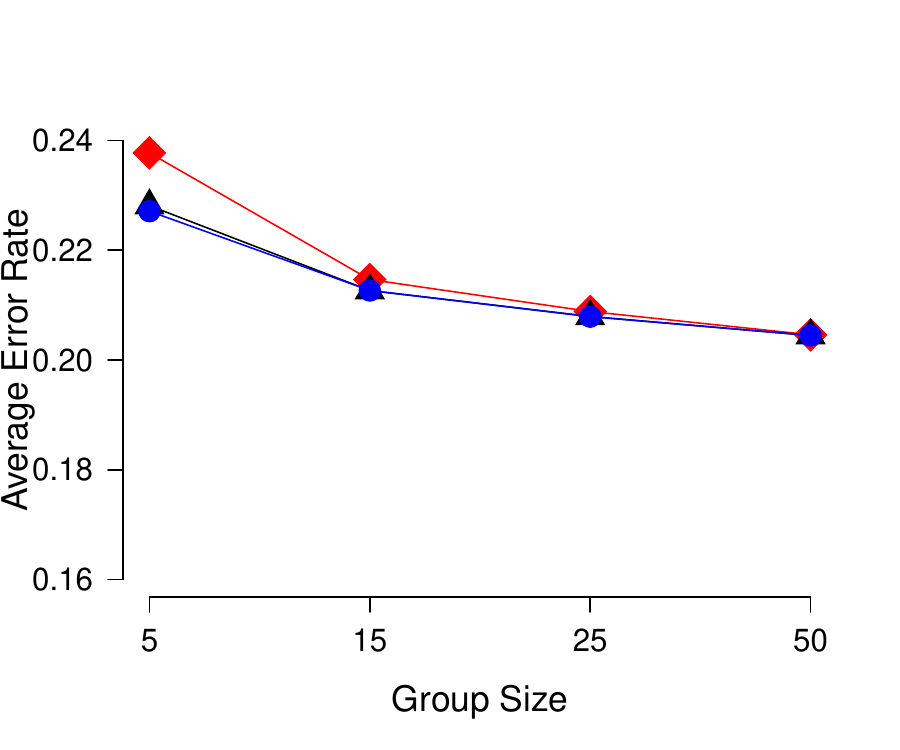}
        \subcaption{$G=50$}
    \end{subfigure}
    
         \vspace{-0.6in}
     \begin{center}
     \includegraphics[scale=.4]{Figures_Tables/Figure05_Legend.pdf}
     \end{center}
      \vspace{-0.6in}
      
    \subcaption*{\textit{Note:} This figure replicates Figure~\ref{fig:prediction} with MLMs fit through PQL. Results across 1000 iterations at each sample size of DGP 1. Comparison of the average test error rates of logistic regression applications of Group-FE, PQL-fit bcRI, and a varying intercepts bcRegFE model with $(\theta, \Omega)$ set to the PQL-fit value. The training and testing datasets were of the same size.}
    \vspace{-0.15in}
    \end{figure}

Figures~\ref{fig:estimates_pql} and~\ref{fig:prediction_pql} replicate the results from Figures~\ref{fig:estimates} and~\ref{fig:prediction}, respectively. In Figure~\ref{fig:estimates_pql}, we see that, just like with MLE-fit models (in Figure~\ref{fig:estimates}), the PQL-fit bcRI and bcRegFE models correct the bias that uncorrected RI and RegFE show in \ref{DGP1} (see Figure~\ref{fig:bias_MLM_FE_pql}). Additionally, in Figure~\ref{fig:prediction_pql}, bcRI and bcRegFE retain the improved predictive accuracy over Group-FE even when fit through PQL. Thus, our lessons from Section~\ref{subsec:insight2} carry over to PQL-fit MLMs: even when fit through PQL, bcMLM and bcRegFE are preferable to uncorrected MLM, uncorrected RegFE, and Group-FE when group-level confounding is present.

\subsubsection*{Results from Section~\ref{subsec:insight3} (Insight 3)}

\begin{figure}[!h]
        \caption{Replicating Figure~\ref{fig:coverage} --- Coverage rates of 95\% confidence intervals for $\beta_1$ in DGP 2 with PQL-fit MLMs}
    \label{fig:coverage_pql}
        \vspace{-.35in}
    \begin{subfigure}{.24\textwidth}
        \includegraphics[scale=0.39]{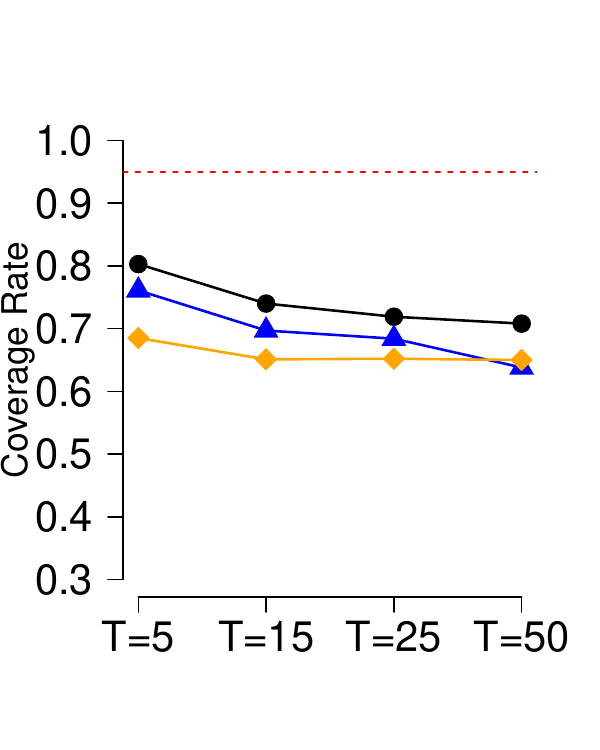}
        \subcaption{RI}\label{fig:coverage_pql_a}
    \end{subfigure}
    \begin{subfigure}{.24\textwidth}
        \includegraphics[scale=0.39]{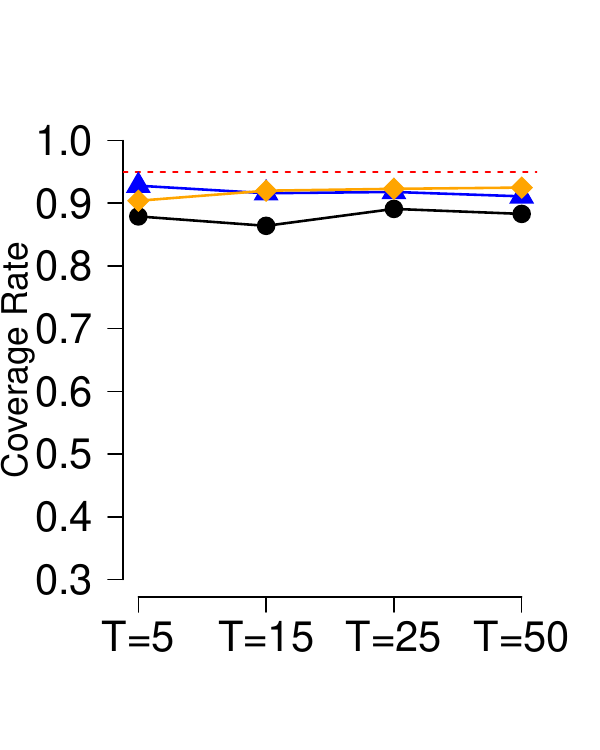}
        \subcaption{RI, Bootstrap}\label{fig:coverage_pql_b}
    \end{subfigure}
    \begin{subfigure}{.24\textwidth}
        \includegraphics[scale=0.39]{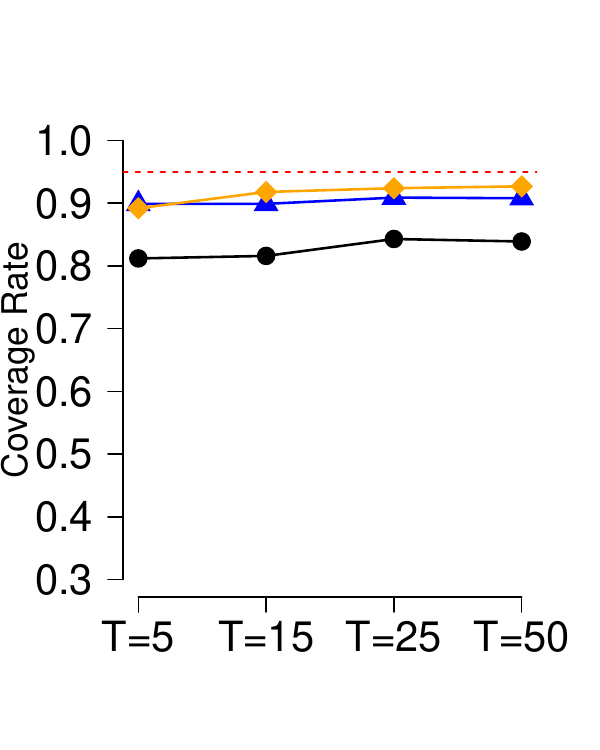}
        \subcaption{Group-FE, CRSE}\label{fig:coverage_pql_c}
    \end{subfigure}
     \begin{subfigure}{.24\textwidth}
        \includegraphics[scale=0.39]{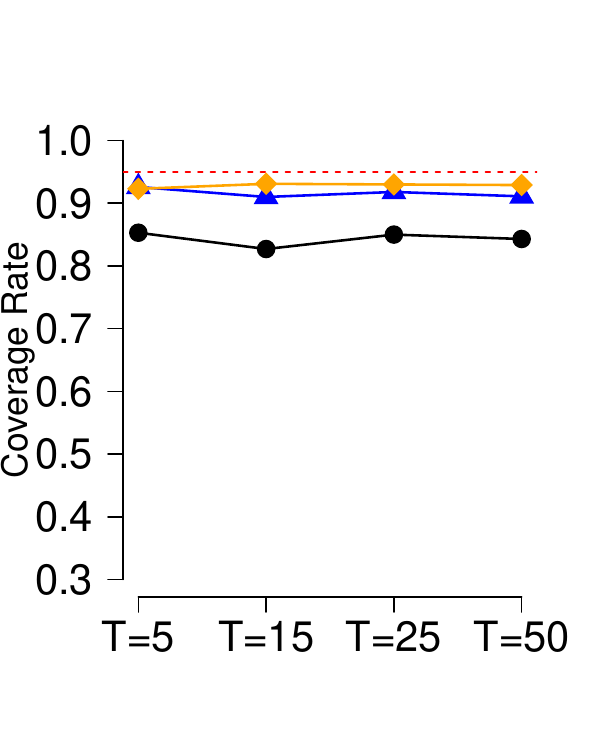}
        \subcaption{RegFE, CRSE}\label{fig:coverage_pql_d}
    \end{subfigure}   
         \vspace{-0.6in}
     \begin{center}
     \includegraphics[scale=.4]{Figures_Tables/Figure06_Legend.pdf}
     \end{center}
      \vspace{-0.6in}
    \subcaption*{\textit{Note:} This figure replicates Figure~\ref{fig:coverage} with MLMs fit through PQL. Results across 1000 iterations at each sample size of DGP 2. Comparison of the coverage rates of 95\% confidence intervals constructed using \textit{(a)} 
    PQL-fit Poisson RI with its default standard errors; \textit{(b)} PQL-fit Poisson RI with a percentile  cluster-bootstrap (using 200 bootstrap samples); \textit{(c)} Poisson Group-FE with CRSEs; and \textit{(d)} Poisson RegFE with CRSEs. The dashed horizontal line shows the target 95\% coverage rate.}
    \end{figure}

Figure~\ref{fig:coverage_pql} replicates the results from Figure~\ref{fig:coverage}. As with MLE-based estimation (see Figure~\ref{fig:coverage_a}), the default standard errors from the PQL-fit RI model are clearly downwardly biased, showing severe undercoverage for $\beta_1$ in DGP 2 (see Figure~\ref{fig:coverage_pql_a}). A cluster-bootstrap with the PQL-fit RI model (Figure~\ref{fig:coverage_pql_b}), and CRSEs with Group-FE (Figure~\ref{fig:coverage_pql_c}) and RegFE (Figure~\ref{fig:coverage_pql_d}) show much improved coverage rates that approach nominal coverage as $G$ increases. These are the same findings as when using MLE-fit MLMs in Figure~\ref{fig:coverage}. Thus, the lessons from Section~\ref{subsec:insight3} carry over to PQL-fit MLMs. 

\newpage
\clearpage 

\subsection{Comparing RegFE and MLM for Poisson Regression }\label{app:pois_gamma_sims}
\begin{figure}[!h]
\caption{Median estimates of $\beta_1$ in (\ref{DGP:Poisson log}) from RI, RegFE, Group-FE, and a GLM without fixed or random effects for Poisson regression with a log link}\label{fig:poisson_log_estimates}

\vspace{-0.1in}

\begin{center}

 \begin{subfigure}{0.48\textwidth}  
     \begin{center}
     \vspace{-.5in}
     \includegraphics[scale=.6]{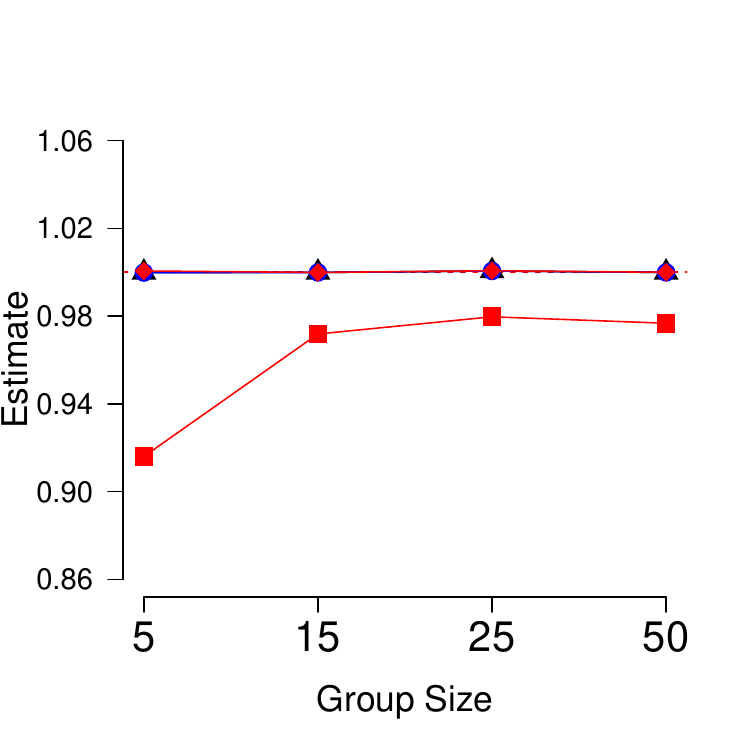}
     \subcaption{$G=15$}
     \end{center}
     \end{subfigure}
 \begin{subfigure}{0.48\textwidth}  
     \begin{center}
     \vspace{-.5in}
     \includegraphics[scale=.6]{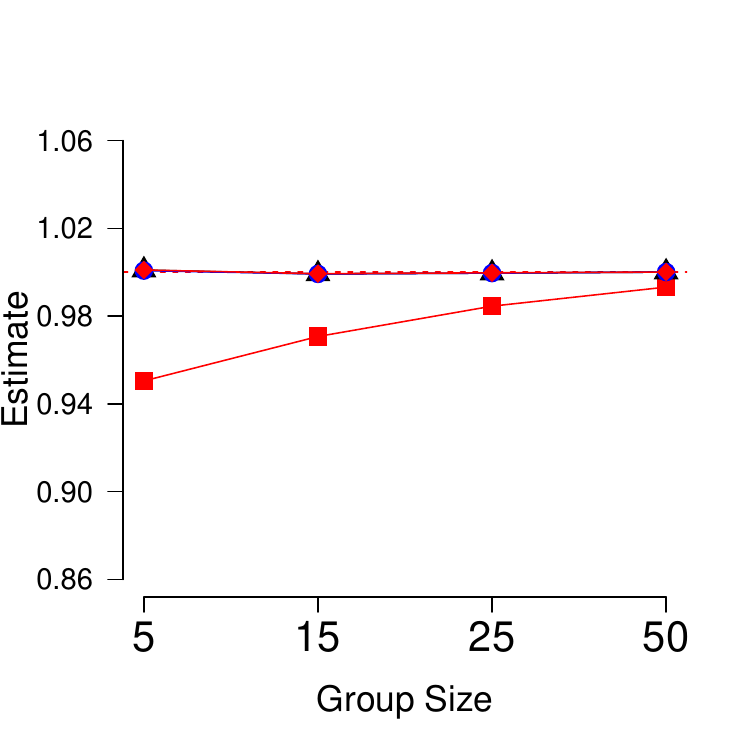}
     \subcaption{$G=50$}
     \end{center}
     \end{subfigure}

     \vspace{-0.5in}
     \begin{center}
     \includegraphics[scale=.4]{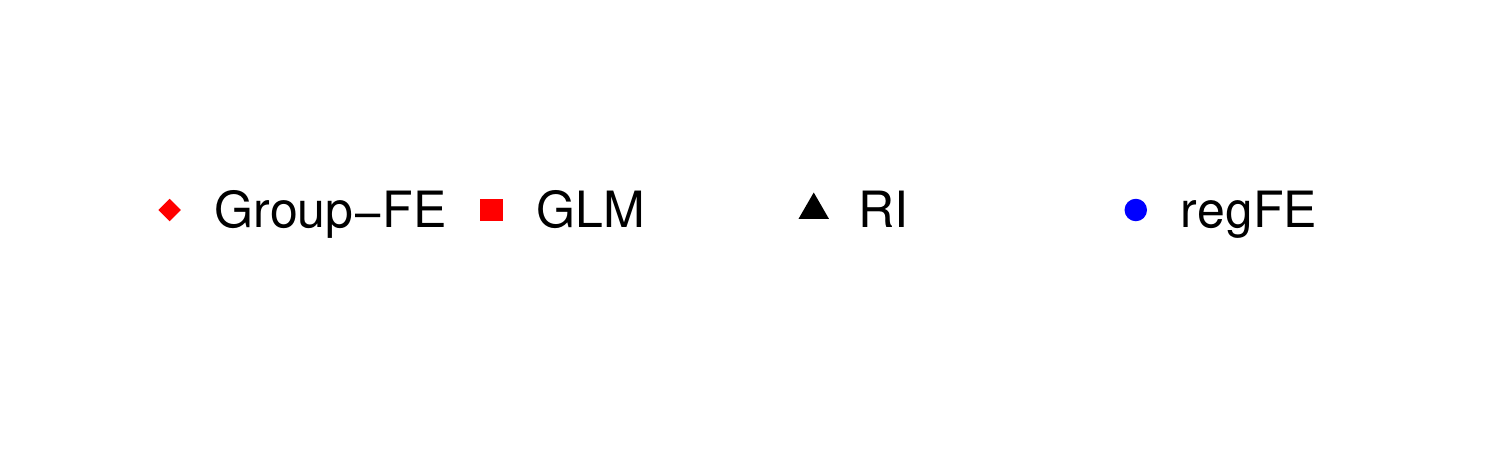}
     \end{center}
      \vspace{-0.5in}
      
\subcaption*{\textit{Note:} Results across 1000 iterations at each sample size of the DGP in (\ref{DGP:Poisson log}) with $\beta_0 = 1$ and $\beta_1 = 1$. The RI model is a Poisson regression RI model with a log link. The RegFE model is a log link Poisson regression RegFE with only varying intercepts, setting $\lambda_{\mathrm{GLM}} =  \frac{1}{2 \hat\omega^2_{\text{RI}}}$. The Group-FE model is a Poisson regression Group-FE model with a log link. The GLM model is a log link Poisson regression that only includes $X_{g[i]}$ as a regressor, and omits fixed and random effects. The graphs plot the median estimates of $\beta_1$. The red dashed line represents the true estimate in the simulation. Median RI estimates are difficult to see in both figures because they are roughly the same as the Group-FE estimates in both figures here. Median RegFE estimates are difficult to see when $G=15$ because they are roughly the same as the Group-FE and RI estimates.}
\vspace{-.3in}

\end{center}
\end{figure}

\begin{figure}[!h]
\caption{Estimates of $\gamma$ in (\ref{DGP:Poisson log}) from RI, RegFE, and Group-FE for Poisson regression with a log link}\label{fig:poisson_re}

\vspace{-0.1in}

\begin{center}

 \begin{subfigure}{0.48\textwidth}  
     \begin{center}
     \vspace{-.5in}
     \includegraphics[scale=.3]{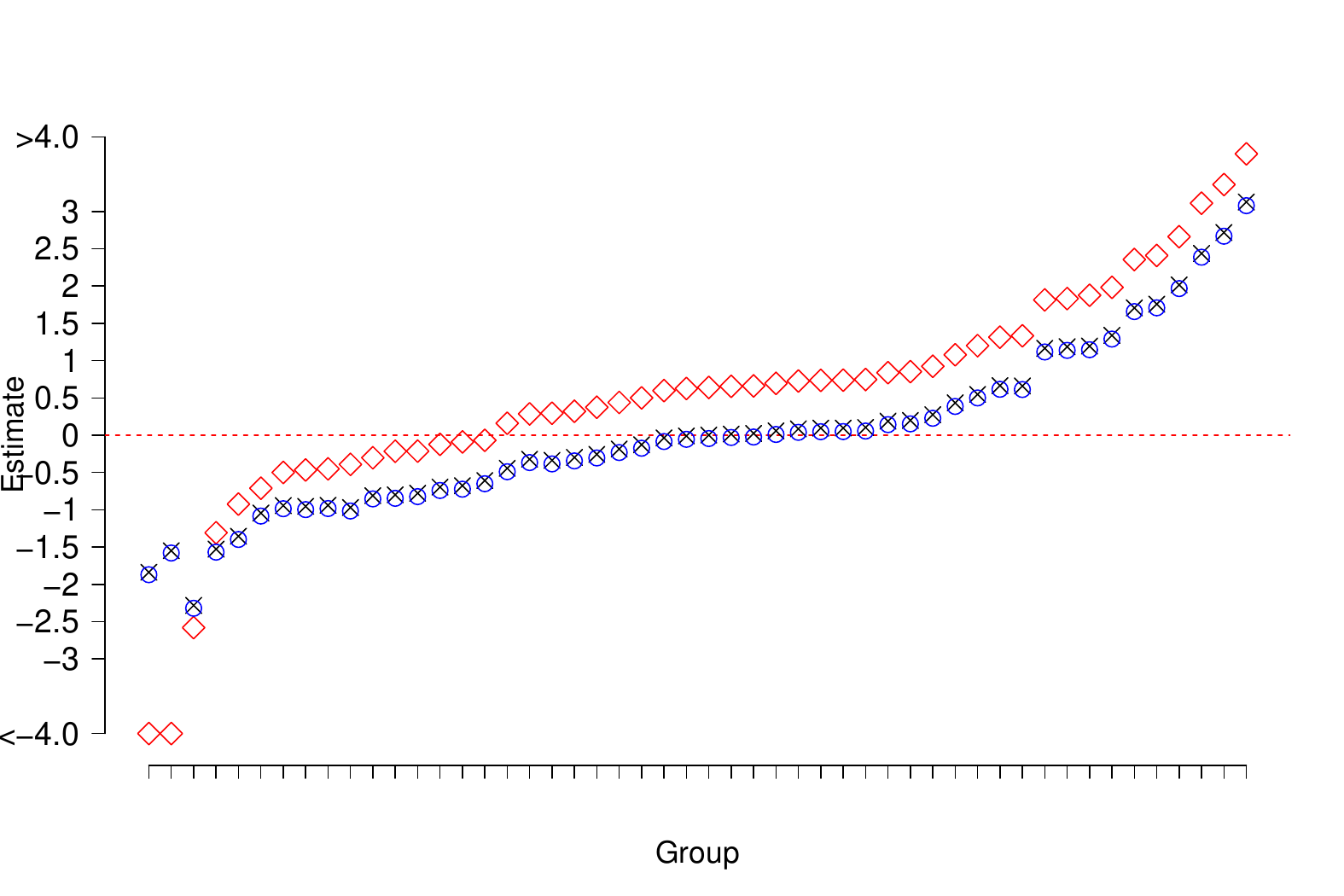}
     \subcaption{$n_g=5$}
     \end{center}
     \end{subfigure}
 \begin{subfigure}{0.48\textwidth}  
     \begin{center}
     \vspace{-.4in}
     \includegraphics[scale=.3]{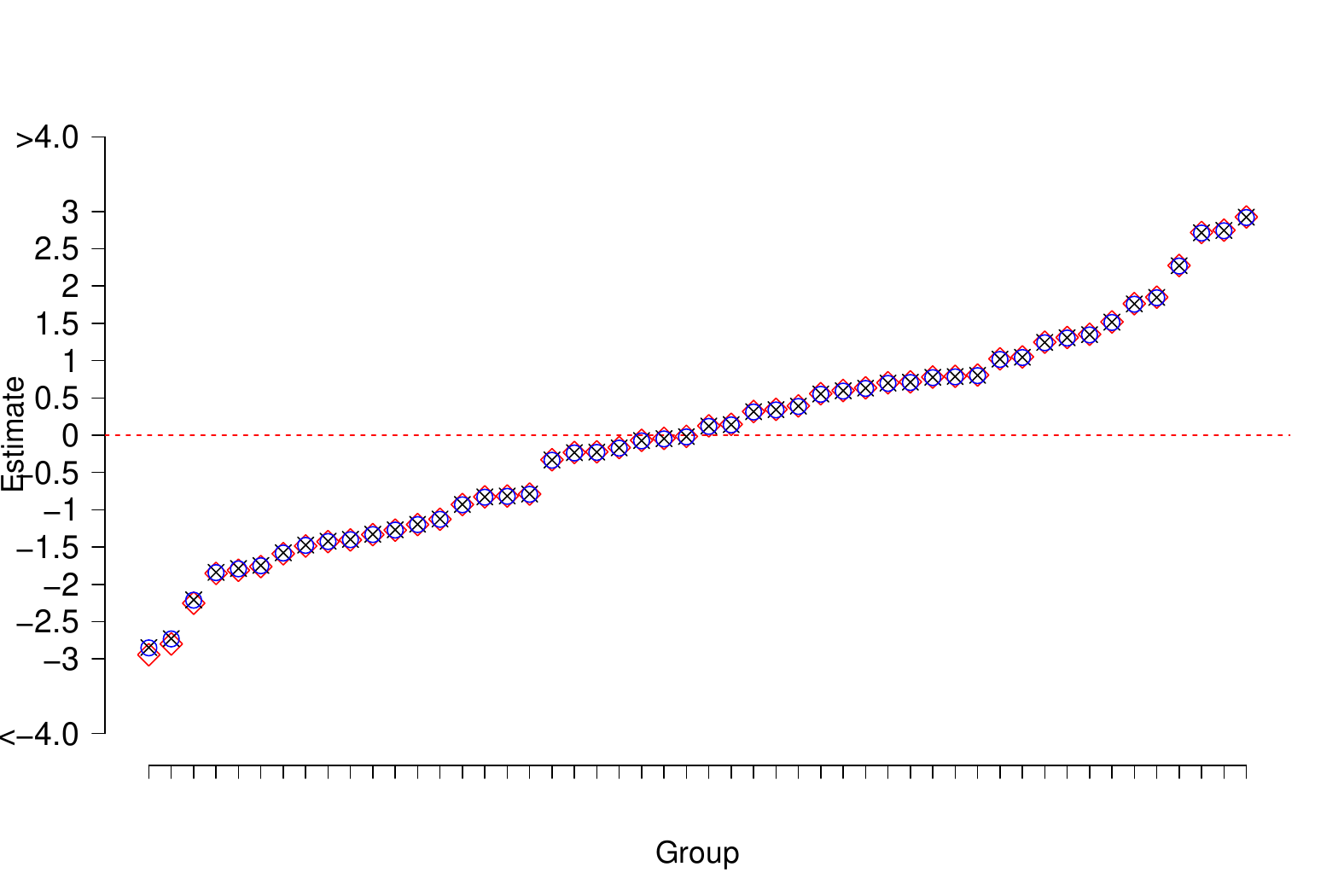}
     \subcaption{$n_g=50$}
     \end{center}
     \end{subfigure}

     \vspace{-0.5in}
     \begin{center}
     \includegraphics[scale=.4]{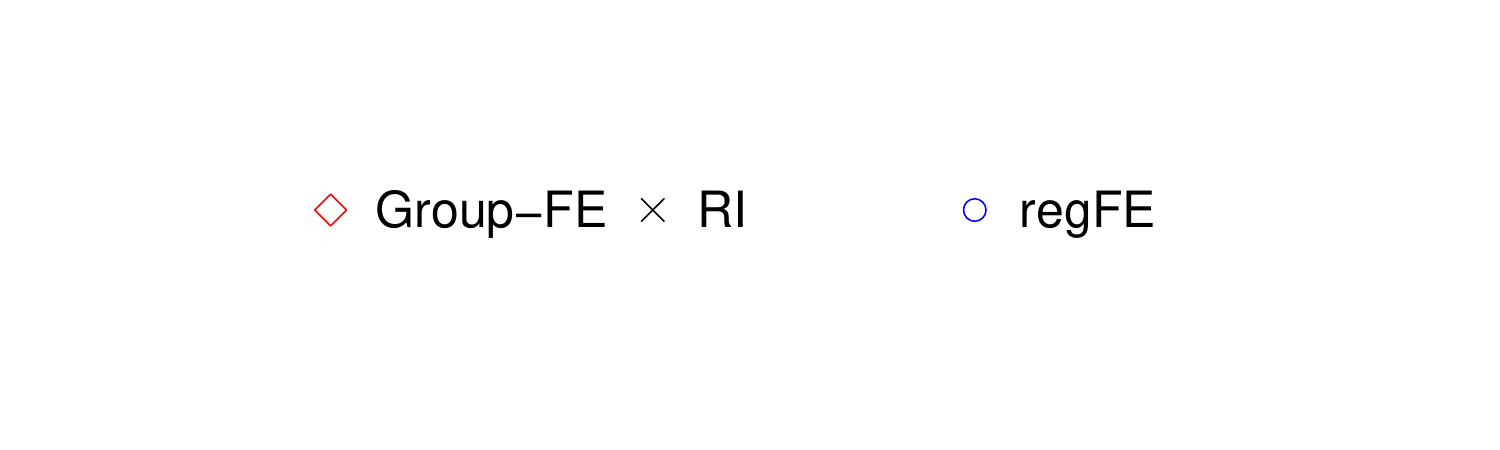}
     \end{center}
      \vspace{-0.5in}
      
\subcaption*{\textit{Note:} Results across one iteration at each sample size of the DGP in (\ref{DGP:Poisson log}) with $G=50$. The RI model is a Poisson regression RI model with a log link. The RegFE model is a log link Poisson regression RegFE with only varying intercepts, setting $\lambda_{\mathrm{GLM}} =  \frac{1}{2 \hat\omega^2_{\text{RI}}}$. The Group-FE model is a Poisson regression Group-FE model with a log link. Estimates of each $\gamma_g$ from Group-FE are found by omitting the intercept term in $X_{g[i]}$ and retaining all group indicators in $Z_{g[i]}$, which yields a different intercept term for each group. The estimates of $\gamma_g$ for Group-FE are then the difference of each of these estimated intercepts from their overall average. The red dashed line represents an estimate of 0.}
\vspace{-.3in}

\end{center}
\end{figure}

In this appendix, we compare parameter estimates of MLM and RegFE for Poisson regression through simulation. Data are generated according to the following DGP:
     \begin{align}\label{DGP:Poisson log}
        & Y_{g[i]}\overset{}{\sim}\text{Poisson}(\lambda = \mathrm{exp} (\beta_0+X_{g[i]}\beta_1+W_g)) \\
        \text{ where } \ \  & W_g \overset{iid}{\sim}N(0, 1.5) \ \ \text{and} \ \ X_{g[i]} \overset{iid}{\sim} N(0,1) \nonumber 
    \end{align}
Figure \ref{fig:poisson_log_estimates} compares the estimates for $\beta_1$ for log-link Poisson regression RI and varying intercepts RegFE with $\lambda_{\mathrm{GLM}}=\frac{1}{2\hat\omega_{\mathrm{RI}}^2}$ as both $G$ and $n_g$ vary. The median estimates for $\beta_1$ from Group-FE, RI, and RegFE are nearly identical in all of the sample sizes tried. Figure~\ref{fig:poisson_re} then compares the estimates of $\gamma$ from RI, RegFE, and Group-FE. The regularization imposed by RegFE and RI is more clear here---when $n_g=5$, the \textit{positive} RI and RegFE estimates of $\gamma$ are shrunken toward 0 from the Group-FE estimates, but when $n_g=50$, the estimates from all three methods are very close. 

\subsection{Bias in Poisson Regression MLM Estimates}\label{app:bias of poison gamma}

\begin{figure}[!h]
    \caption{
    Estimates of $\beta_1$ in (\ref{DGP:Poisson bias})  from log link Poisson regression for uncorrected RI, bcRI, RegFE, bcRegFE, Group-FE, and a GLM without fixed or random effects}
    \label{fig:poisson_estimates}
    \vspace{-.20in}
    \begin{subfigure}{1\textwidth}
        \begin{center}
        \includegraphics[scale=0.50]{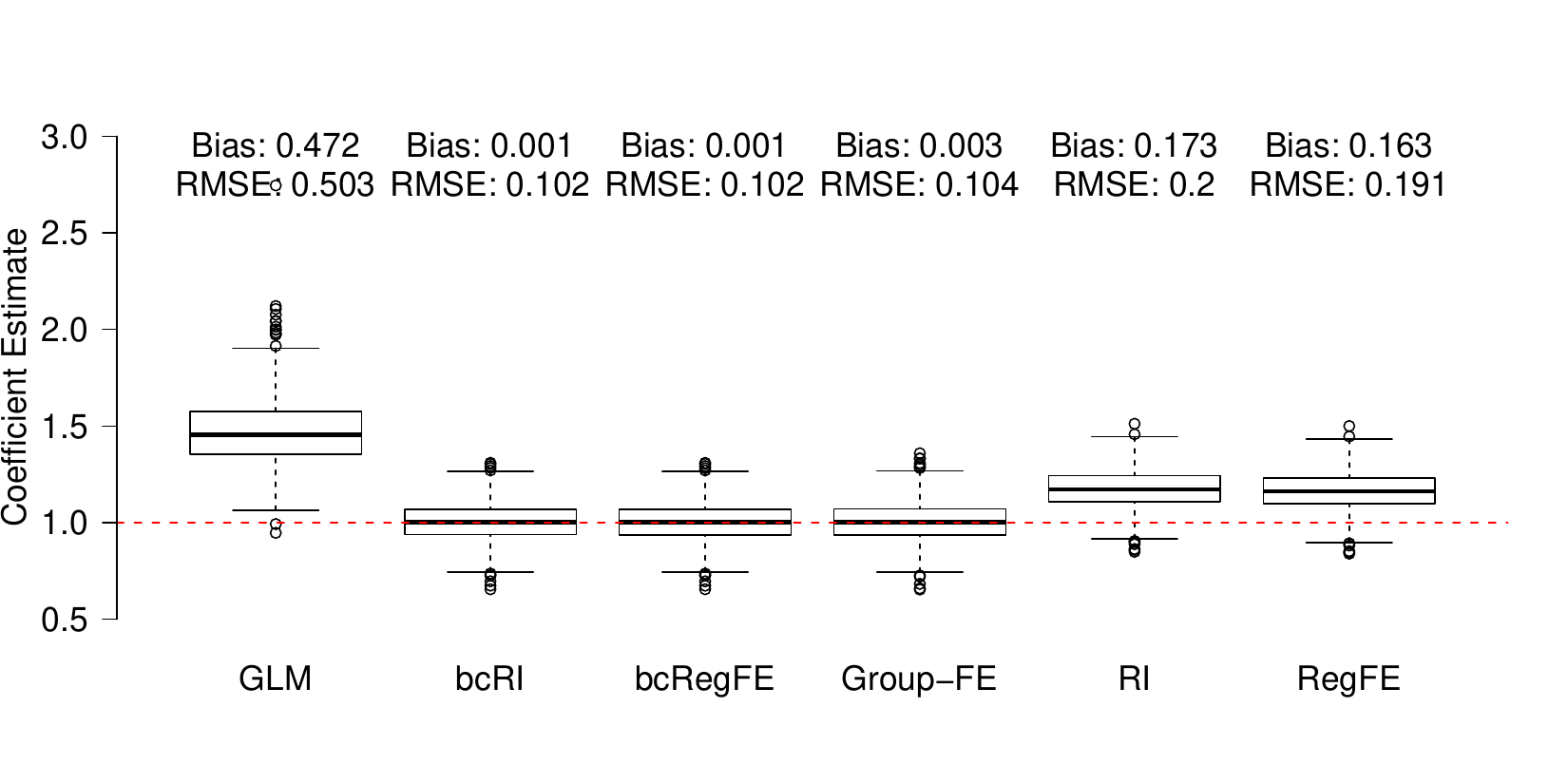}
        \vspace{-.10in}
        \subcaption{$G=50$ and $n_g=5$}
        \end{center}
    \end{subfigure}
    \begin{subfigure}{1\textwidth}
        \begin{center}
        \includegraphics[scale=0.50]{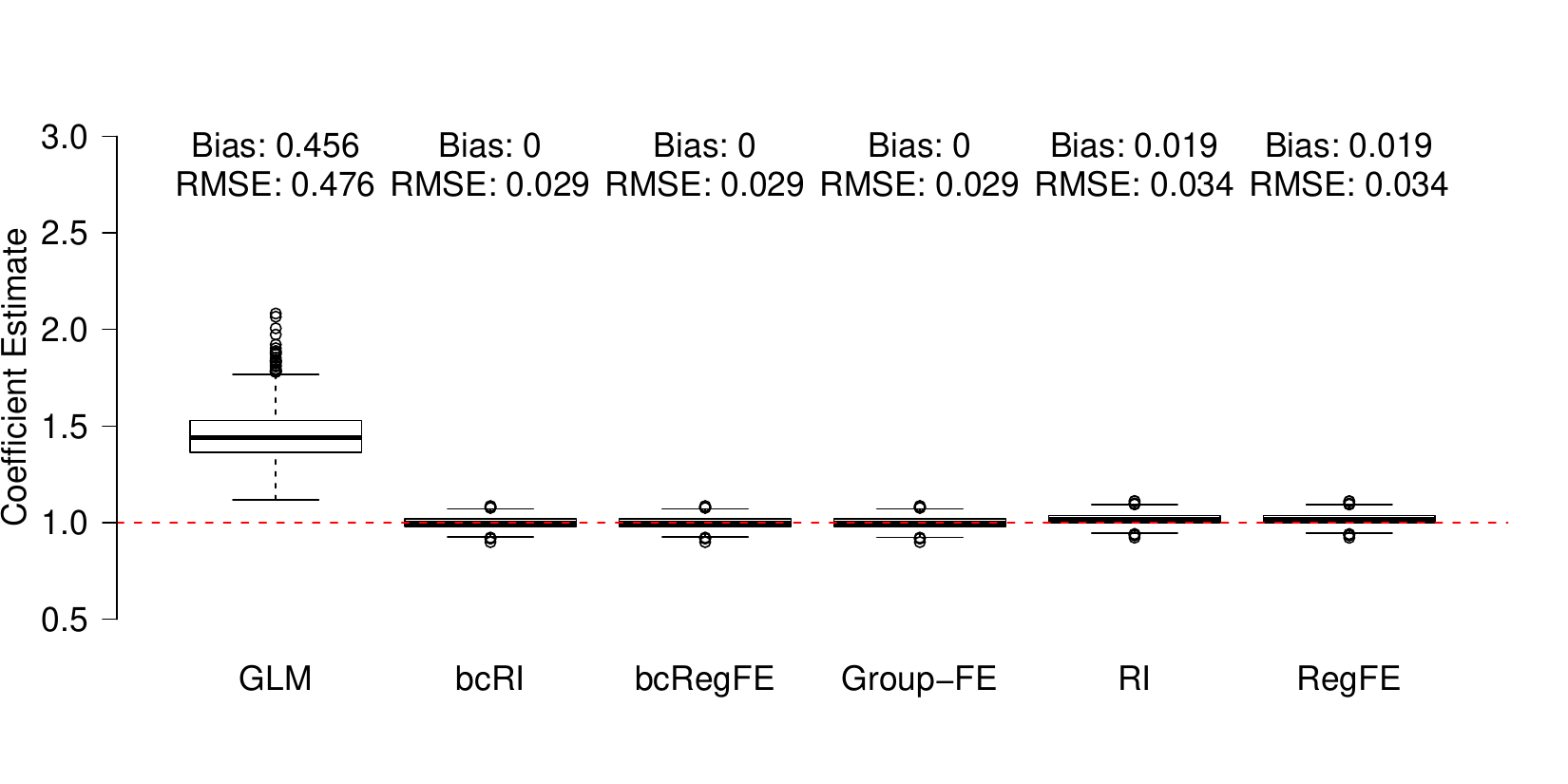}
        \vspace{-.10in}
        \subcaption{$G=50$ and $n_g=50$}
        \end{center}
    \end{subfigure}

    \vspace{.15in}
    \subcaption*{\textit{Note:} Results across 1000 iterations at each sample size of (\ref{DGP:Poisson bias}). Distributions of estimates for $\beta_1$ by log link Poisson regression applications of Group-FE, bcRI, uncorrected RI, varying intercepts RegFE and bcRegFE, and a GLM without fixed or random effects. The dashed horizontal line represents the true parameter value, $\beta_1 = 1$. }
\end{figure}

In this appendix, we demonstrate through simulation that uncorrected MLM's parameter estimates in a Poisson regression can be biased when group-level confounding is present, but this bias can be corrected by using bcMLM or bcRegFE. Data are generated according to the following DGP:
     \begin{align}\label{DGP:Poisson bias}
        & Y_{g[i]}{\sim}\text{Poisson}(\lambda = \mathrm{exp} (\beta_0+X_{g[i]}\beta_1+W_g^{(1)}+W_g^{(2)}))  \\ 
        \text{ where } \ \  & [W_g^{(1)} \ W_g^{(2)}]^{\top} \overset{iid}{\sim}\mathcal{N}(\Vec{0}, \frac{1}{4} I_2) \ \ \text{and} \ \ X_{g[i]}{\sim}N(W_g^{(1)},0.5) \nonumber 
    \end{align}
where the $W_g^{(j)}$ are unobserved. As in \ref{DGP1}, $X_{g[i]}$ is correlated with the random effect $W_g^{(1)}$, which acts as a confounder. Thus, MLM should be expected to produce biased estimates of $\beta_1$. Figure \ref{fig:poisson_estimates} confirms this hypothesis---RI reports bias in both small in large groups, though the bias is greatly decreased in larger groups. bcRI, bcRegFE, and Group-FE, on the other hand, are effectively unbiased at both sample sample sizes tried. Note that Group-FE's result here differs slightly from the logistic regression case in \ref{DGP1}, where Group-FE showed noticeable bias when group sizes were small. 

\begin{figure}[!h]
    \vspace{.05in}
    \caption{Average test mean squared error of log link Poisson regression applications of bcRI, Group-FE, and bcRegFE in (\ref{DGP:Poisson bias})}
    \label{fig:poisson_prediction}
        \vspace{-.35in}
    \begin{subfigure}{.48\textwidth}
        \includegraphics[scale=0.45]{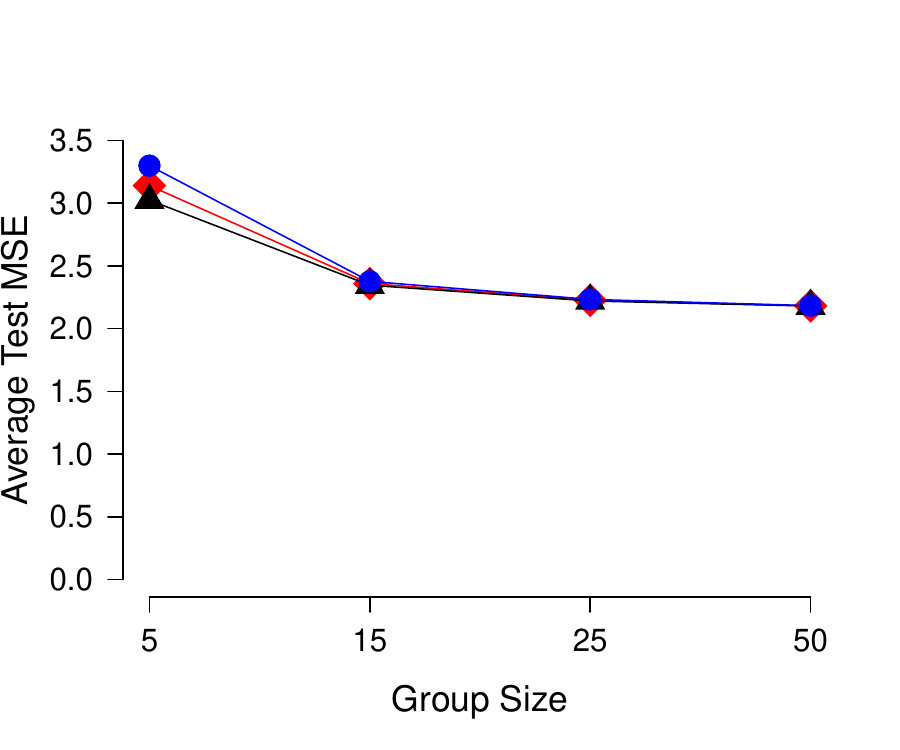}
        \subcaption{$G=15$}
    \end{subfigure}
    \begin{subfigure}{.48\textwidth}
        \includegraphics[scale=0.45]{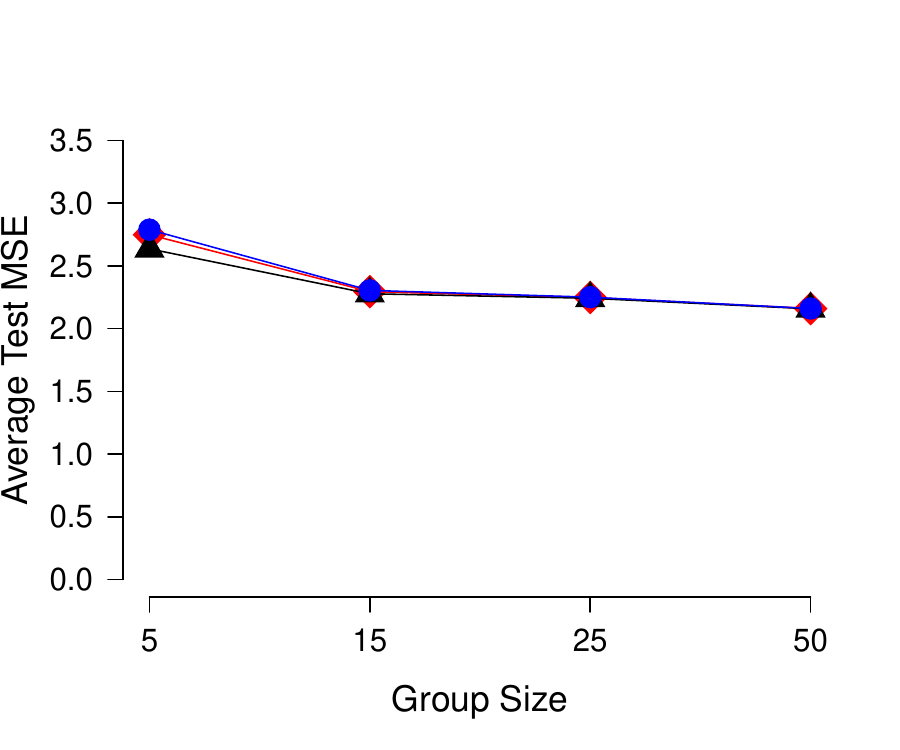}
        \subcaption{$G=50$}
    \end{subfigure}
    
         \vspace{-0.6in}
     \begin{center}
     \includegraphics[scale=.4]{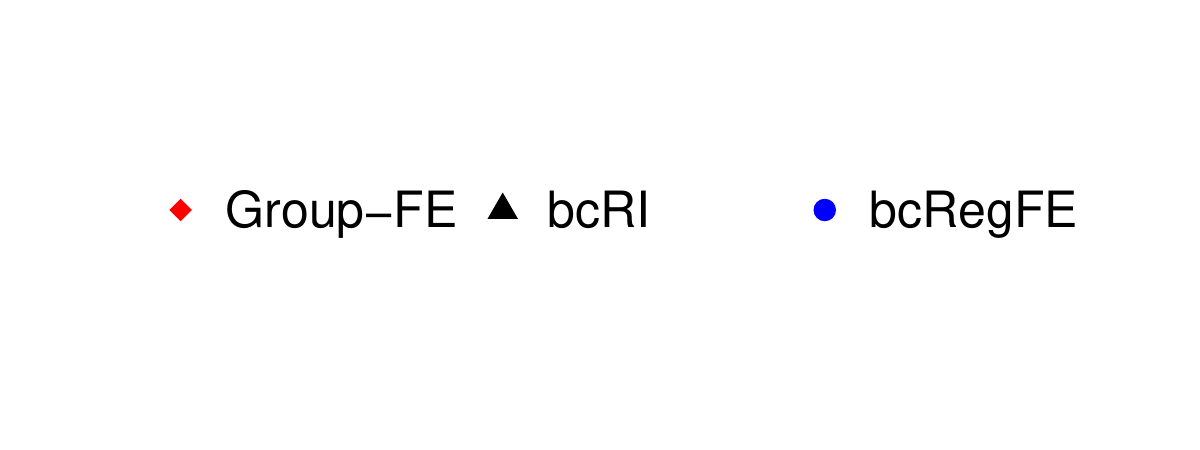}
     \end{center}
      \vspace{-0.6in}
      
    \subcaption*{\textit{Note:} Results across 1000 iterations at each sample size of (\ref{DGP:Poisson bias}). Comparison of the average mean squared error on test data of log link Poisson regression applications of Group-FE, bcRI, and bcRegFE. The training and testing datasets were of the same size. }
    \vspace{-0.15in}
\end{figure}

Finally, Figure~\ref{fig:poisson_prediction} shows that bcRI, Group-FE, and bcRegFE all have very similar predictive accuracy on test data in this DGP. When $n_g=5$, bcRI's test error is slightly lower than that of Group-FE, and bcRegFE's test error is slightly higher than those of Group-FE and bcRI (likely because Group-FE already does well in this DGP, and our implementation of bcRegFE here does not use cross-validation to choose its level of regularization). When $n_g > 5$, however, the test mean squared errors from these three models are indistinguishable.  

\subsection{GLM bcMLM and bcRegFE with Random Slopes}\label{app:rs}

\begin{figure}[!h]
    \caption{
    Estimates of $\beta_1$ in (\ref{dgp_rs_bcmlm})  from logistic regression for uncorrected MLM, bcMLM, bcRegFE, and FE}
    \label{fig:rs_estimates}
    \vspace{-.20in}
    \begin{subfigure}{.50\textwidth}
        \includegraphics[scale=0.40]{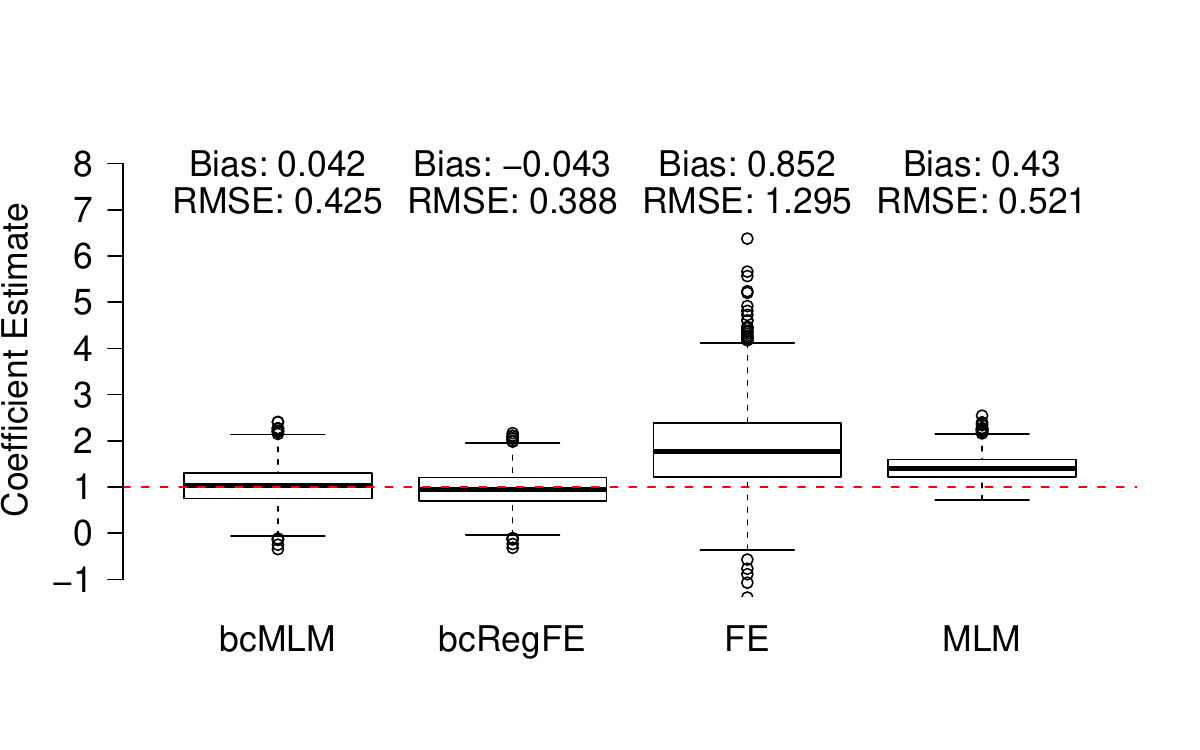}
        \vspace{-.20in}
        \subcaption{$G=50$ and $n_g=5$}
    \end{subfigure}
    \begin{subfigure}{.50\textwidth}
        \includegraphics[scale=0.40]{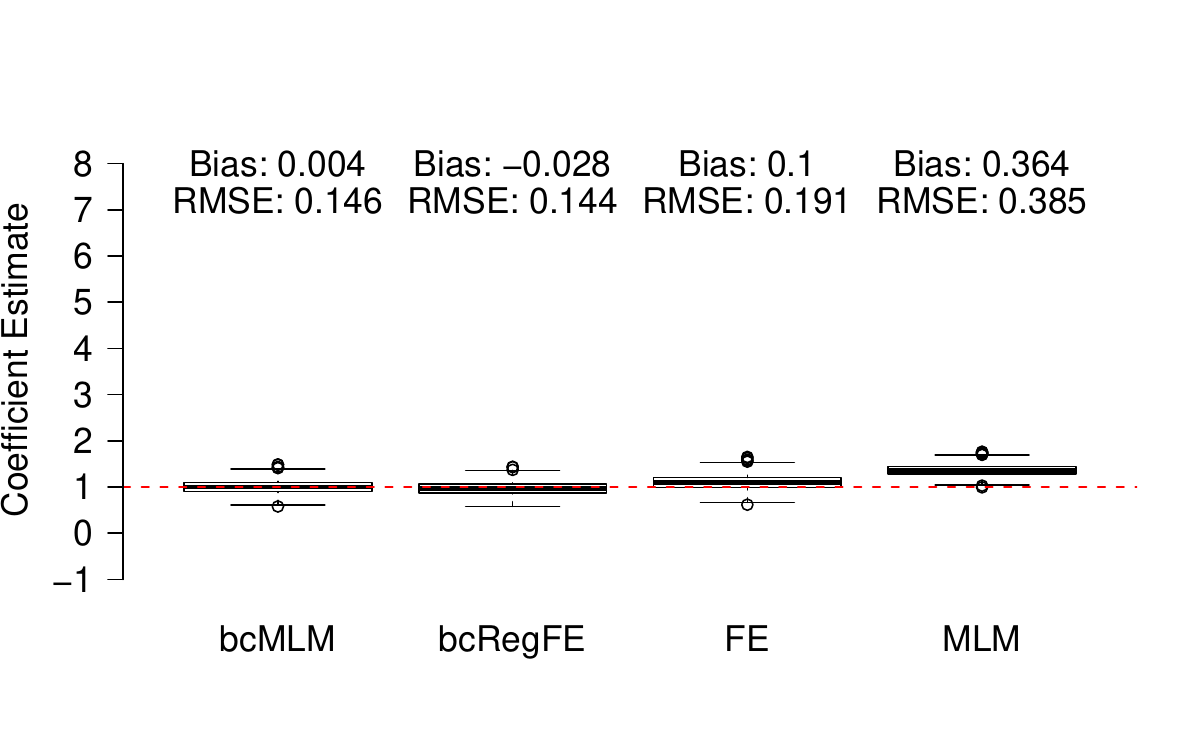}
        \vspace{-.20in}
        \subcaption{$G=50$ and $n_g=25$}
    \end{subfigure}

    \vspace{.15in}
    \subcaption*{\textit{Note:} Results across 1000 iterations at each sample size of (\ref{dgp_rs_bcmlm}). Distributions of estimates for $\beta_1$ by logistic regression applications of FE, bcMLM, bcRegFE, and uncorrected MLM, where each model allows a group-varying intercept and slope for $X_{g[i]}^{(2)}$. The dashed horizontal line represents the true parameter value, $\beta_1 = 1$. }
\end{figure}

\begin{figure}[!h]
    \vspace{.05in}
    \caption{Average test error rates of logistic regression applications of bcMLM, bcRegFE, and FE in (\ref{dgp_rs_bcmlm})}
    \label{fig:rs_prediction}
        \vspace{-.35in}
    \begin{subfigure}{.48\textwidth}
        \includegraphics[scale=0.45]{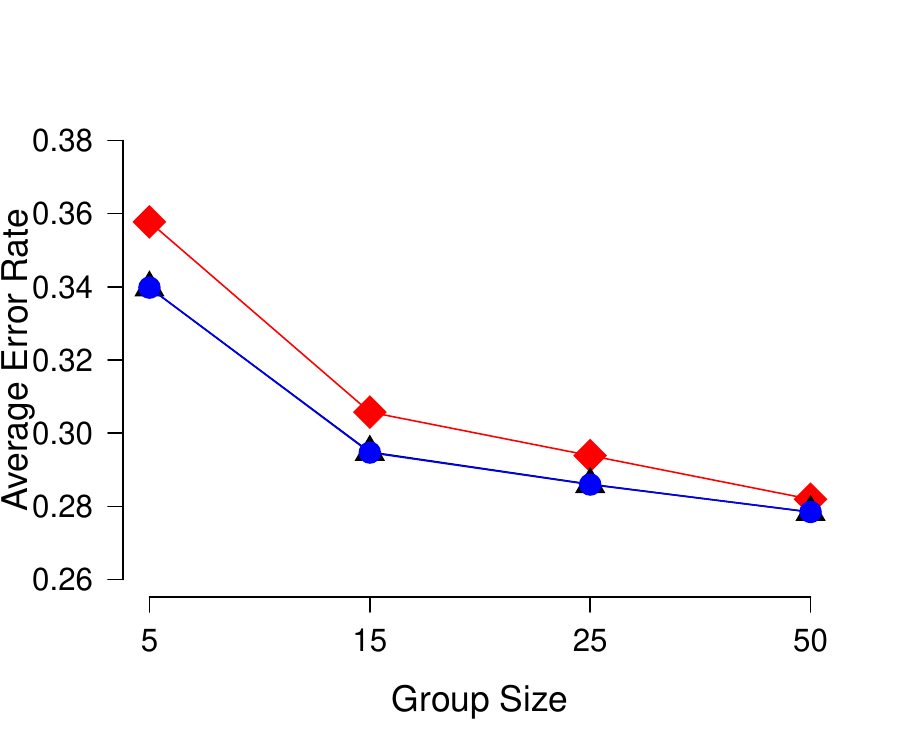}
        \subcaption{$G=15$}
    \end{subfigure}
    \begin{subfigure}{.48\textwidth}
        \includegraphics[scale=0.45]{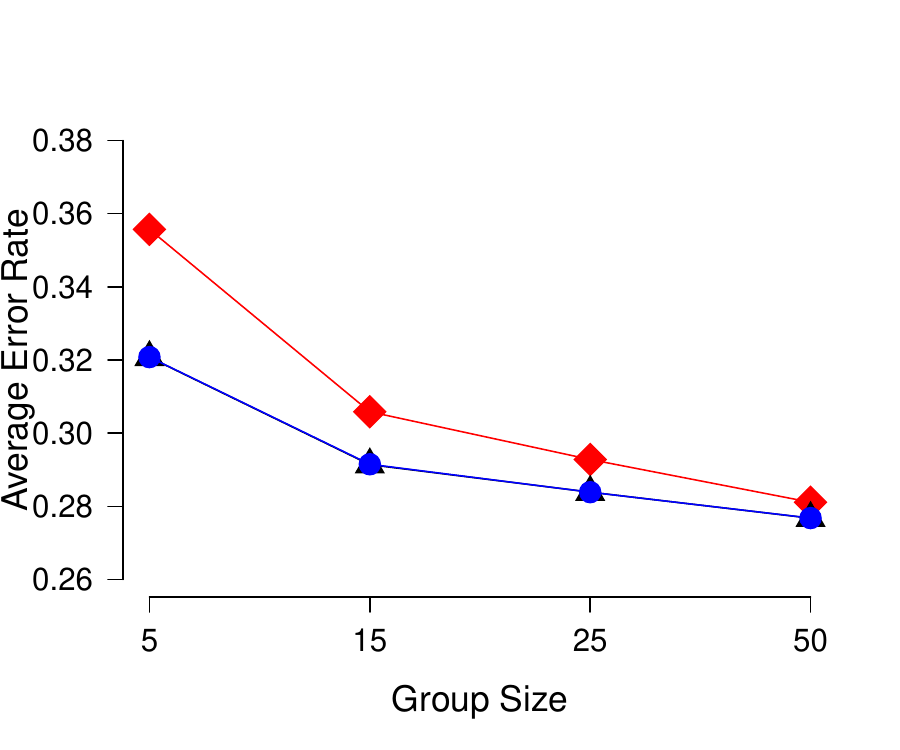}
        \subcaption{$G=50$}
    \end{subfigure}
    
         \vspace{-0.6in}
     \begin{center}
     \includegraphics[scale=.4]{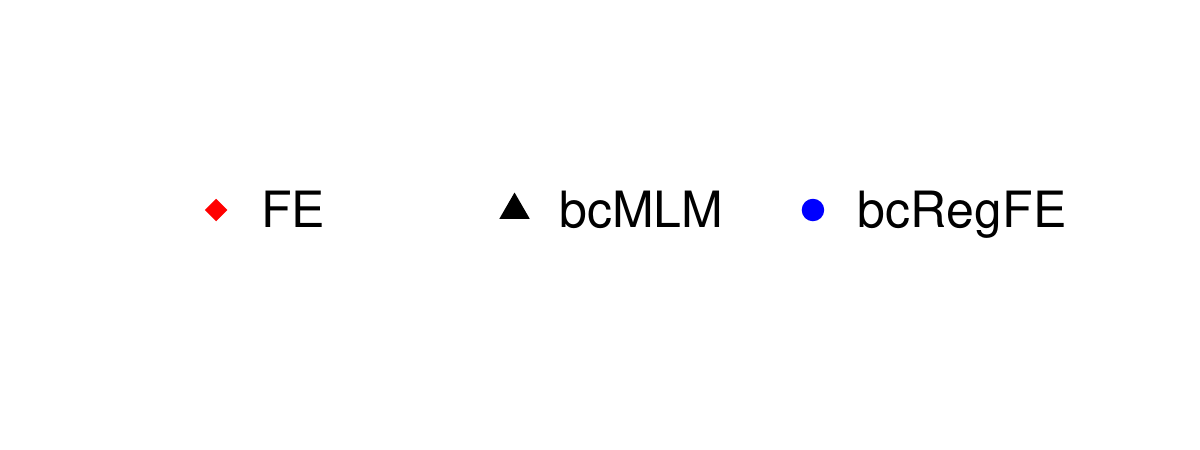}
     \end{center}
      \vspace{-0.6in}
      
    \subcaption*{\textit{Note:} Results across 1000 iterations at each sample size of (\ref{dgp_rs_bcmlm}). Comparison of the average error rates of logistic regression applications of FE, bcMLM, and bcRegFE which allow group-varying intercepts and slopes for $X_{g[i]}^{(2)}$. The training and testing datasets are of the same size. }
    \vspace{-0.15in}
\end{figure}

In this appendix, we demonstrate through simulation how general bcMLM and bcRegFE with a random slopes can fix MLM's bias problem, and retain superior predictive accuracy to FE. Consider the following DGP in the logistic regression setting:
 \begin{align}\label{dgp_rs_bcmlm}
    & Y_{g[i]}\overset{}{\sim}\text{Bernoulli}(\operatorname{logit}^{-1}(\beta_0+X^{(1)}_{g[i]}\beta_1+ X^{(2)}_{g[i]}(\beta_2 + W_g^{(2)}) + W_g^{(1)})) \\ 
    \text{ where } \ \  & W_g^{(1)} \overset{iid}{\sim} N(0, 1) \ \ \text{and} \ \  W_g^{(2)} \overset{iid}{\sim} \chi^2_1 - 1 \nonumber \\
    \text{and} \ \ & X^{(2)}_{g[i]}{\sim} N(0,0.5) \ \ \text{and} \ \  X^{(1)}_{g[i]} =  X^{(2)}_{g[i]} W_g^{(2)} + N(0, 0.5) \nonumber 
\end{align}
where $W_g^{(1)}$ and $W_g^{(2)}$ are unobserved. Here, $W_g^{(1)}$ is a random intercept, and  $W_g^{(2)}$ is a random slope on $X^{(2)}_{g[i]}$. Further, $X^{(2)}_{g[i]}$ and $W_g^{(2)}$ act as confounders for $X^{(1)}_{g[i]}$, and $X^{(1)}_{g[i]}$ is correlated with the random effect contribution, $X^{(2)}_{g[i]} W_g^{(2)}$. Thus, uncorrected MLM should show bias. This is confirmed in Figure~\ref{fig:rs_estimates}---uncorrected MLM with a random intercept and random slope for $X^{(2)}_{g[i]}$ shows large amounts of bias at each sample size. bcMLM and bcRegFE, however, are effectively unbiased in small and large groups. FE shows large bias and high variance when $n_g=5$, and still shows some bias at $n_g=25$, though the bias has shrunken greatly, and is much less than uncorrected MLM. At both sample sizes, bcMLM and bcRegFE have the lowest bias, and the lowest RMSE. Further, Figure~\ref{fig:rs_prediction} shows that bcMLM and bcRegFE have consistently higher predictive accuracy than does FE in this DGP.

\newpage
\clearpage 

\subsection{Comparing Conditional Logistic Regression and Firth's Correction to bcMLM, bcRegFE, and FE in \ref{DGP1}}\label{app:firth_condlr}

In this section, we compare bcRI, Group-FE, and bcRegFE to Conditional Logistic Regression (\citealp{breslow1978estimation}) and Firth's correction (\citealp{firth1993}), which has been extended to other GLMs (e.g., \citealp{kosmidis2009bias}) in \ref{DGP1}. Table~\ref{tab.bias} reports the bias for each method in estimating $\beta_1$ in \ref{DGP1}, and Table~\ref{tab.rmse} reports the RMSE. In terms of absolute bias and RMSE, bcRI, bcRegFE, Conditional Logistic Regression, and Firth's correction perform remarkably similarly, except at the smallest sample size ($G=15$ and $n_g=5$). At the smallest sample size, Firth's correction and bcRegFE have the lowest bias and RMSE, followed by bcRI, and then Conditional Logistic Regression.

\begin{table}[!h]
\caption{Bias for $\beta_1$ in \ref{DGP1}}
\label{tab.bias}
\scriptsize
\begin{center}
\begin{tabular}{| c | c | c | c | c | c | c | c | c | }
	\hline
		\textbf{Groups} ($G$) & \textbf{Group Size} ($n_g$) & \textbf{GLM} & \textbf{RI} & \textbf{Group-FE} & \textbf{bcRI} & \textbf{bcRegFE} & \textbf{Cond-LR} & \textbf{Firth} \\
	\hline
		15 & 5 & 0.643 & 0.817 & 0.520 & 0.154 & 0.082 & 0.166 & 0.006 \\
        \hline
		50 & 5 & 0.523 & 0.684 & 0.344 & 0.044 & -0.032 & 0.054 & -0.014 \\
        \hline
		15 & 15 & 0.534 & 0.509 & 0.098 & 0.016 & -0.022 & 0.018 & -0.008 \\
        \hline
		50 & 15 & 0.502 & 0.499 & 0.084 & 0.006 & -0.036 & 0.007 & -0.005 \\
        \hline
		15 & 25 & 0.535 & 0.410 & 0.060 & 0.013 & -0.015 & 0.015 & 0.001 \\
        \hline
		50 & 25 & 0.506 & 0.396 & 0.048 & 0.004 & -0.026 & 0.004 & -0.002 \\
        \hline
		15 & 50 & 0.520 & 0.250 & 0.022 & -0.001 & -0.018 & 0.000 & -0.006 \\
        \hline
		50 & 50 & 0.502 & 0.248 & 0.022 & 0.000 & -0.018 & 0.000 & -0.002 \\
        \hline
\end{tabular}
\end{center}
\normalsize
    \subcaption*{\textit{Note:} Results across 1000 iterations at each sample size of \ref{DGP1}. Comparison of bias in estimating $\beta_1$ from a base logistic regression model (GLM) that does not include any group-varying intercepts ($\gamma_g$); uncorrected RI (RI); uncorrected Group-FE (Group-FE); bcRI; varying intercepts bcRegFE (bcRegFE); conditional logistic regression that stratifies by group (Cond-LR); and a Group-FE model with Firth's bias correction (Firth).  } 

\end{table}

\begin{table}[!h]
\caption{RMSE for $\beta_1$ in \ref{DGP1}}
\label{tab.rmse}
\scriptsize
\begin{center}
\begin{tabular}{| c | c | c | c | c | c | c | c | c | }
	\hline
		\textbf{Groups} ($G$) & \textbf{Group Size} ($n_g$) & \textbf{GLM} & \textbf{RI} & \textbf{Group-FE} & \textbf{bcRI} & \textbf{bcRegFE} & \textbf{Cond-LR} & \textbf{Firth} \\
	\hline
		15 & 5 & 0.819 & 1.039 & 1.394 & 0.915 & 0.840 & 0.959 & 0.766 \\
        \hline
		50 & 5 & 0.573 & 0.739 & 0.663 & 0.422 & 0.386 & 0.435 & 0.400 \\
        \hline
		15 & 15 & 0.618 & 0.607 & 0.436 & 0.391 & 0.374 & 0.391 & 0.380 \\
        \hline
		50 & 15 & 0.530 & 0.528 & 0.245 & 0.211 & 0.205 & 0.212 & 0.210 \\
        \hline
		15 & 25 & 0.602 & 0.498 & 0.324 & 0.303 & 0.293 & 0.303 & 0.298 \\
        \hline
		50 & 25 & 0.528 & 0.423 & 0.179 & 0.165 & 0.161 & 0.165 & 0.164 \\
        \hline
		15 & 50 & 0.578 & 0.328 & 0.217 & 0.210 & 0.207 & 0.211 & 0.209 \\
        \hline
		50 & 50 & 0.520 & 0.274 & 0.117 & 0.112 & 0.112 & 0.113 & 0.112 \\
        \hline
\end{tabular}
\end{center}
\normalsize
   \subcaption*{\textit{Note:} Results across 1000 iterations at each sample size of \ref{DGP1}. Comparison of RMSE in estimating $\beta_1$ from a base logistic regression model (GLM) that does not include any group-varying intercepts ($\gamma_g$); uncorrected RI (RI); uncorrected Group-FE (Group-FE); bcRI; varying intercepts bcRegFE (bcRegFE); conditional logistic regression that stratifies by group (Cond-LR); and a Group-FE model with Firth's bias correction (Firth). } 
\end{table}

\newpage
\clearpage

\includepdf[pages=-]{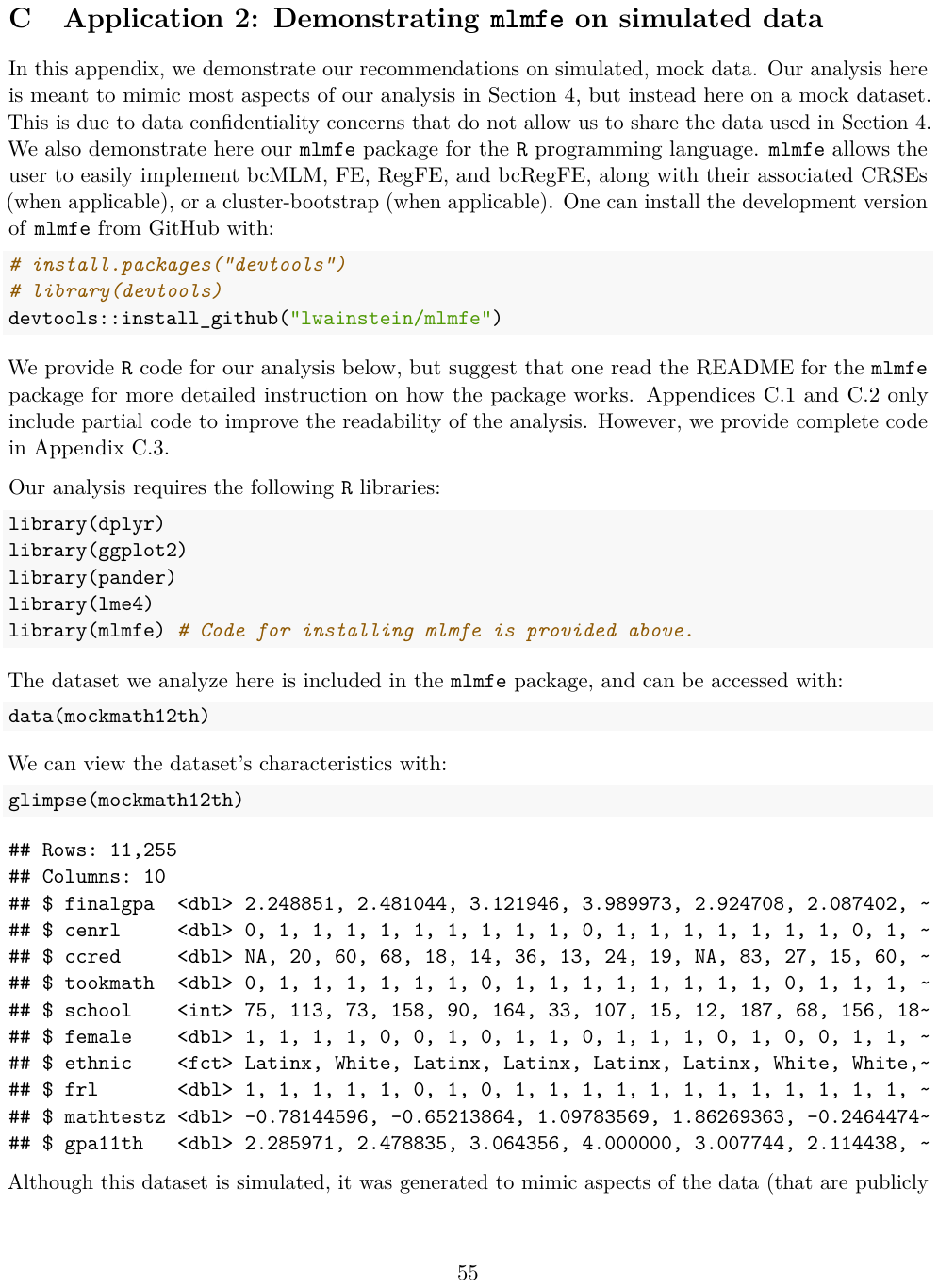}

\end{document}